\pdfoutput=1
\documentclass[preprint,fleqn,11pt,numbers,sort&compress,3p]{elsarticle}
\usepackage{graphicx}
\usepackage{amssymb,amsmath}
\usepackage{natbib}
\usepackage{hyperref}
\usepackage{mathtools,cuted}
\usepackage{commath}
\usepackage{siunitx}
\hypersetup{pdftitle = The title of my PDF, pdfauthor = My name, pdfsubject= The subject, pdfkeywords = keyword1 keyword2 keyword3}
\journal{New Astronomy}
\begin{document}
\begin{frontmatter}
\title{Unveiling the basins of convergence in the pseudo-Newtonian planar circular restricted four-body problem}
\author[mss]{Md Sanam Suraj\corref{cor1}}
\ead{mdsanamsuraj@gmail.com}
\author[eez]{Euaggelos E. Zotos}
\author[ra]{Rajiv Aggarwal}
\author[am]{Amit Mittal}
\cortext[cor1]{Corresponding author}
\address[mss]{Department of Mathematics,
Sri Aurobindo College, University of Delhi, Delhi, India}
\address[eez]{Department of Physics, School of Science,
Aristotle University of Thessaloniki, GR-541 24, Thessaloniki, Greece}
\address[ra]{Department of Mathematics,
Deshbandhu College, University of Delhi, Delhi, India}
\address[am]{Department of Mathematics, ARSD College,
University of Delhi, New Delhi, India}
\begin{abstract}
The dynamics of the pseudo-Newtonian restricted four-body problem has been studied in the present paper, where the primaries have equal masses. The parametric variation of the existence as well as the position of the libration points are determined, when the value of the transition parameter $\epsilon \in [0, 1]$. The stability of these libration points has also been discussed. Our study reveals that the Jacobi constant as well as transition parameter $\epsilon$ have substantial effect on the regions of possible motion, where the fourth body is free to move. The multivariate version of Newton-Raphson iterative scheme is introduced for determining the basins of attraction in the configuration $(x,y)$ plane. A systematic numerical investigation is executed to reveal the influence of the transition parameter on the topology of the basins of convergence. In parallel, the required number of iterations is also noted to show its correlations to the corresponding basins of convergence. It is unveiled that the evolution of the attracting regions in the pseudo-Newtonian restricted four-body problem is a highly complicated yet worth studying problem.
\end{abstract}
\end{frontmatter}
\section{Introduction}\label{Introduction}
Over the past decades, the few-body problem and more precisely the planar circular restricted four-body problem remains one of the most fascinated as well as important problems in celestial mechanics and dynamical astronomy. This problem has various applications in the research field, such as planetary physics, space sciences, astrodynamics, or even in stellar and galactic dynamics.

In order to obtain a more realistic description of the dynamics of the test particle, various perturbing terms have been included to the effective potential in the classical restricted four-body problem. The classical Newtonian three and four-body problems with perturbations are studied by many scientists. The existence of equilibrium points in the restricted problem of three bodies (e.g., \cite{kum86}), their stability (e.g., \cite{abo12}), the existence of periodic orbits around the equilibrium points (e.g., \cite{aboandsha12}), the stability of the libration points in linear and non-linear sense with heterogeneous primaries (e.g., \cite{Sur14}, \cite{sha17}), the restricted three-body problem by taking smaller primary as an ellipsoid (e.g., \cite{jav13}), and the basins of convergence associated with the libration points in the photogravitational restricted problem of three bodies (e.g., \cite{zot16b}) are some characteristic examples. The effect of the oblateness of the primaries (e.g., \cite{sha75}), the radiation due to the primaries (e.g., \cite{bha79}) and the effect of small perturbations in the Coriolis and centrifugal forces (e.g., \cite{bha78, bha83}), the copenhagen problem with repulsive Manev potential (e.g., \cite{sur18b})  are the most important perturbations that have been considered.

In the same vein, the restricted four-body problem has also investigated by many scientists including various perturbations (e.g., \cite{cha15a, cha15b, cha16, cha17},  \cite{pap13}, \cite{pap16}, \cite{she13}, \cite{Sur14b, Sur17a, Sur17b, Sur18a}, \cite{zot16b, zot17a}, \cite{che17}), with variable mass (e.g., \cite{mit16}, \cite{mit18}, \cite{agg18}).

A series of research papers (e.g., \cite{kre67}; \cite{bru72}; \cite{con76}) are available on the $1^{st}$ order post-Newtonian equations of motion for the restricted problem of three bodies which are deduced by using the Einstein--Infeld--Hoffmann theory (e.g., \cite{ein38}, \cite{dub17b}). \cite{dub17a} studied the dynamics of the planar circular restricted problem of three-bodies in the context of a pseudo-Newtonian approximation by using the Fodor--Hoenselaers--Perj\'{e} procedure, while \cite{hua14} examined the influence of the separation between the primaries. It is concluded that the post-Newtonian dynamics substantially differ from the corresponding classical Newtonian dynamics provided the distance between the primaries is sufficiently small.

The most intrinsic attributes of a dynamical system are revealed by the attracting domains, which make the study of basins of convergence associated with the libration points extremely important. The multivariate version of the Newton-Raphson iterative scheme is, without any doubt, an important method to find the basins of convergence. We scan sets of initial conditions in order to unveil to which attractor these initial condition converge. In the past few years, a series of research papers emerged on the study of Newton-Raphson basins of convergence in various dynamical system, such as the restricted three-body problem including oblateness and radiation pressure (e.g., \cite{zot16a}, \cite{zot17d}), the restricted four-body problem (e.g., \cite{zot17a}; \cite{Sur17a}; \cite{bal11a}), the electromagnetic Copenhagen problem (e.g., \cite{zot17b}), pseudo-Newtonian three-body problem (e.g., \cite{zot17c}), the ring problem of $N+1$ bodies (e.g., \cite{cro07}), and  the restricted 2+2 body problem (e.g., \cite{cro13}).

Very recently, \cite{zot17c} discussed the basins of convergence associated with the libration points in the pseudo-Newtonian planar restricted three-body problem using the multivariate version of the Newton-Raphson iterative scheme. Furthermore, it was revealed that the transition parameter strongly influences the topology of the basins of convergence. It is observed that the total number of the libration points in the pseudo-Newtonian restricted problem of three bodies, with equal masses,  strongly depends on the value of the transition parameter $\epsilon$.

In the present work, we model a system composed by three pseudo-Newtonian primaries pinpointed at the vertices of an equilateral triangle while performing circular orbits around their common center of mass. The fourth body, of infinitesimal mass, is moving in the same plane of motion of the primaries and it is assumed that it does not influence the motion of the primaries. The dynamics of the fourth body is studied by introducing the new parameter $\epsilon$, which allows us to unveil the transition from the Newtonian to the pseudo-Newtonian system.

It is the first time when the pseudo-Newtonian planar circular restricted four-body problem is investigated numerically in a systematic manner to reveal the influence of the transition parameter $\epsilon$ on the existence and the stability of the libration points as well as on the topology of the basins of convergence corresponding to these points and on the regions of possible motion. On this basis, the presented outcomes are novel and this is precisely the contribution of our work.

The paper is organized as follows: In Section \ref{Description of the mathematical model}, we present the description of the mathematical model and the equations of motion of the fourth body moving under the gravitational effect of three equal primaries. The parametric evolution of the position of the libration points and their linear stability are investigated in Section \ref{Parametric evolution of the libration points and their stability}. Section \ref{Regions of possible motion} deals with the zero velocity curves of the proposed model. In Section \ref{Newton-Raphson Basin of attractions}, we present a systematic numerical exploration by unveiling the Newton-Raphson basins of attraction in the restricted four-body problem and how they are affected by the transition parameter. This paper ends with  Section \ref{Discussion}, where the discussion and conclusion of the problem are presented.
\section{Description of the mathematical model}\label{Description of the mathematical model}
\begin{figure}[ht]
\begin{center}
\resizebox{0.6\hsize}{!}{\includegraphics*{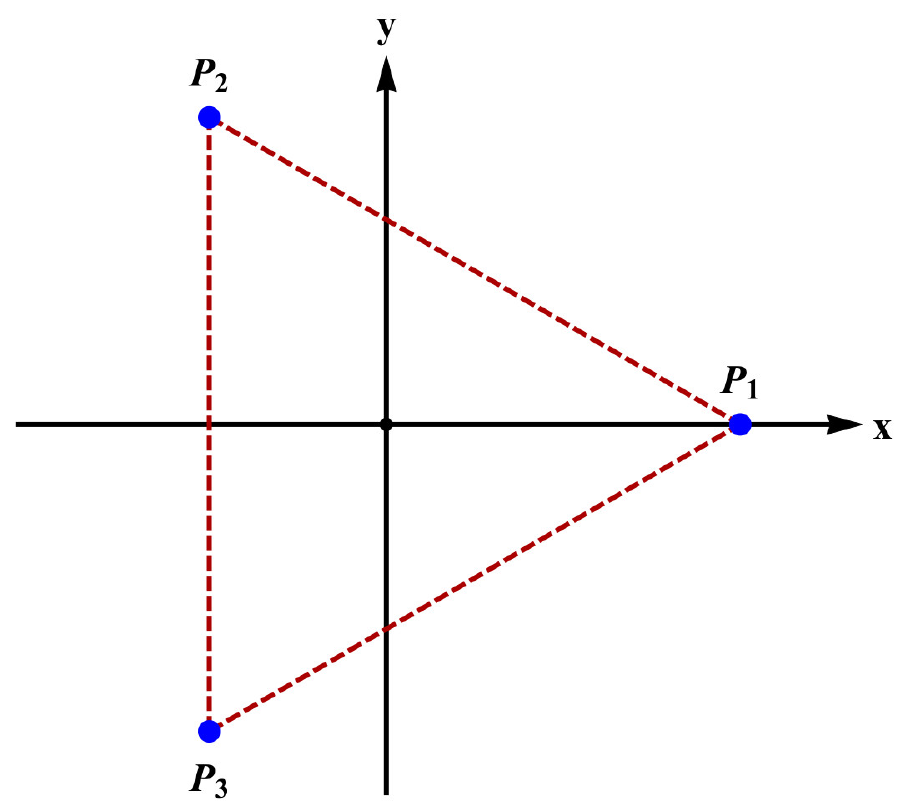}}
\end{center}
\caption{The configuration of the planar circular restricted four-body problem (CR4BP). The positions of the centers of the three primary bodies, with equal masses, are located at the vertices of an equilateral triangle and indicated by blue dots.} 
\label{fig:1}
\end{figure}

We adopt a rotating rectangular system by taking the origin as the center of mass of the primaries, which rotates with a uniform angular velocity, so that the centers of the three primaries remain fixed on the $(x,y)$ plane. Without loss of the generality, we assign the primary of mass $m_1$ on the positive $x-$axis at $P_1(x_1, y_1)$. The remaining two primaries, with masses $m_2$ and $m_3$, respectively, are situated at $P_2(x_2, y_2) $ and $P_3(x_3, y_3) $, where $x_1=\sqrt{3}\mu$, $x_2=-\frac{\sqrt{3}(1-2\mu)}{2}=x_3$, $y_1=0$, and $y_2=-y_3=\frac{1}{2}$  while $\mu$ denotes the mass parameter (see Fig.\ref{fig:1}). We scale the units by taking the sum of the masses and the distance between the primaries both equal to unity. Therefore, $m_1=1-2\mu$ and $m_2=m_3=\mu$ with $m_1+m_2+m_3=1$. Also, the scale of the time is chosen so that the gravitational constant is unity.
Regarding the value of the mass parameter, there are three limiting cases (see  \cite{zot16b}). In the present study, we shall discuss the case when $\mu=\frac{1}{3} $ i.e., the symmetric case of three primaries with equal masses $m=\frac{1}{3}$.

The time-independent effective potential function of the planar circular restricted problem of four bodies in the context of a pseudo-Newtonian approximation, with only the first correction terms, is\footnote{The speed of light $c$ is explicitly introduced in the pseudo-Newtonian potential in order to reveal its contributions. However, we set $c=1$ in the numerical simulations.}
\begin{eqnarray}\label{Eq:1}
\Omega&=&\frac{1}{2}(x^2+y^2)+\sum_{i=1}^{3}\frac{m_i}{r_{i}}-\frac{\epsilon}{2c^4}\sum_{i=1}^{3}\frac{m_i^3}{r_{i}^3},
\end{eqnarray}
where the $(x,y)$ are the coordinates of the infinitesimal mass $m_4$ on the configuration plane with
\begin{equation}
r_{i}=\sqrt{(x-x_i)^2+(y-y_i)^2}, (i=1, 2, 3),
\end{equation}
the distances of the infinitesimal mass $m_4$ from the three primary bodies $m_i, i=1, 2, 3$ respectively.

The scaled equations of motion depicting the dynamics of the infinitesimal mass in the synodical coordinates $(x, y)$ are
\begin{eqnarray}\label{Eq:2}
\ddot{x}-2\dot{y}&=&\Omega_{x},\nonumber\\
\ddot{y}+2\dot{x}&=&\Omega_{y},
\end{eqnarray}
where
\begin{eqnarray}
\label{Eq:3}
  \Omega_x &=& x -\sum_{i=1}^{3}\frac{m_i (x- x_i)}{r_i^3}+\frac{3\epsilon}{2c^4}\sum_{i=1}^{3}\frac{m_i^3 (x-x_i)}{r_i^5},\\
  \label{Eq:4}
  \Omega_y &=& y-\sum_{i=1}^{3}\frac{m_i (y-y_i)}{r_i^3}+\frac{3\epsilon}{2c^4}\sum_{i=1}^{3}\frac{m_i^3 (y-y_i)}{r_i^5}.
\end{eqnarray}\\
Similarly, the second order partial derivatives, read as:
\begin{align}
\label{Eq:5}
  \Omega_{xx}&=1-\sum_{i=1}^{3}\frac{m_i(r_i^2-3(x-x_i)^2)}{r_i^5}+\frac{3\epsilon}{2c^4}\sum_{i=1}^{3}\frac{m_i^3(r_i^2-5(x-x_i)^2)}{r_i^7},\nonumber\\
  &\\
  \Omega_{yy} &=1-\sum_{i=1}^{3}\frac{m_i(r_i^2-3(y-y_i)^2)}{r_i^5}+\frac{3\epsilon}{2c^4}\sum_{i=1}^{3}\frac{m_i^3(r_i^2-5(y-y_i)^2)}{r_i^7},\nonumber\\
  &\\
  \Omega_{xy} &=3\sum_{i=1}^{3}\frac{m_i(x-x_i)(y-y_i)}{r_i^5}+\frac{15\epsilon}{2c^4}\sum_{i=1}^{3}\frac{m_i^3(x-x_i)(y-y_i)}{r_i^7}=\Omega_{yx}.
\end{align}
For the equations (\ref{Eq:2}), there exists
exactly one integral of motion known as Jacobi integral
which is constituted by the following Hamiltonian:
\begin{equation}\label{Eq:9}
J(x, y, \dot{x}, \dot{y})=2\Omega(x, y)-(\dot{x}^2+\dot{y}^2) =C,
\end{equation}
where the velocities are represented by $\dot{x}$ and $\dot{y}$, while the  Jacobi constant is represented by $C$. This Jacobian constant is conserved and defines a 3-dimensional invariant manifold in the total 4-dimensional phase-space. Thus, the test particle with a given value of its orbital energy is confined to move inside the regions in which $2\Omega(x,y)\geq C$, while the rest of the region on the configuration $(x,y)$ plane is energetically prohibited for the test particle. In the canonical co-ordinates, the value of Hamiltonian corresponds to Jacobian integral is known as the total orbital energy $E$, which is related to Jacobian constant by $C=-2E$.

\section{Parametric evolution of the libration points and their stability}\label{Parametric evolution of the libration points and their stability}
\begin{figure*}[!t]
\begin{center}
(a)\resizebox{0.3\hsize}{!}{\includegraphics*{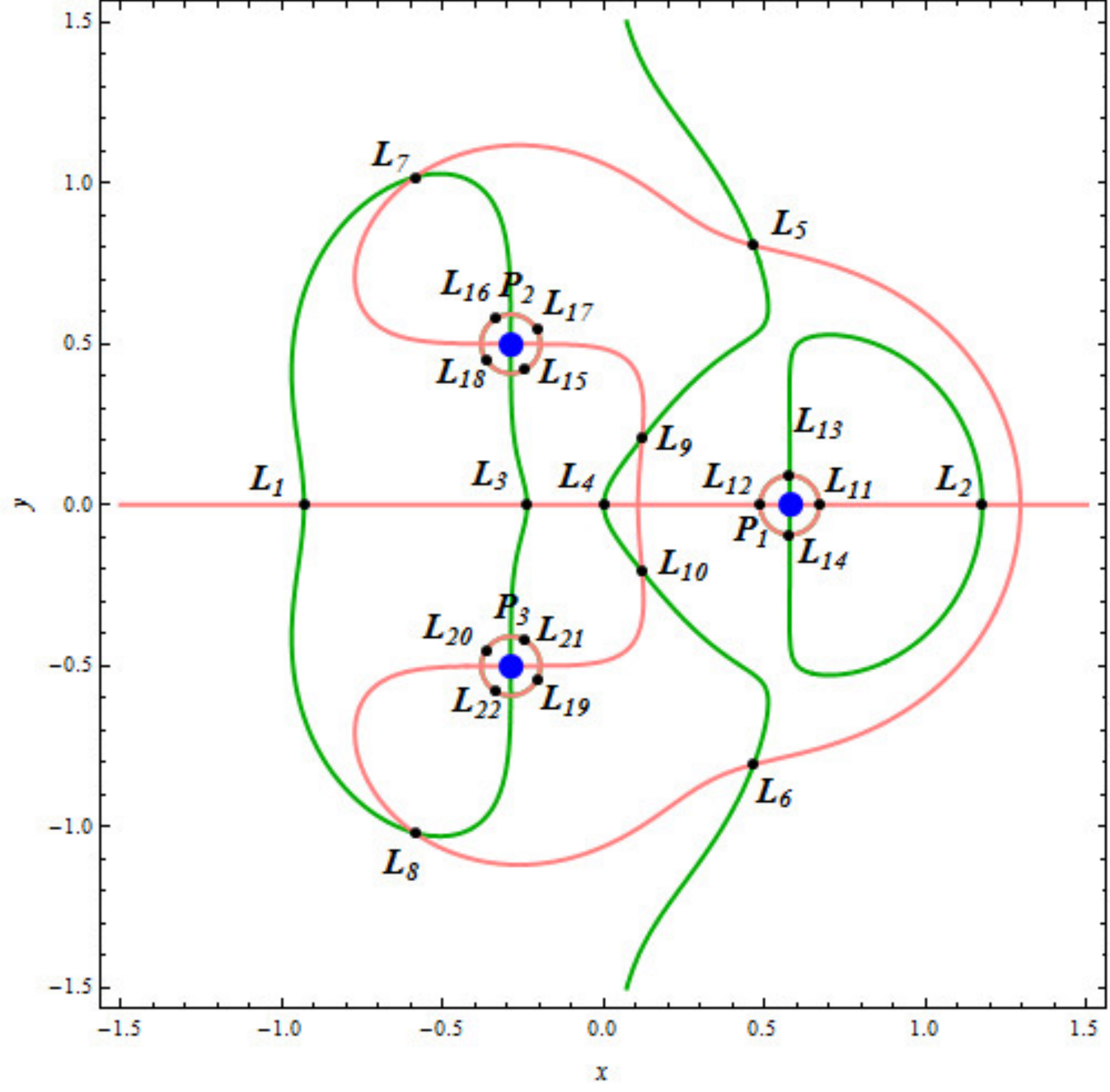}}
(b)\resizebox{0.3\hsize}{!}{\includegraphics*{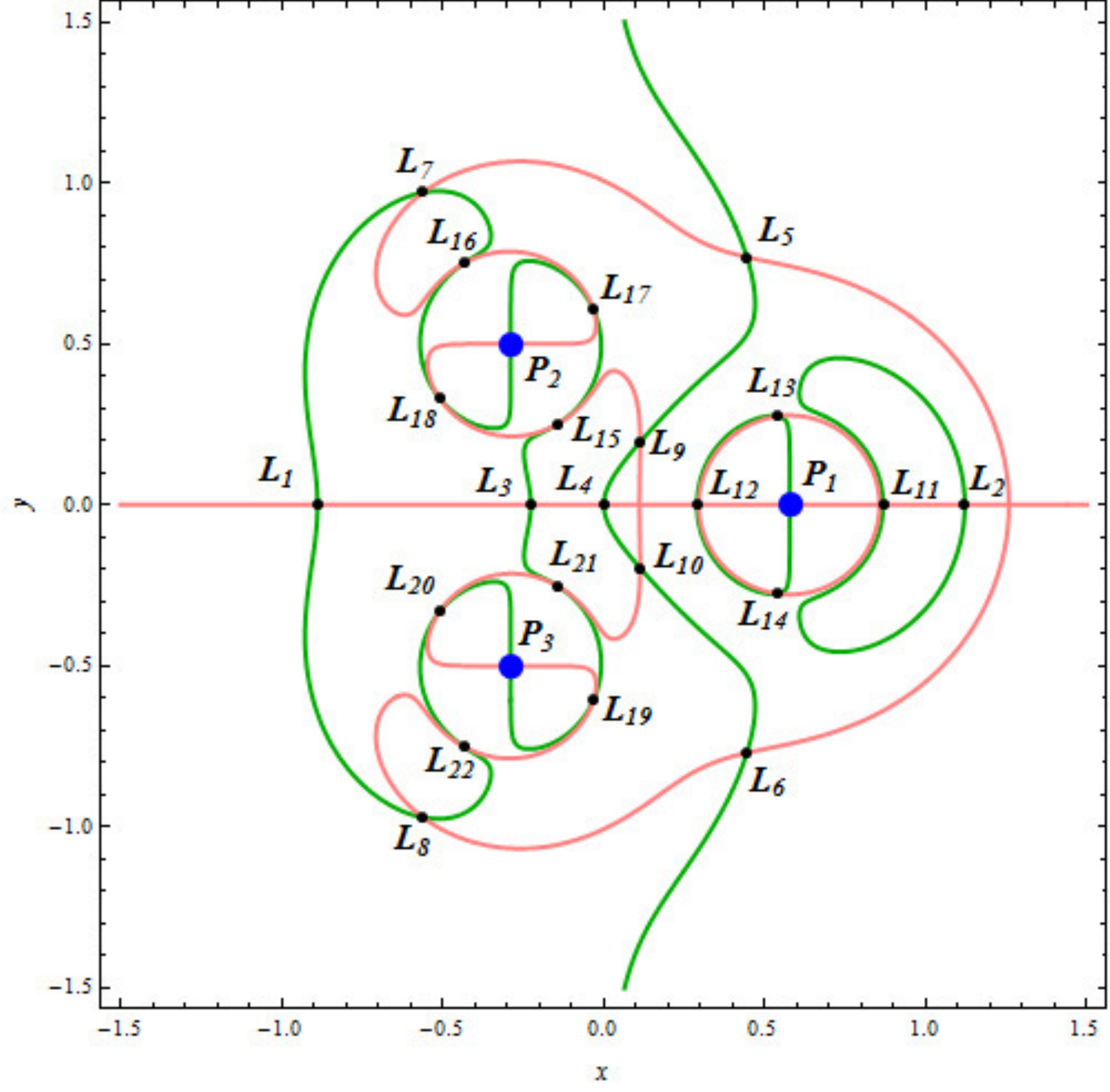}}
(c)\resizebox{0.3\hsize}{!}{\includegraphics*{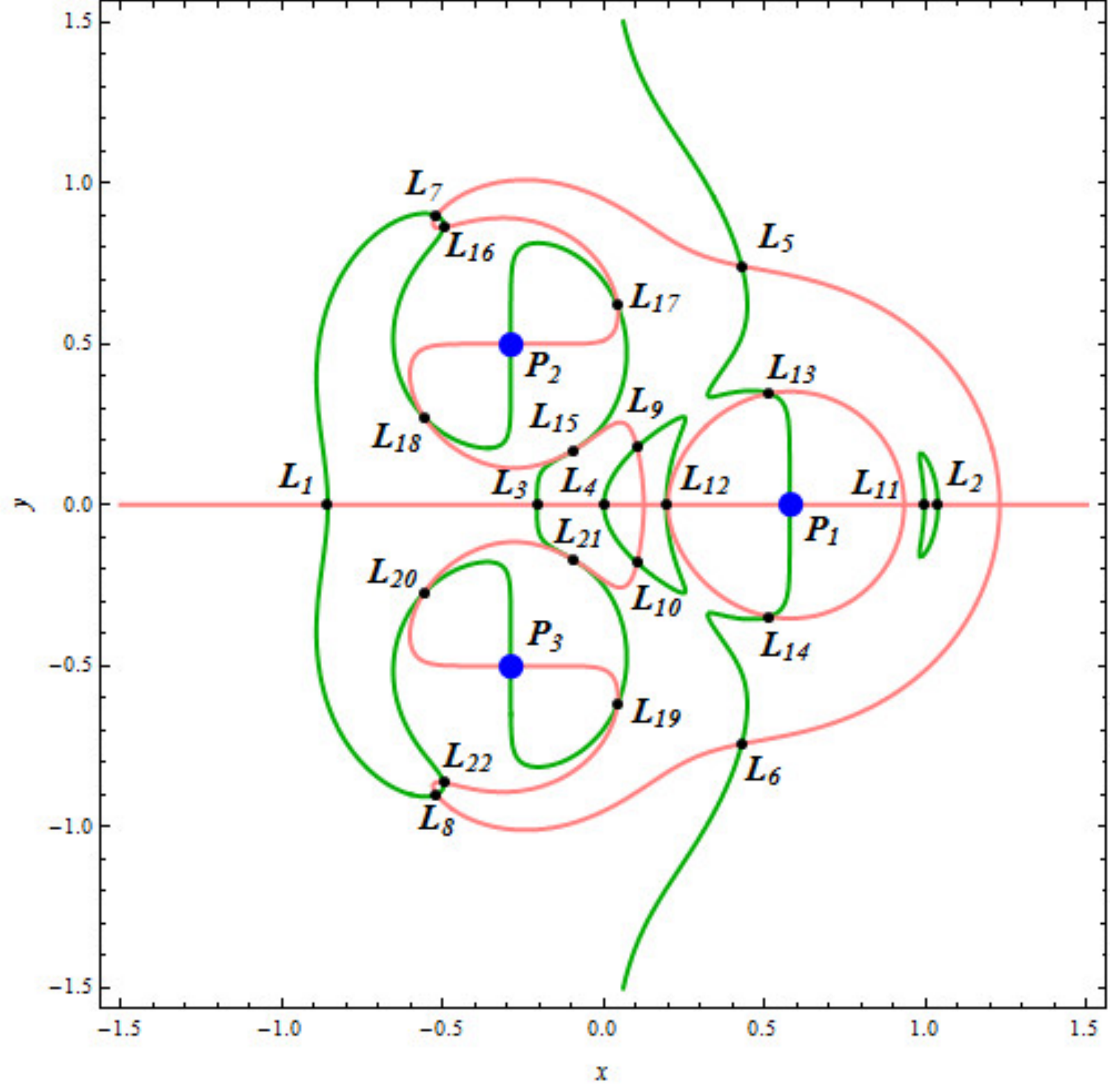}}\\
(d)\resizebox{0.3\hsize}{!}{\includegraphics*{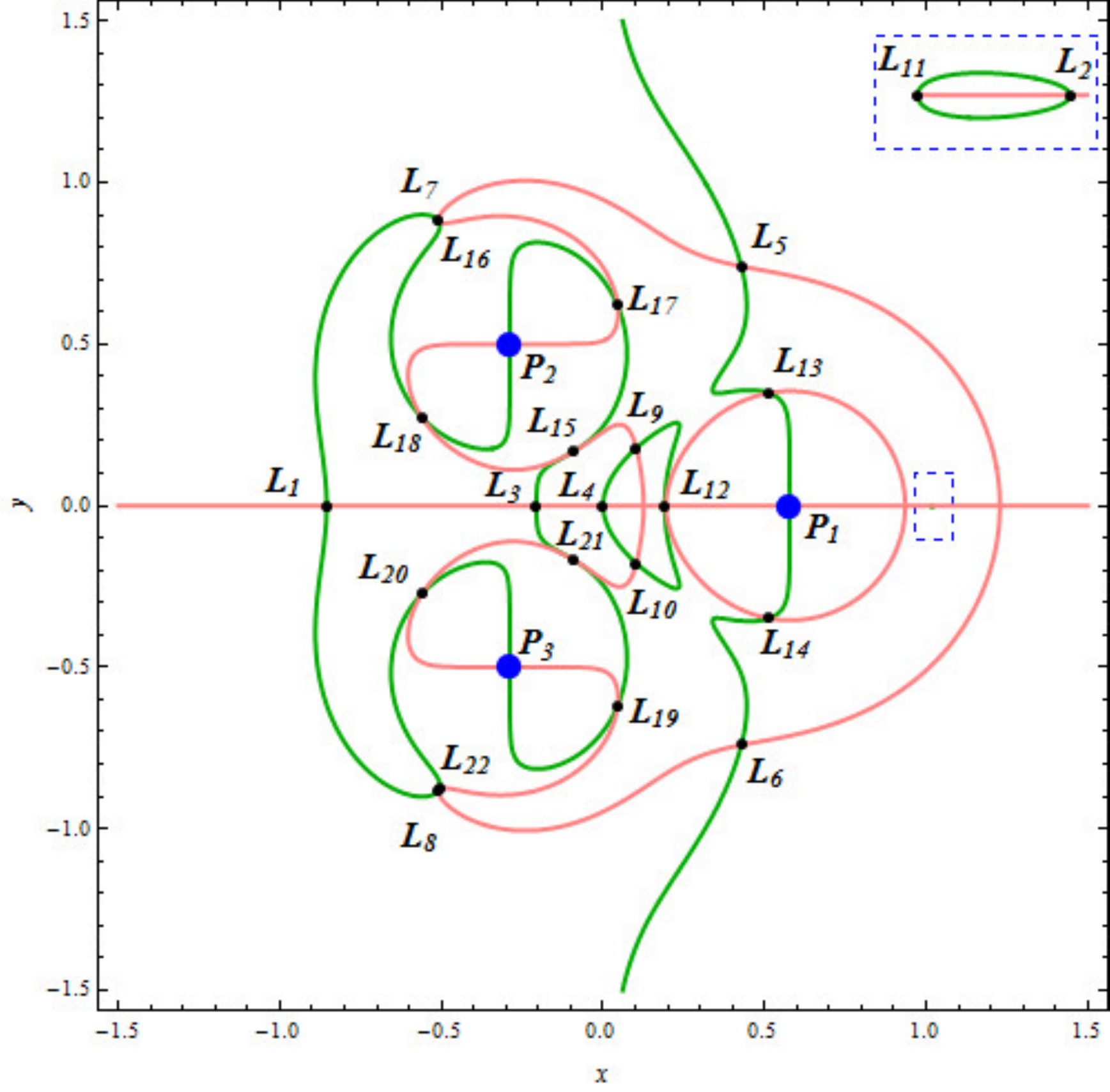}}
(e)\resizebox{0.3\hsize}{!}{\includegraphics*{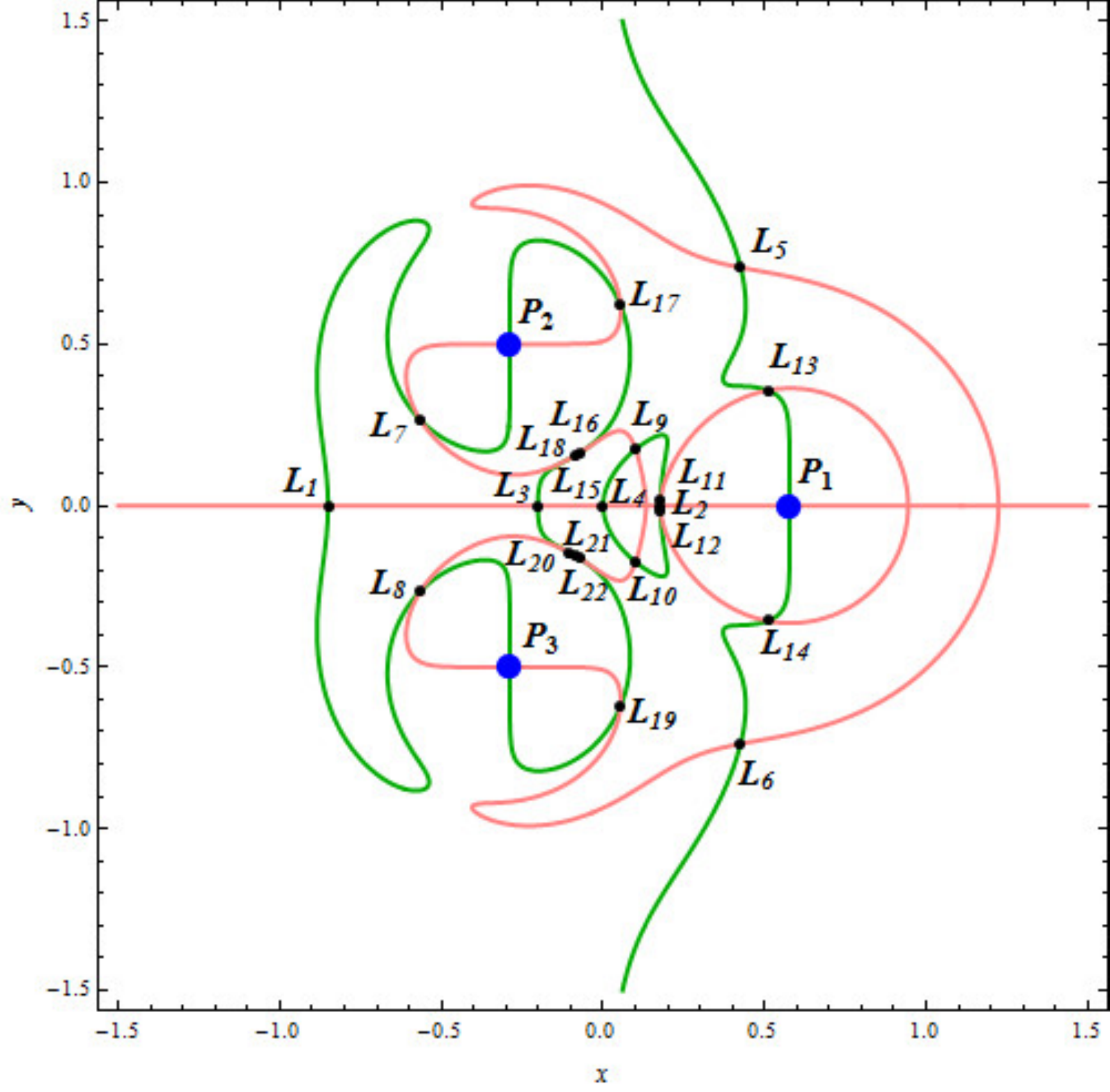}}
(f)\resizebox{0.3\hsize}{!}{\includegraphics*{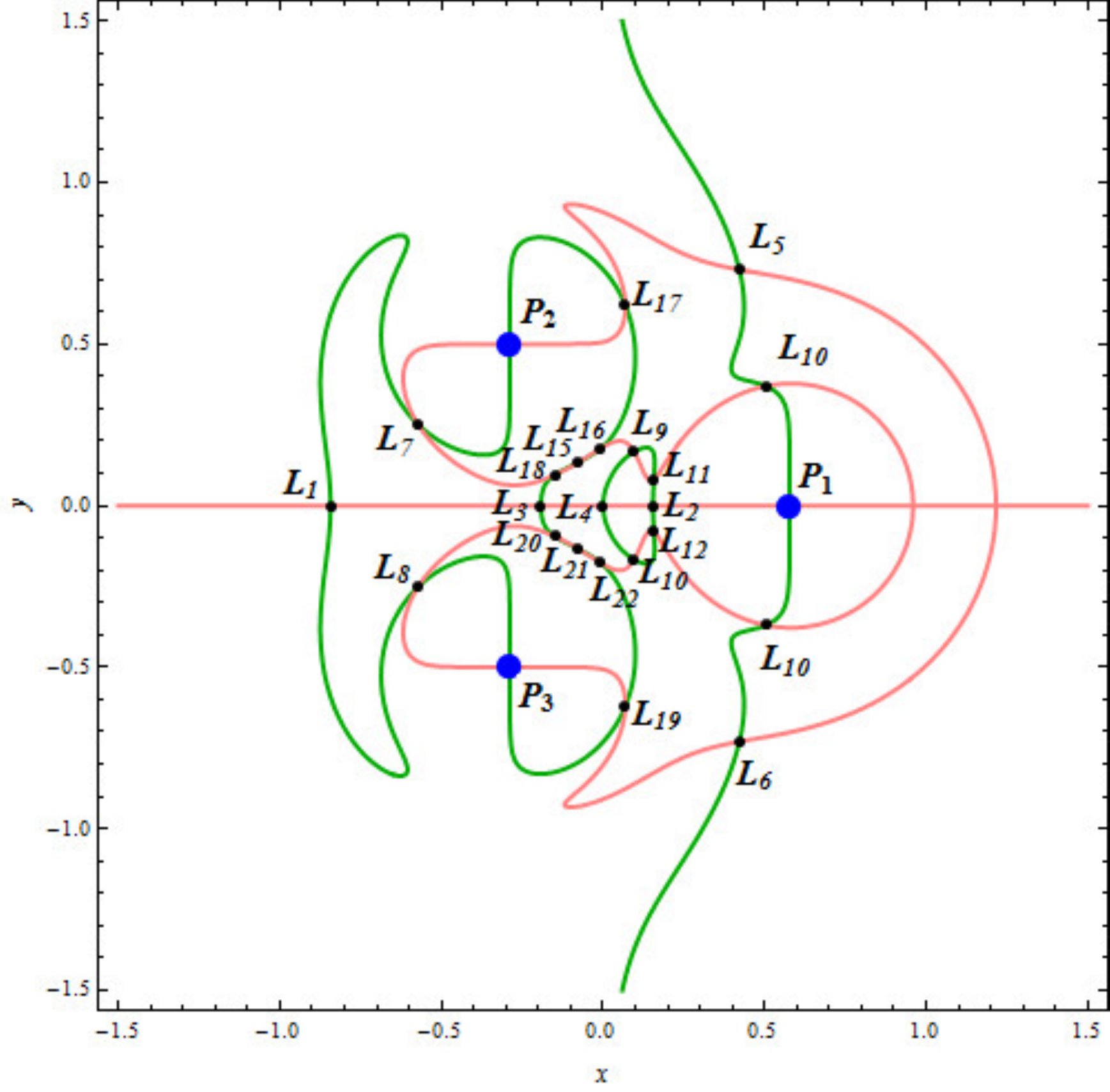}}\\
(g)\resizebox{0.3\hsize}{!}{\includegraphics*{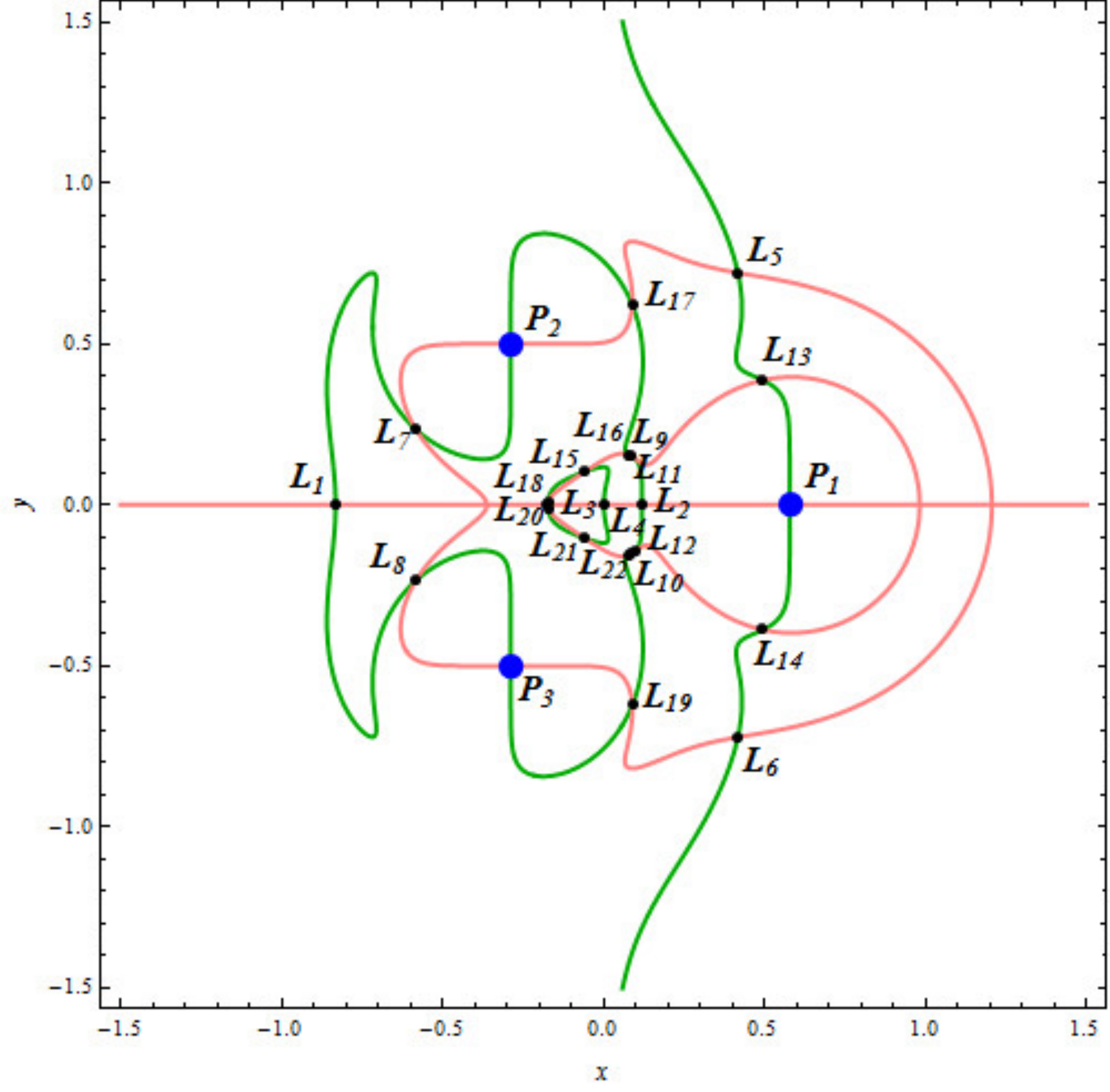}}
\end{center}
\caption{Locations of the positions and numbering of the libration points ($L_i, i=1,2,...,22$) through
the intersections of $\Omega_x = 0$ (green) and $\Omega_y = 0$ (pink), when (a): $\epsilon=0.05$, (b): $\epsilon=0.44$,
 (c): $\epsilon=0.67$,  (d): $\epsilon=0.67752839$, (e): $\epsilon=0.705528$, (f): $\epsilon=0.751528$,  (g): $\epsilon=0.812528$.
  The black dots pinpoint the positions of the libration points, while the blue dots denote the centers of the three primaries.
  (Color figure online).} 
\label{fig:2}
\end{figure*}

\begin{figure*}[!t]
\begin{center}
(a)\resizebox{0.3\hsize}{!}{\includegraphics*{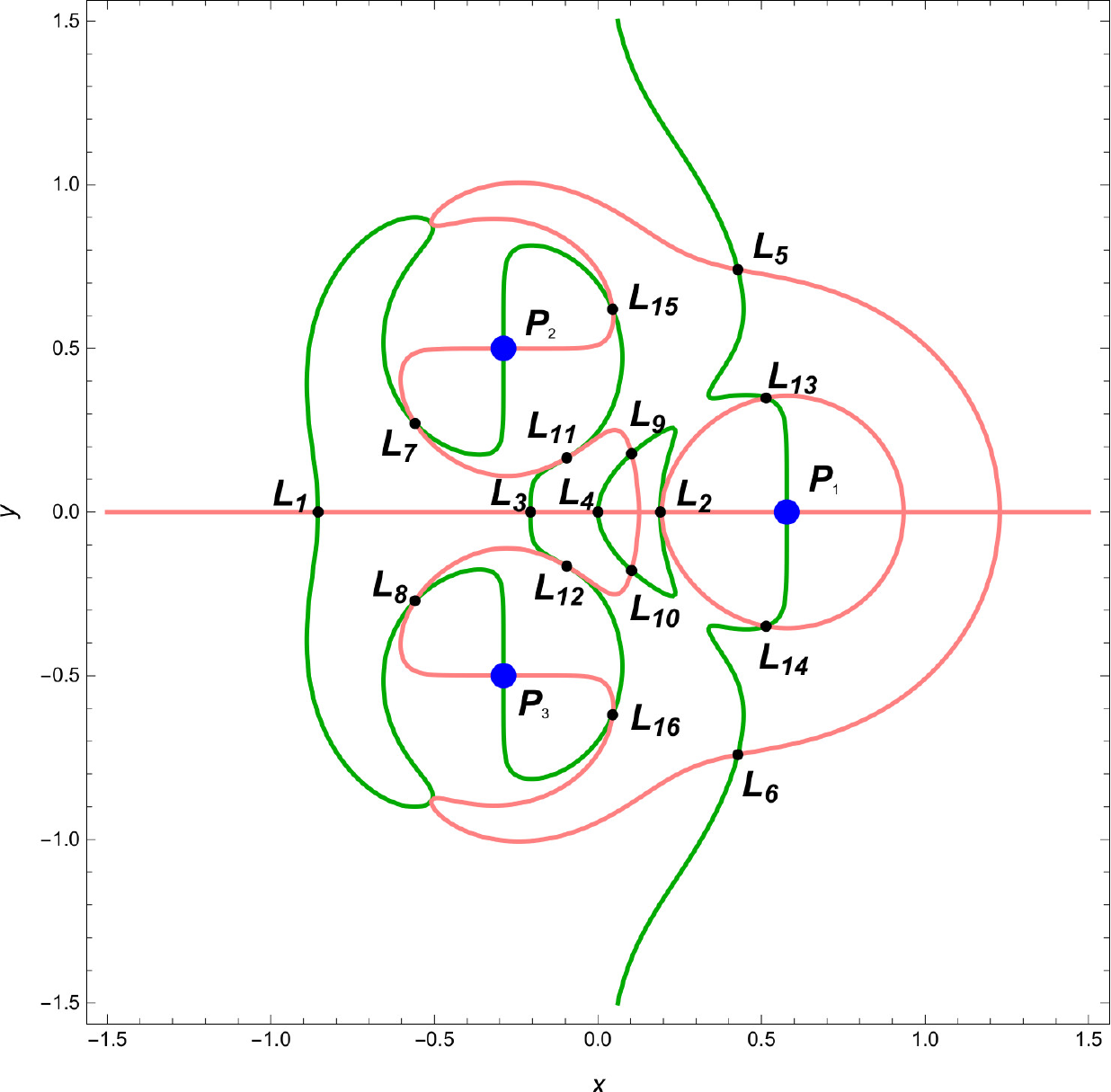}}
(b)\resizebox{0.3\hsize}{!}{\includegraphics*{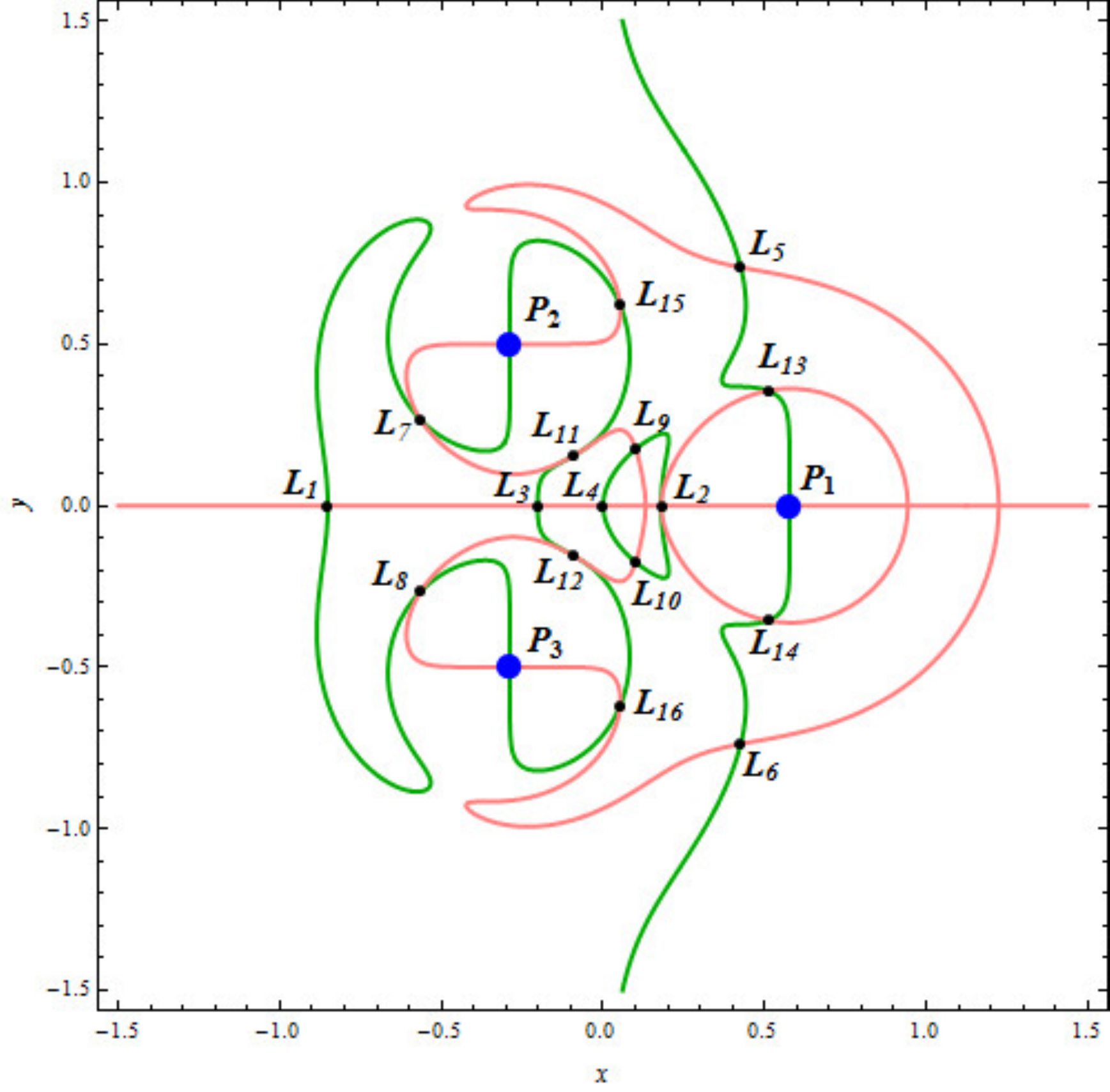}}
(c)\resizebox{0.3\hsize}{!}{\includegraphics*{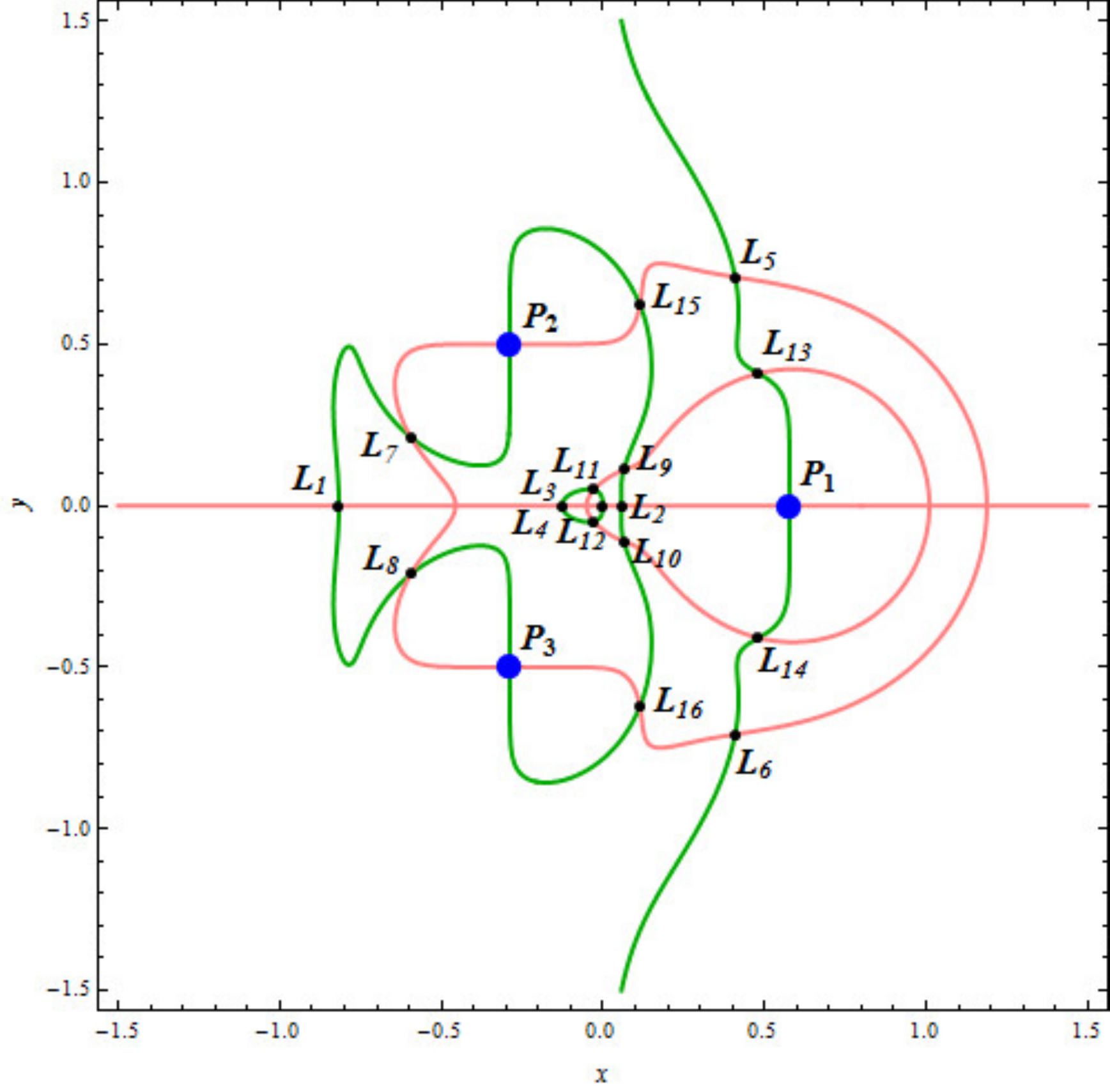}}
\end{center}
\caption{Locations of the positions and numbering of the libration points ($L_i, i=1,2,...,16$) through
the intersections of $\Omega_x = 0$
(green) and $\Omega_y = 0$ (pink),
when (a): $\epsilon=0.677618$, (b): $\epsilon=0.701528$, (c): $\epsilon=0.888528$. The black dots
 pinpoint the position of the libration points, while the blue dots denote the centers of the three primaries.
 (Colored figure online).} 
\label{fig:3}
\begin{center}
(a)\resizebox{0.4\hsize}{!}{\includegraphics*{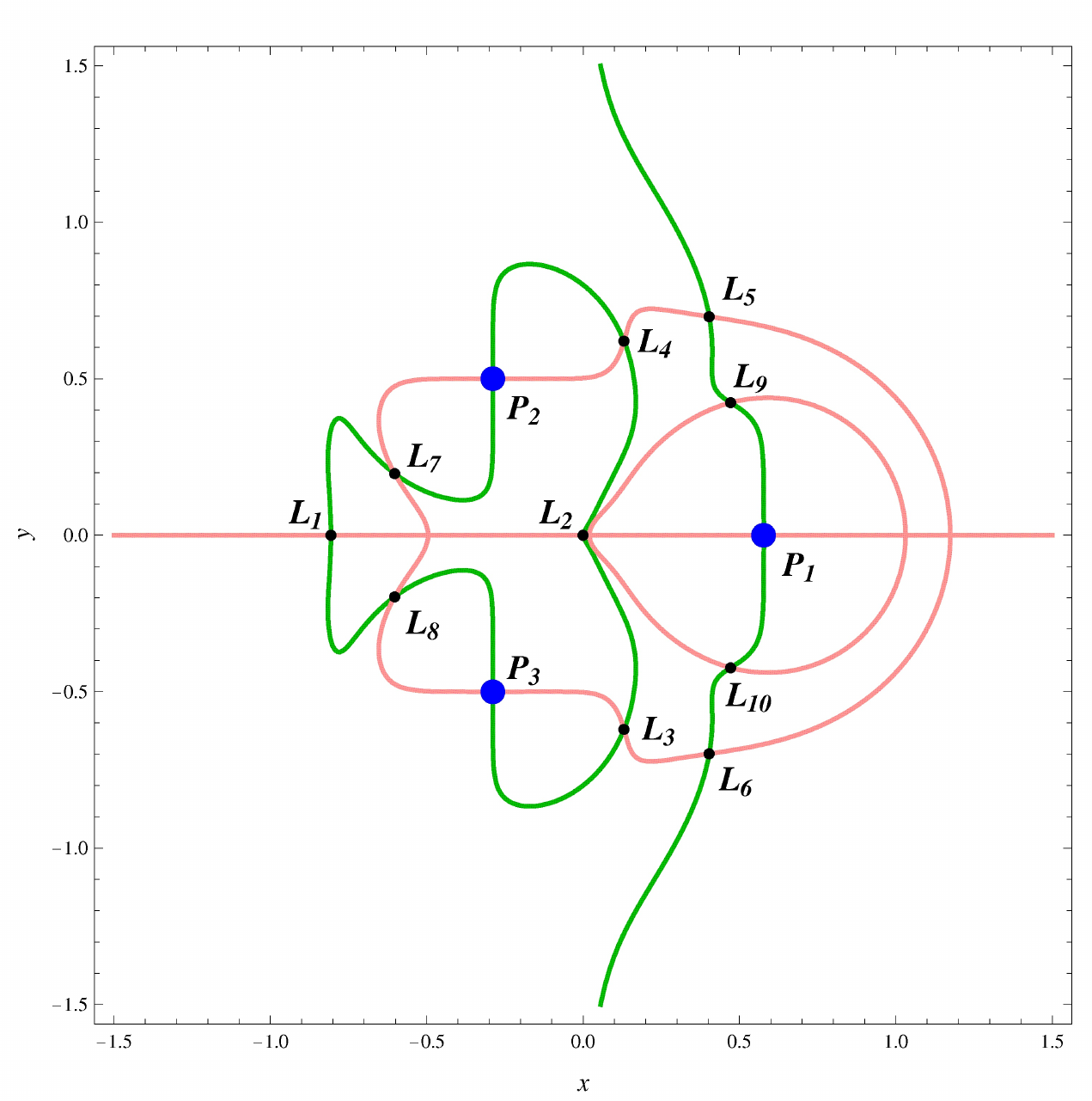}}
(b)\resizebox{0.4\hsize}{!}{\includegraphics*{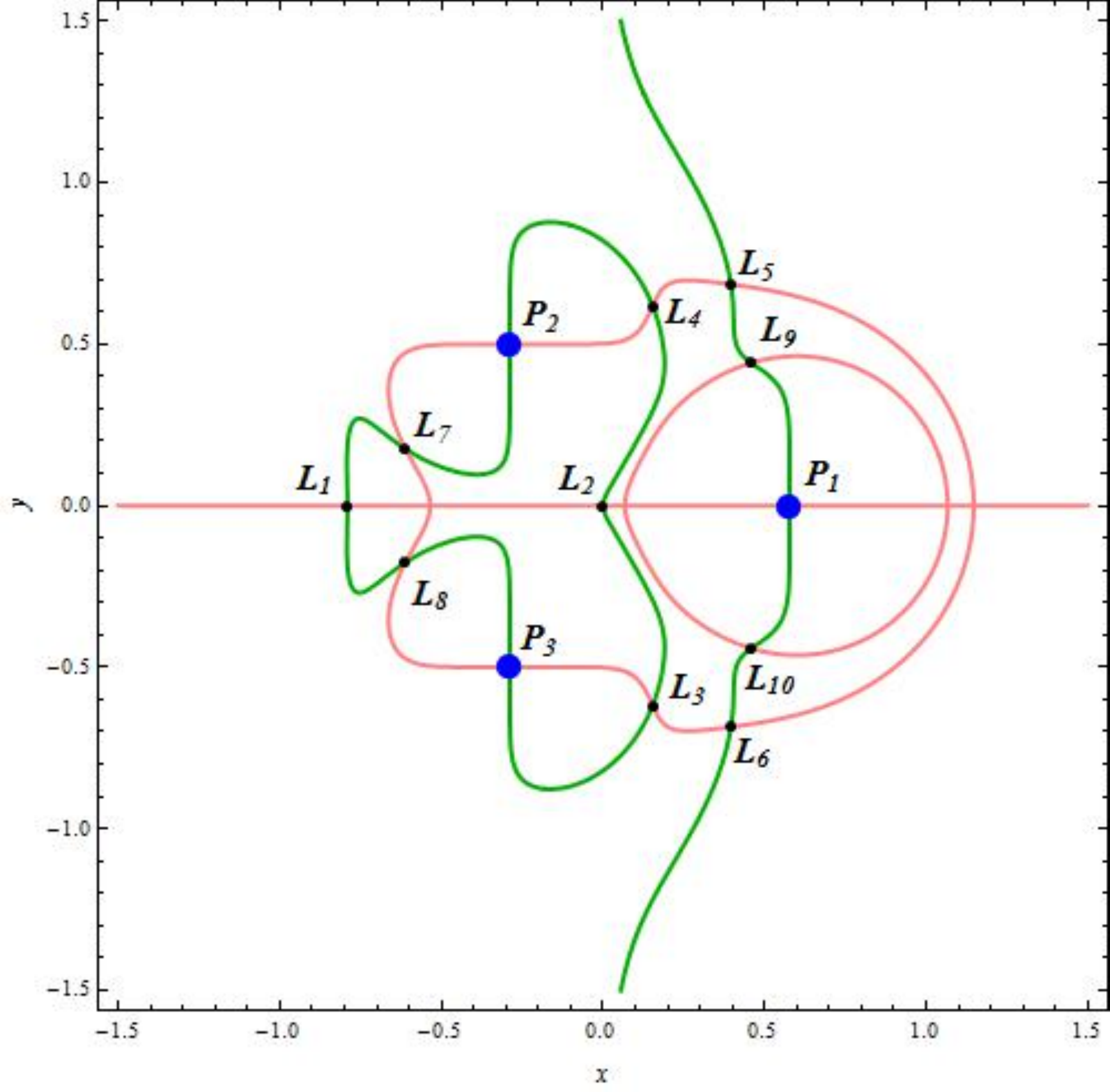}}
\end{center}
\caption{Locations of the positions and numbering of libration points ($L_i, i=1,2,...,10$) through
the intersections of $\Omega_x = 0$
(green) and $\Omega_y = 0$ (pink),
when (a): $\epsilon=0.937$, (b): $\epsilon=0.999$. The black dots pinpoint the position of the libration points, while the
blue dots denote the centers of the three primaries. (Colored figure online).} 
\label{fig:4}
\end{figure*}

The exact positions of the libration points are solutions of the system
\begin{align}
\dot{x} &=\dot{y}=\ddot{x}=\ddot{y}=0, \nonumber\\
\label{Eq:10}
\Omega_x &=\Omega_y=0.
\end{align}
The intersections of Eqs. $\Omega_x = \Omega_y = 0$
define the positions of the libration points. For various values of the transition parameter $\epsilon$ these libration points are presented in Figs. \ref{fig:2}, \ref{fig:3}, \ref{fig:4}.

It is observed that the total number of the libration points in the pseudo-Newtonian circular restricted four-body problem, with three equal masses, is not constant but it strongly depends on the value of the transition parameter $\epsilon$. More precisely
\begin{description}
  \item[*] When $\epsilon=0$, the problem corresponds to the case of the Copenhagen restricted four-body problem, so there exist ten libration points in which six are non-collinear libration points, while four are collinear.
  \item[*] When $\epsilon \in (0,  0.67752839]$ there exist twenty-two libration points
  \item[*] When $\epsilon \in (0. 0.67752839, 0.704528]$ there exist sixteen libration points.
  \item[*] When $\epsilon \in (0.704528, 0.812528]$ there exist twenty-two libration points.
  \item[*] When $\epsilon \in (0.812528, 0.929528]$ there exist sixteen libration points.
\item[*] When $\epsilon \in (0.929528, 1]$ there exist ten libration points.
\end{description}
The values $\epsilon=0.67752839$, $\epsilon=0.704528$, $\epsilon=0.812528$, and  $\epsilon=0.929528$ are the critical values of the transition parameter, since these values define the starting or end the points of the various intervals which contain the different number of the libration points. In Fig. \ref{fig:2}, we can observe how the intersections of the equations $\Omega_x=0$ and $\Omega_y=0$  define the positions of the libration points when the transition parameter $\epsilon$ varies only in those intervals for which 22 libration points exist. The exact positions of the libration points are marked by $L_i, i=1,2,...,22$. In Figs. \ref{fig:3} and \ref{fig:4}, we present the number and the exact positions of 16 and 10 libration points, respectively. The exact positions of the libration points are marked by $L_i, i=1,2,...,16$ and $i=1,2,...,10$ as the value of the transition parameter $\epsilon$ varies in the interval for which 16 and 10 libration points exist, respectively. It is interesting to note that all the libration points lie to the left of the primary $P_1$, when the transition parameter $\epsilon >  0.67752839$.

\begin{figure*}[!t]
\centering
\resizebox{\hsize}{!}{\includegraphics{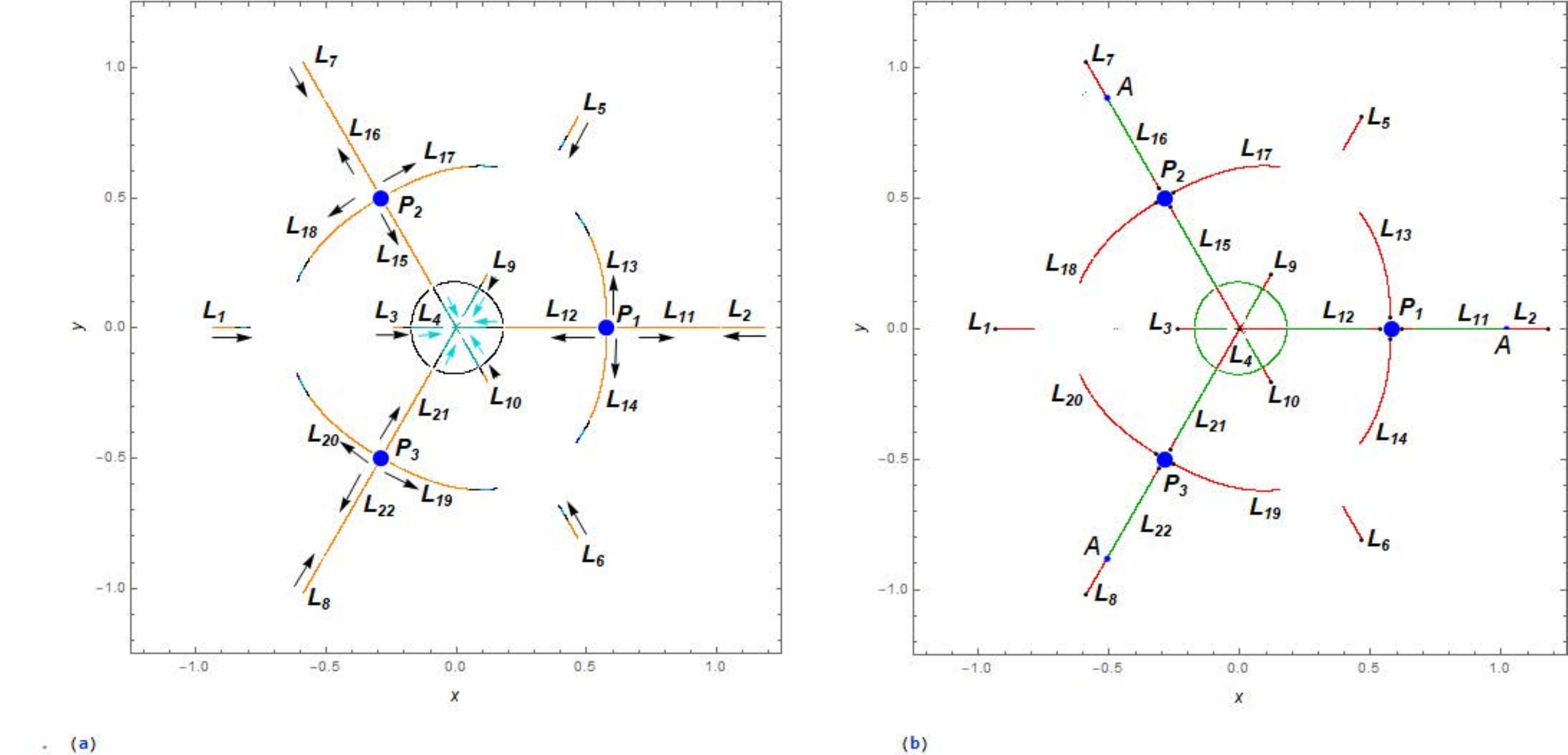}}
\caption{The parametric evolution of (a-left): the positions of 22 (\emph{orange} and \emph{black} color), 16 (\emph{cyan} color) and 10 (\emph{blue} color) libration points and (b-right):
the linear stability (green) or instability (red) of the libration points in the
pseudo-Newtonian planar circular restricted four-body problem with
equal masses, when $\epsilon \in (0, 1]$. The arrows indicate the movement direction
of the libration points as the value of the transition parameter
$\epsilon$ increases. The big blue dots pinpoint the fixed centers of three primaries, the small black dots show the twenty-two libration points for $\epsilon=0.05$, while the small blue dot (point A) correspond to the
value $\epsilon=0.677$. (Color figure online).} 
\label{fig:5}
\end{figure*}

It is clear that the number of the libration points and their existence in the pseudo-Newtonian four-body problem depend on the value of the transition parameter $\epsilon$. Therefore, we believe that it is necessary to describe the exact evolution of the locations of the libration points when $\epsilon \in (0,1]$. In Fig.(\ref{fig:5}), we have presented the parametric evolution of all the libration points, on the configuration $(x, y)$ plane. It is observed that for $\epsilon >0$, twelve libration points, in three sets of four, come forth from the centers of the primaries $P_i, i=1,2,3$. At this point, it is worth mentioning that in the Newtonian four-body problem when $\epsilon=0$, there exist four collinear libration points and six non-collinear libration points in total, while in the pseudo-Newtonian four-body problem when $\epsilon \in (0,  0.67752839],$ there exist six collinear libration points namely $L_{1,2,3,4,11,12}$ and twenty-two libration points in total.

When the value of the transition parameter $\epsilon \in (0,  0.67752839]$, the libration points $L_{11, 12, 13, 14}$ , $L_{15, 16, 17, 18}$  and $L_{19, 20, 21, 22}$ move away from the centers of the primaries $P_1, P_2,$ and $P_3$, respectively, while on the other hand the libration points $L_{2, 7, 8}$  move towards the centers of the primaries $P_1, P_2,$ and $P_3$, respectively. In particular, $L_{1, 3, 5, 6, 9}$ and $L_{10}$, move towards the central libration point $L_4$.

When the value of transition parameter $\epsilon \in (0.704528,  0.812528]$, there exist twenty-two libration points in which four are collinear, while twelve libration points exist on the circumference of a circle (see fig.\ref{fig:5}a in \emph{black} color) centered at $L_4$.\\
As the value of transition parameter $\epsilon \in ( 0.812528, 0.929528]$, there exist sixteen libration points in which four are collinear  (see fig. \ref{fig:5}a in persian cyan color). For $\epsilon = 0.929528$, the libration points $L_{2,3,11,12,15,16}$ collide with the libration point $L_4$ and decimated completely and the phenomenon of the introduction of new set of libration points originated for $\epsilon \in ( 0.929528, 1]$. Imminently as $\epsilon > 0.929528$, ten libration points exist in which two are collinear. However, the centers of the primaries $P_i, i=1,2,3$ remain unchanged by the change in the transition parameter $\epsilon$.

Moreover, to study the stability of the libration point in the linear sense, the origin of the reference plane is shifted at the exact location of the libration point $(x_0, y_0)$  by applying the transformation
\begin{equation}\label{Eq:11a}
x=x_0+\xi, \quad y=y_0+\eta.
\end{equation}
Now, expand the system of Eqs. (\ref{Eq:2}) and neglecting the terms of $O(\xi)>1$, and $O(\eta)>1$. 
The variational equations with variations $\xi$ and $\eta$ read
\begin{eqnarray}\label{Eq:11}
\ddot{\xi}-2\dot{\eta}&=&\Omega_{xx}^0\xi+\Omega_{xy}^0\eta,\nonumber\\
\ddot{\eta}+2\dot{\xi}&=&\Omega_{xy}^0\xi+\Omega_{xx}^0\eta,
\end{eqnarray}
where the superscript "0",  denotes the corresponding values are calculated at the libration point $(x_0, y_0)$.\\
The characteristic equation which is quadratic in $\Delta=\lambda^2$ corresponding to the linear system (\ref{Eq:11}) reads
\begin{equation}\label{Eq:13}
a_1\Delta^2+a_2\Delta+a_3=0,
\end{equation}
where
\begin{eqnarray}
  a_1 &=& 1, \\
  a_2 &=& 4-\Omega_{xx}^0- \Omega_{yy}^0,\\
  a_3 &=& \Omega_{xx}^0\Omega_{yy}^0-\Omega_{xy}^0\Omega_{yx}^0.
\end{eqnarray}
The libration points are said to be stable if the characteristic equation has only purely imaginary roots. This happens if the conditions
\begin{eqnarray}
  a_2>0, \quad
  a_3 >0,\quad
  a_2^2-4a_1a_3 > 0,
\end{eqnarray}
are satisfied simultaneously.

This leads to the fact that the characteristic equation (\ref{Eq:13}) has two real negative roots $\Delta_{1,2}$, which consequently ensures four purely imaginary roots for $\lambda$.

Therefore, through the nature of the four roots of the characteristic equation (\ref{Eq:13}), we can examine the linear stability of the libration points for the known positions of them. Our numerical study reveals that when $\epsilon \in (0,1]$ then the libration points are mostly either stable or unstable. In fig. \ref{fig:5}b, we have presented  the evolution of the stability of all the libration points, when $\epsilon \in (0,1]$.
It is observed that the central libration point $L_4$ is always stable, while the libration points $L_{2, 3, 9, 10}$ are stable when the sixteen libration points exist. $L_{11, 12, 15, 16, 21, 22}$ are stable when twenty-two libration points exist for $\epsilon \in (0.05, 0.67)$ and none of the libration point is stable except $L_4$ when ten libration points exist.

\section{Regions of possible motion}\label{Regions of possible motion}
The regions of motion on the configuration $(x, y)$ plane can be determined by using the surface $2\Omega(x, y)=C$ where the fourth body can move freely for a given value of the Jacobi constant $C$. The projection of this surface on the $(x, y)$ plane describes the zero velocity curves (ZVCs) of the restricted problem of four bodies. The ZVCs in the classical restricted four-body problem have been studied by many scientists for various values of the Jacobi constant $C$. In our case, the Jacobi constants calculated at the libration points $L_i$, $i=1,2,...,22$ for $22$ or $16$ or $10$ are critical values.
\begin{figure*}[!t]
\begin{center}
\resizebox{0.6\hsize}{!}{\includegraphics*{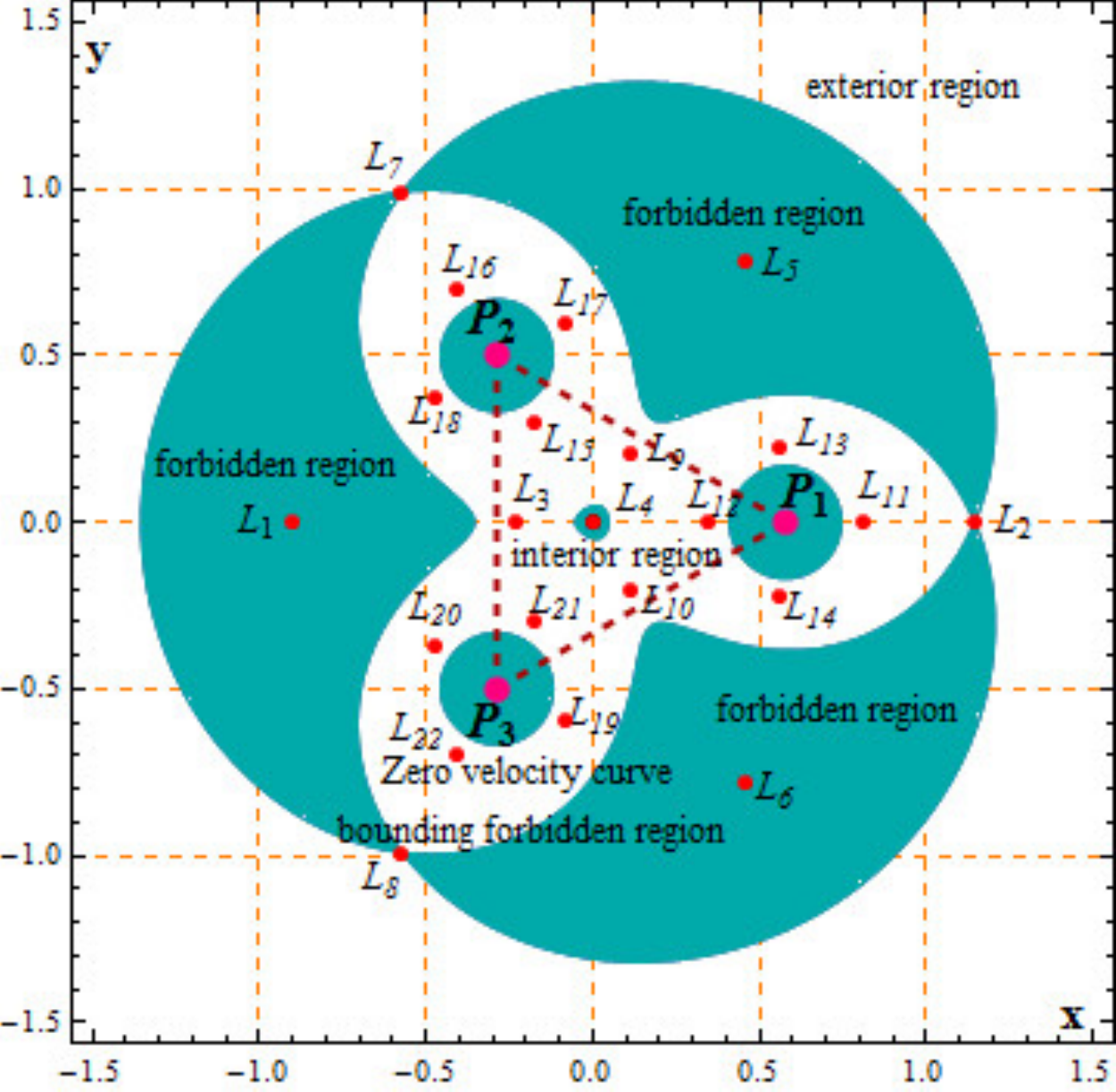}}
\end{center}
\caption{The zero velocity curves when 22 libration points exist. The interior and the exterior regions shown in \emph{white} color are the regions where the motion is possible, while the shaded regions shown in \emph{persian green } color correspond to the  forbidden regions. The equilateral triangle formed by the centres of three primaries is shown   in \emph{maroon dashed} line. The red dots indicate the 22 libration points, while\emph{ pink dots } show the centres of the three primaries. (Color figure online).} 
\label{fig:4a}
\end{figure*}
In Fig. \ref{fig:4a}, we have depicted the Hill's regions by projecting the $4-$dimensional phase-space onto the configuration $(x, y)$ plane which is divided into three regions: the interior, the exterior and the forbidden regions. The ZVCs are the boundaries of these Hill's regions, where the kinetic energy is zero. The interior as well as the exterior regions represent those regions where the infinitesimal mass is free to move, while the energetically forbidden regions correspond to the regions where the motion of the infinitesimal mass is not allowed. These regions strongly depend upon the value of the Jacobi constant as well as on the transition parameter.

\subsection{Case I: Twenty-two libration points}
\begin{figure*}[!t]
\begin{center}
(a)\resizebox{0.3\hsize}{!}{\includegraphics*{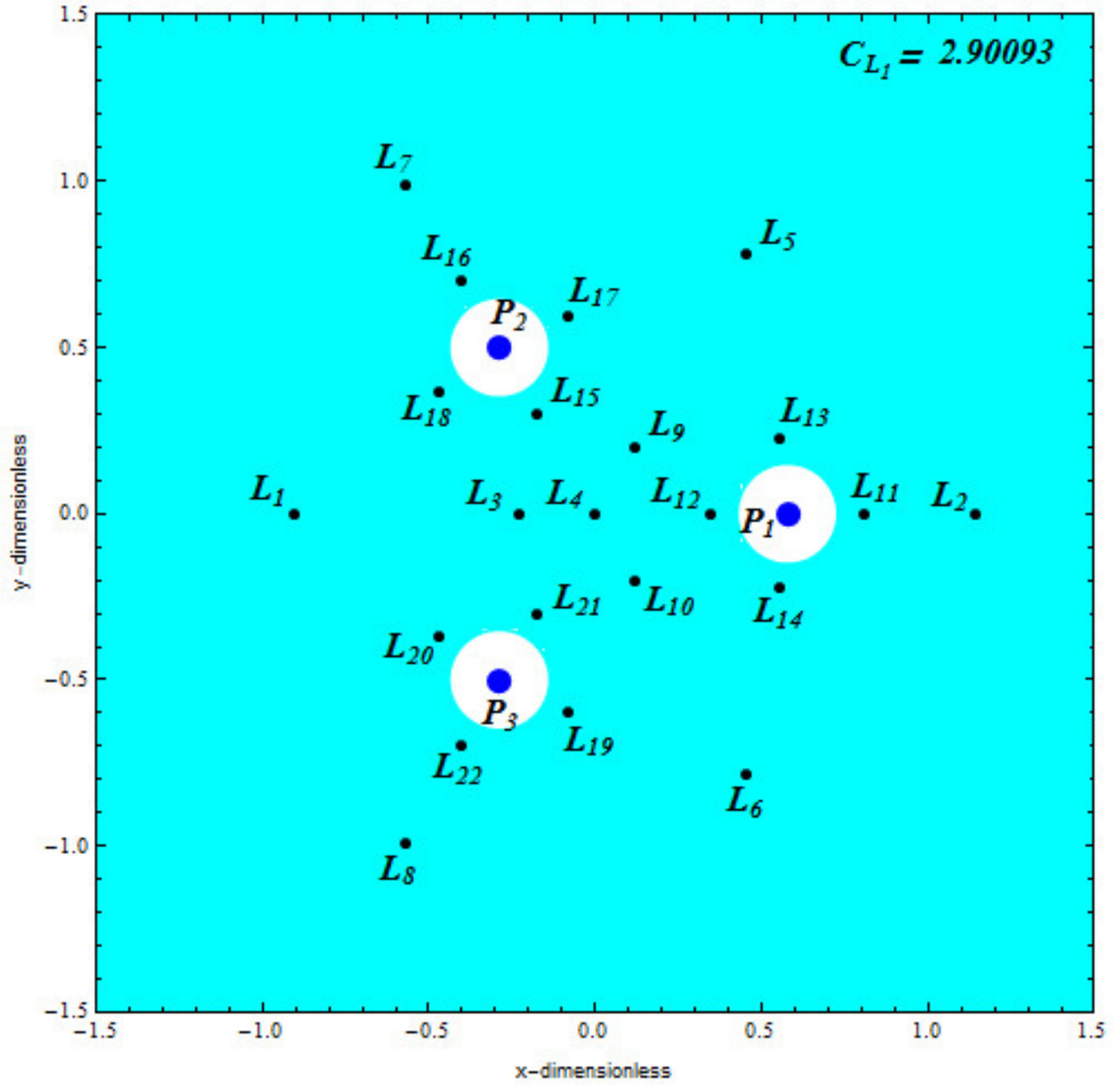}}
(b)\resizebox{0.3\hsize}{!}{\includegraphics*{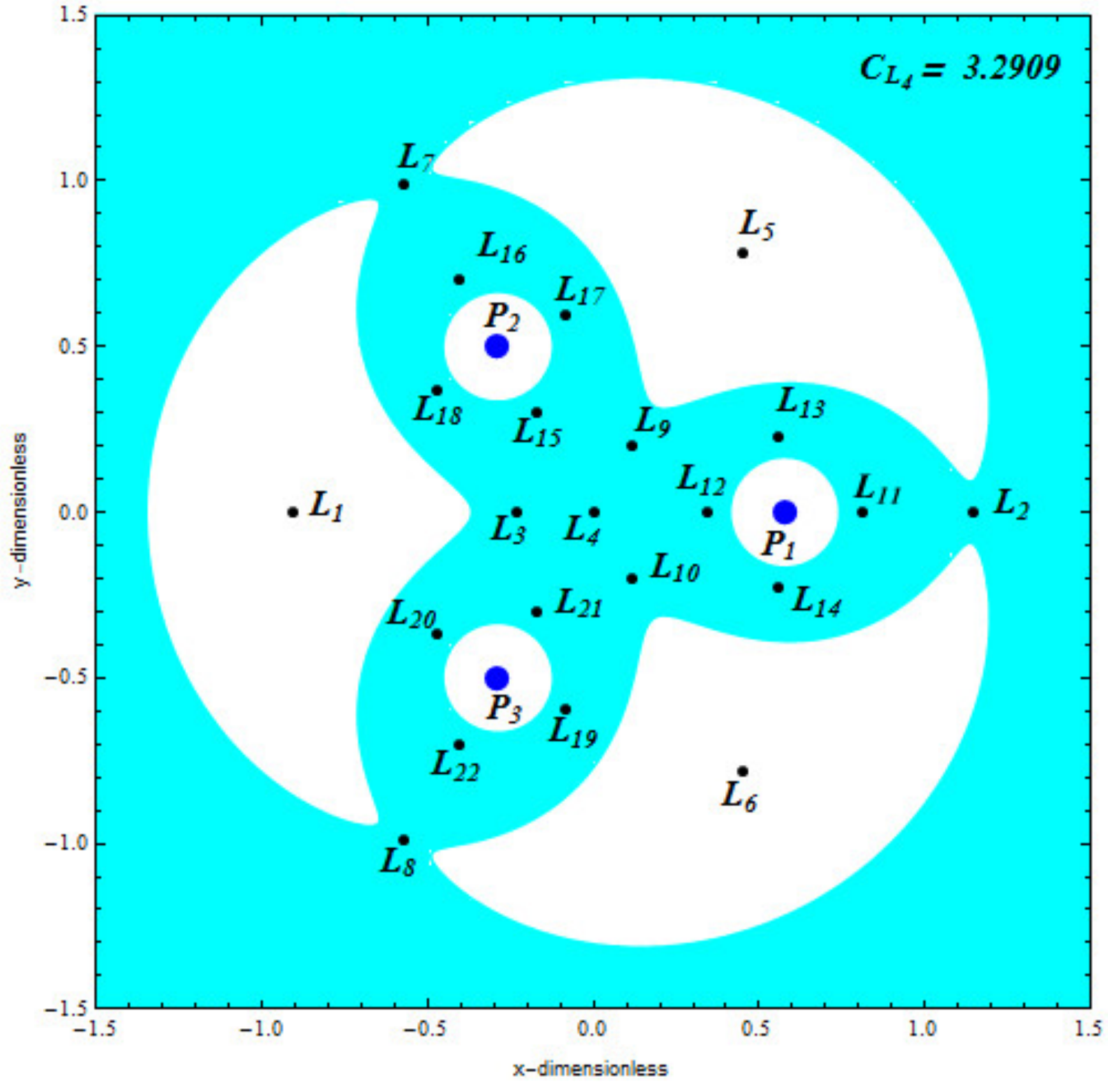}}
(c)\resizebox{0.3\hsize}{!}{\includegraphics*{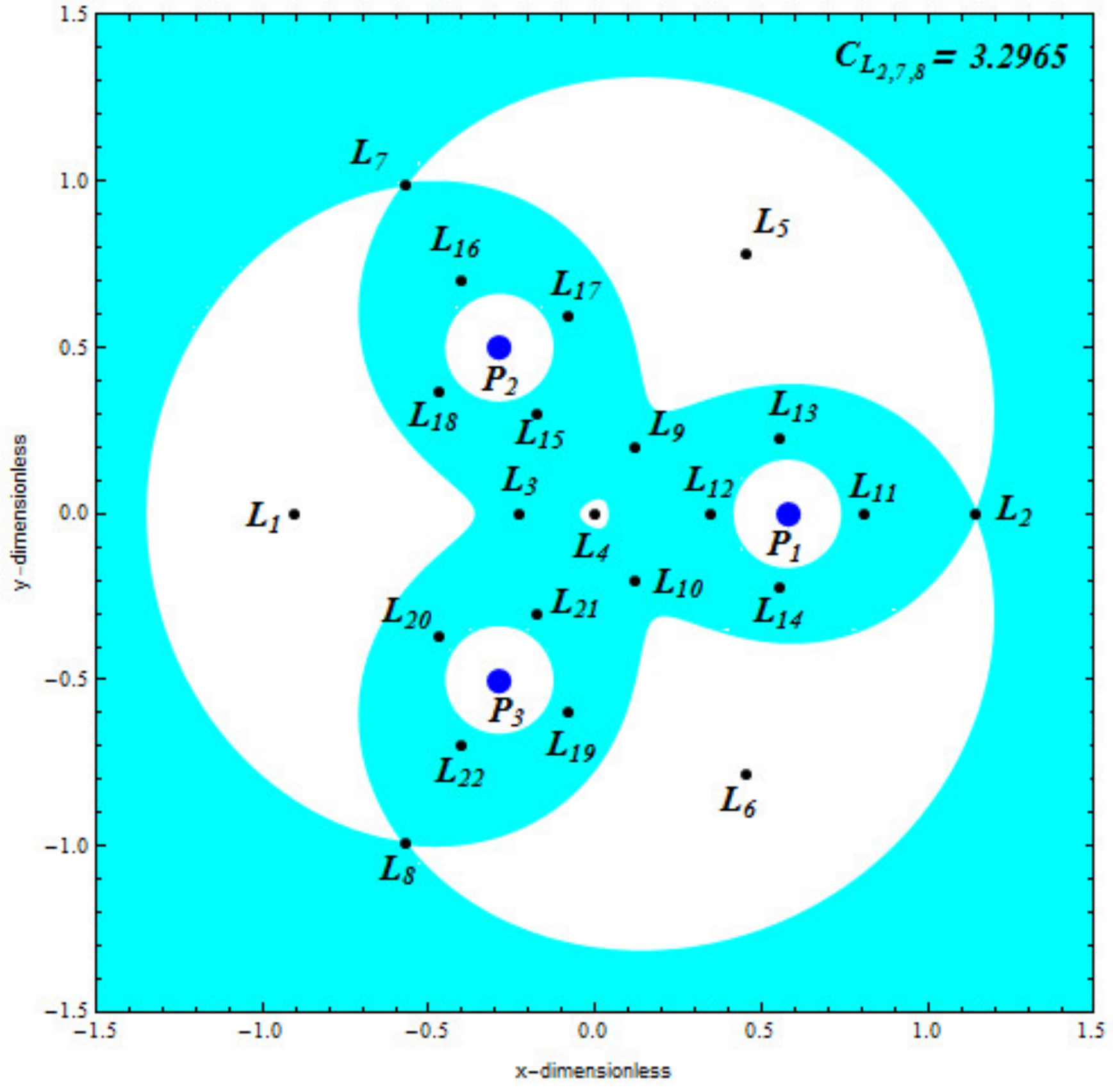}}
(d)\resizebox{0.3\hsize}{!}{\includegraphics*{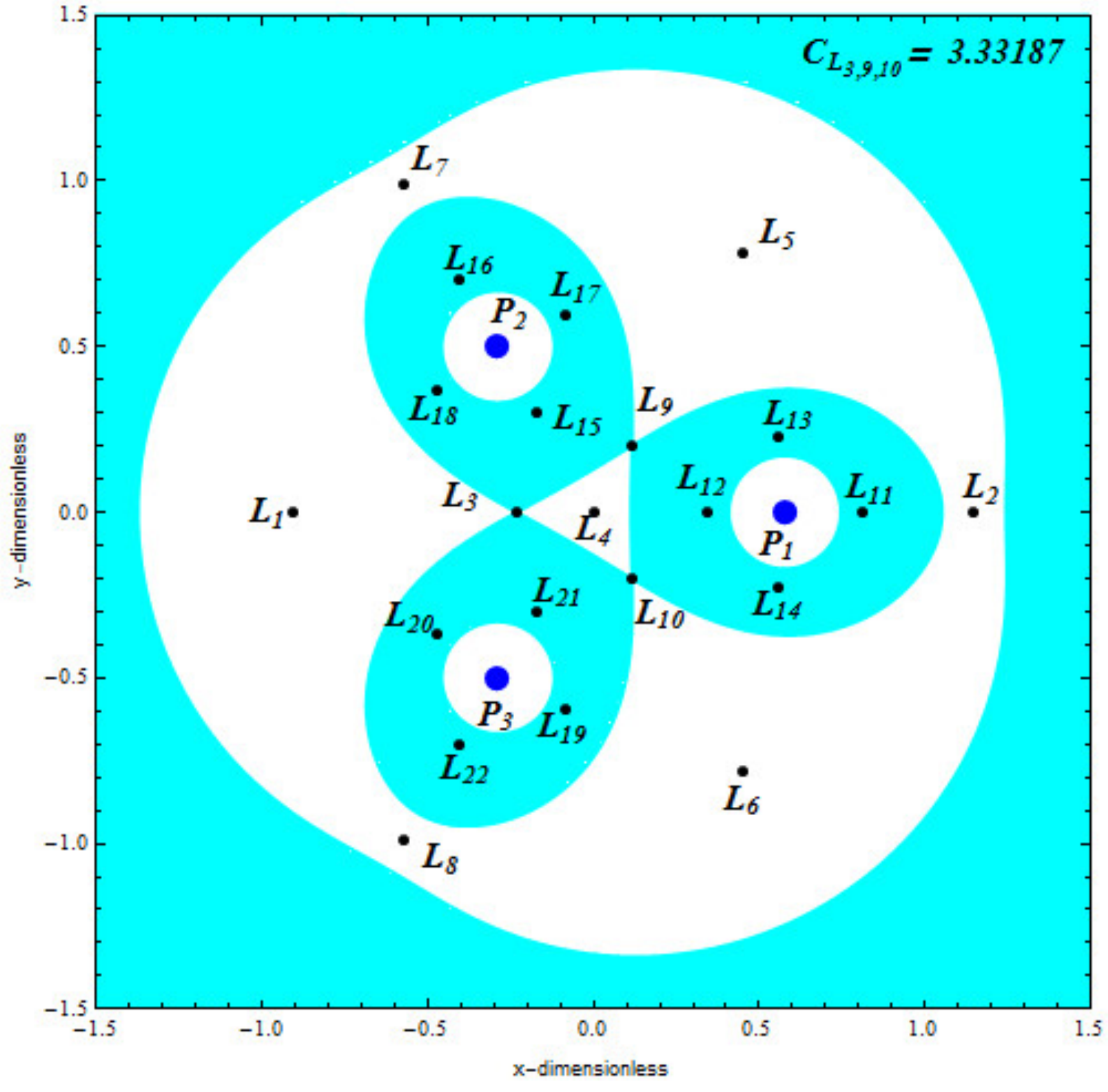}}
(e)\resizebox{0.3\hsize}{!}{\includegraphics*{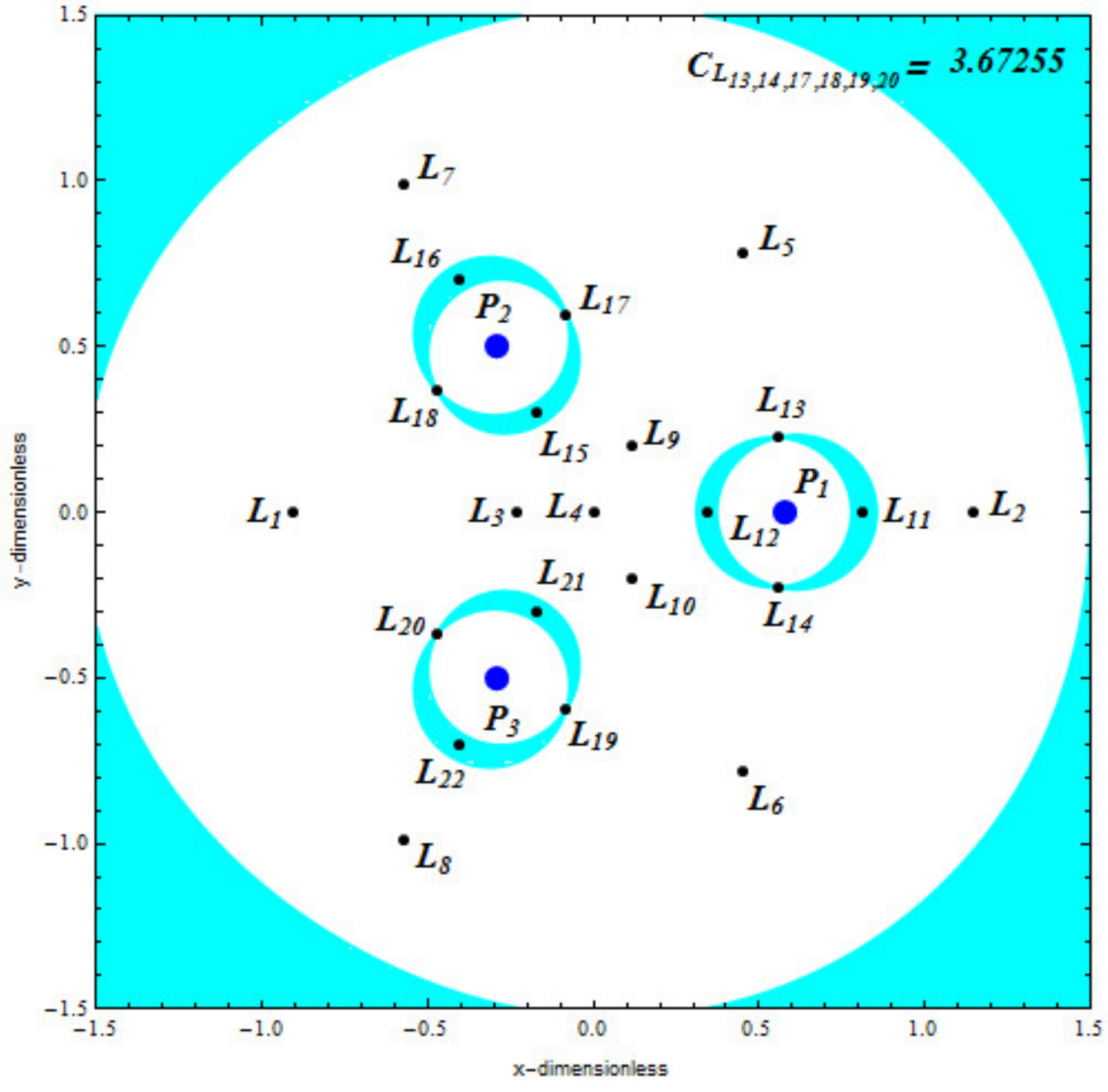}}
(f)\resizebox{0.3\hsize}{!}{\includegraphics*{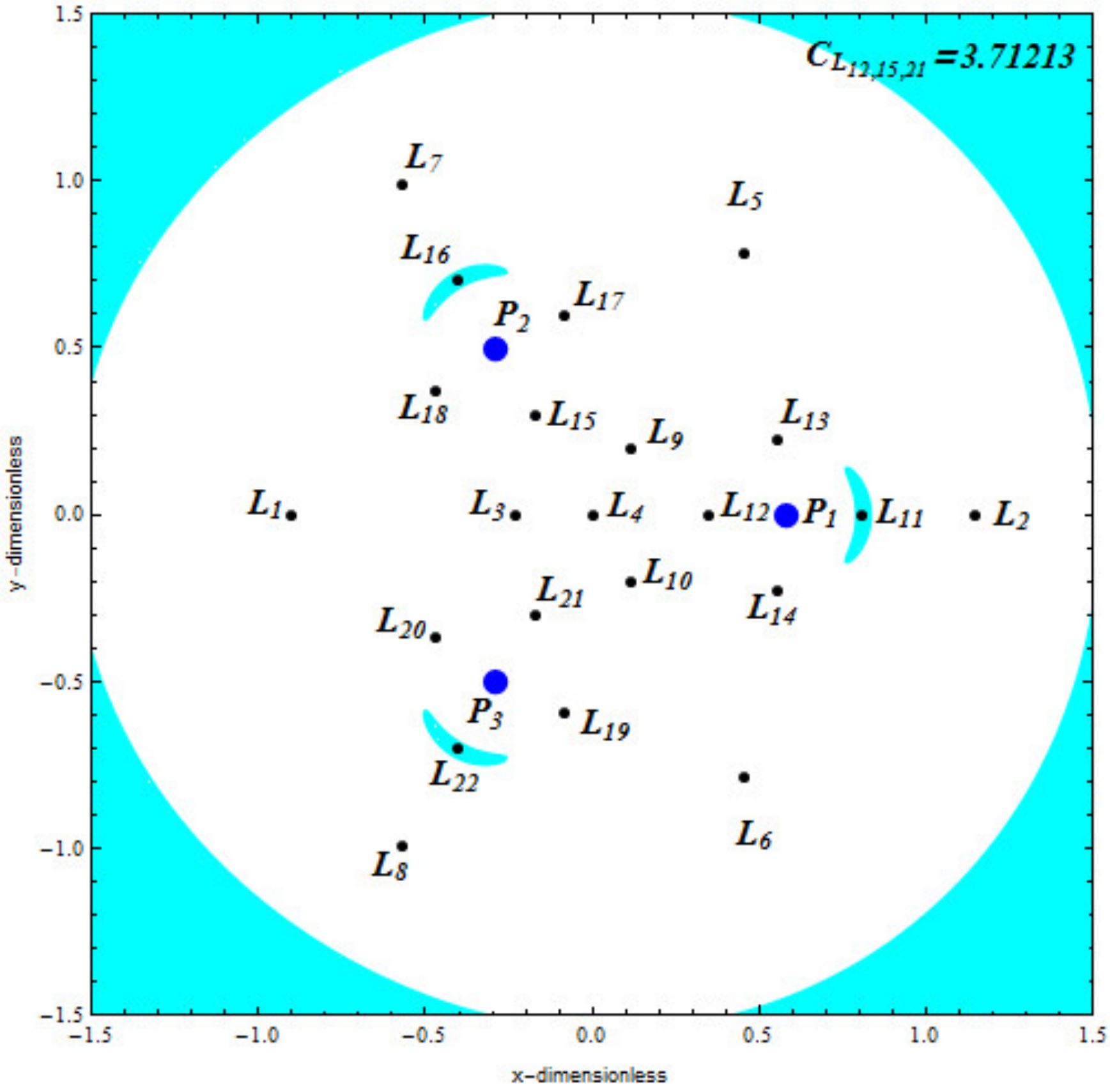}}
\end{center}
\caption{Evolution of the structure of the Hill's regions configuration for $\epsilon=0.3$ when 22 libration points exist
 for the  pseudo-Newtonian restricted four-body problem. The cyan shaded regions correspond to the Hill's regions where the motion
 of the test particle is possible, white domains indicate the forbidden regions, while the boundaries shown by thick cyan lines
 depict the zero velocity curves. The black dots pinpoint the position of the libration points, while the blue dots indicate the
  positions of the centers of the three primaries.} 
\label{fig:7}
\end{figure*}

In the pseudo-Newtonian four-body problem, when the primaries have equal masses, we have $C_{L_1}=2.90093$, $C_{L_4}=3.2909$,
$C_1=C_{L_2}=C_{L_7}=C_{L_8}=3.2965$, $C_2=C_{L_3}=C_{L_9}=C_{L_{10}}=3.33187$, $C_3=C_{L_{13}}=C_{L_{14}}=C_{L_{17}}=C_{L_{18}}=C_{L_{19}}=C_{L_{20}}=3.67255$, and $C_4=C_{L_{12}}=C_{L_{15}}=C_{L_{21}}=3.71213$.
In Fig. (\ref{fig:7}), we present the evolution of the geometry of the Hill's regions, for various values of the Jacobi constant. Here, the white regions show the forbidden regions, while on the other hand the colored regions indicate the regions of possible motion
 of the test particle. It is clear that as the value of the Jacobi constant $C$ decreases, various thresholds appear and allow the
  test particle to move in several possible regions of motion. For $C=C_4$,  in panel-(f), the test particle is prohibited to move
   inside the white region except the three tadpole shaped regions containing the libration points $L_{11, 16, 22}$  in the interior
    region. In panel-(e) for $C=C_3$, the interior region contains three branches of permissible region around each of the primaries.
    However, the test particle is not allowed to move from one primary to the other as well as from the interior to the exterior region and
    vice-versa. For $C=C_1$, three limiting situations occur at the libration points $L_{2, 7, 8}$, which provide the passage for
     the test particle to move from the interior region to the exterior region for $C<C_1$. Further, it is observed that  three circular
     white regions, around the center of each of the primaries, still exist so that the test particle can not move from one primary to other
      and vice-versa. Our analysis suggests that the possible regions of motion substantially increase, with the decrease of the Jacobi
       constant.
\subsection{Case II: Existence of sixteen and ten libration points}
\begin{figure*}[!t]
\begin{center}
(a)\resizebox{0.3\hsize}{!}{\includegraphics*{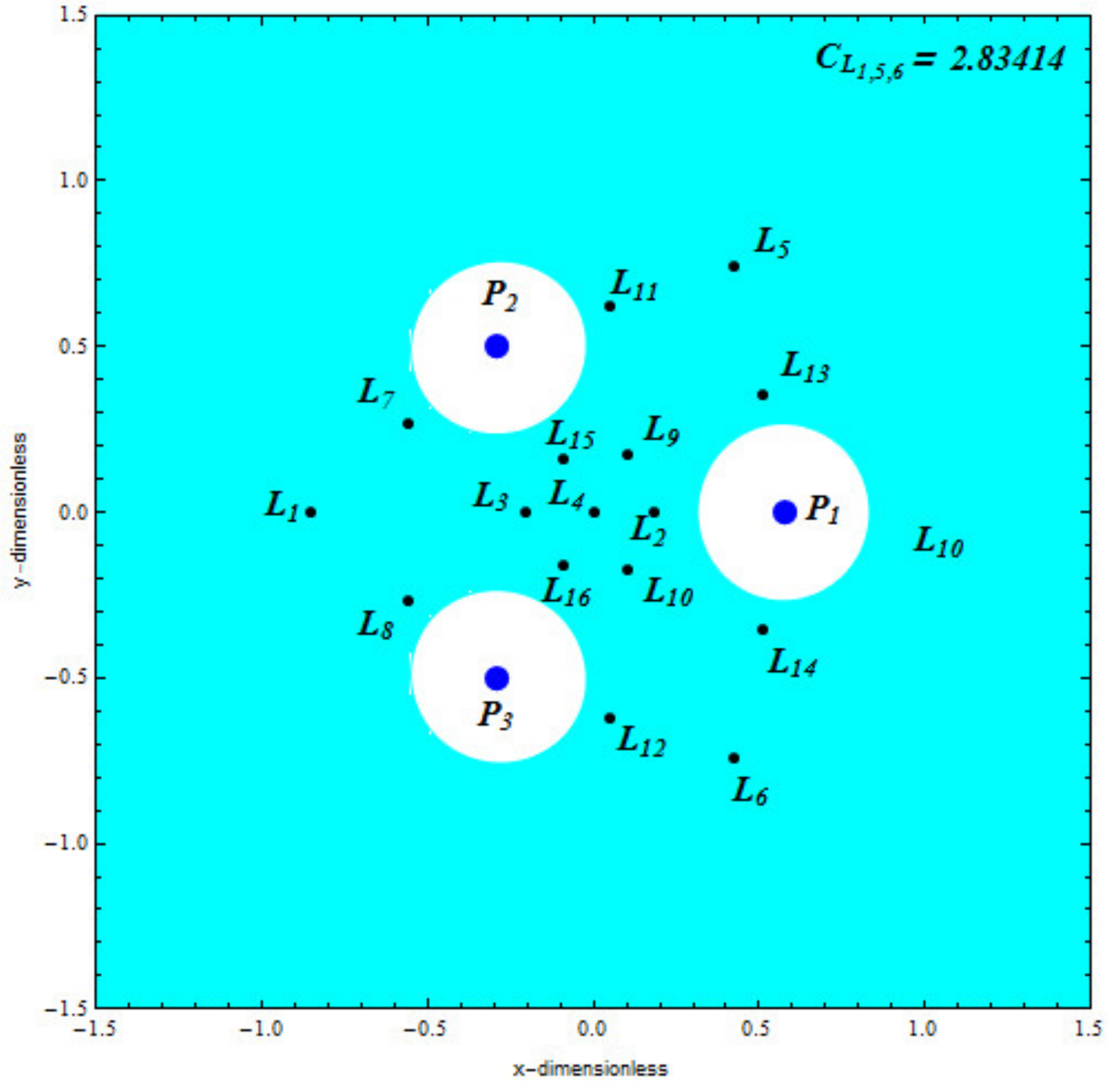}}
(b)\resizebox{0.3\hsize}{!}{\includegraphics*{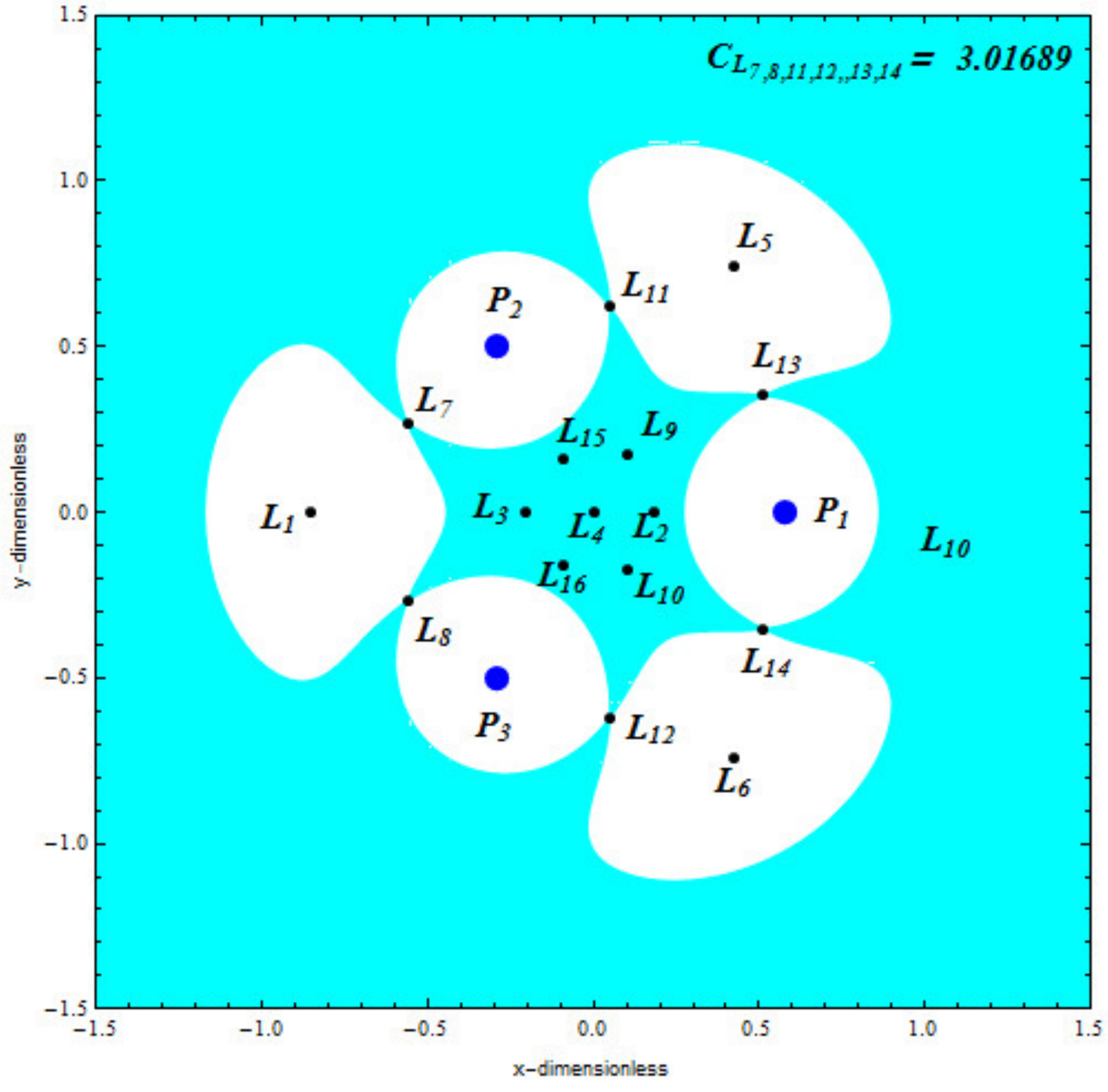}}
(c)\resizebox{0.3\hsize}{!}{\includegraphics*{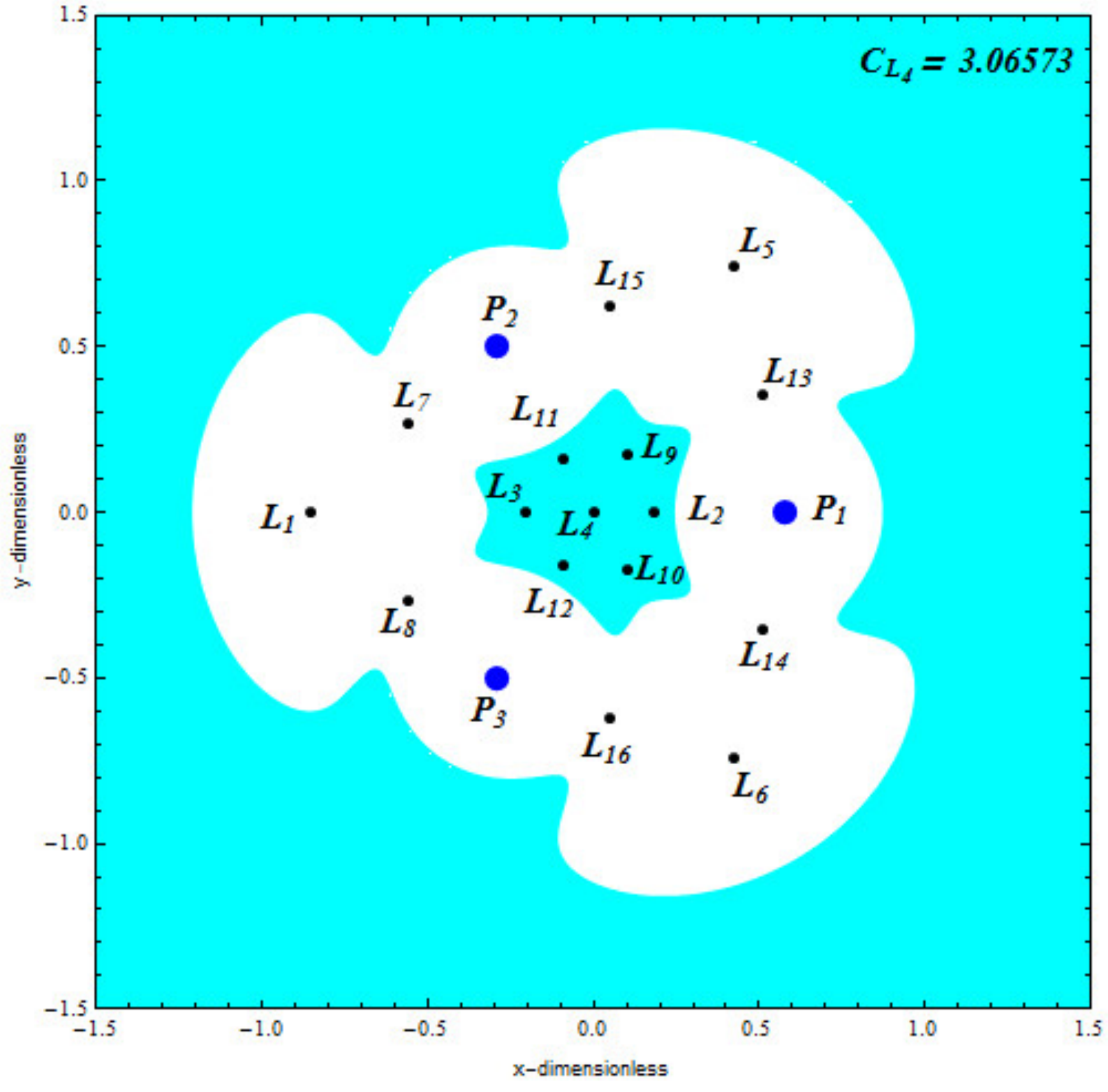}}
(d)\resizebox{0.3\hsize}{!}{\includegraphics*{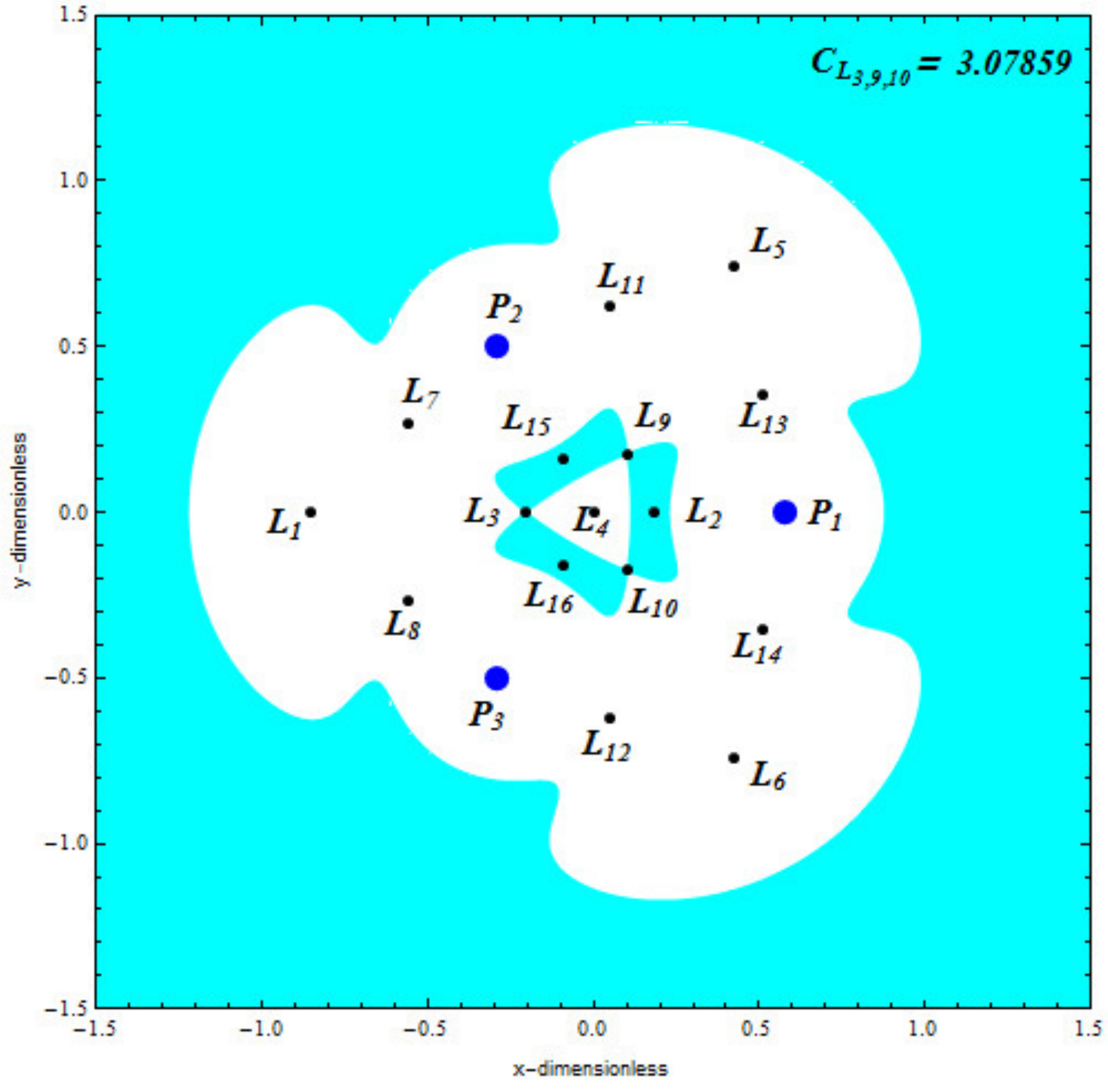}}
(e)\resizebox{0.3\hsize}{!}{\includegraphics*{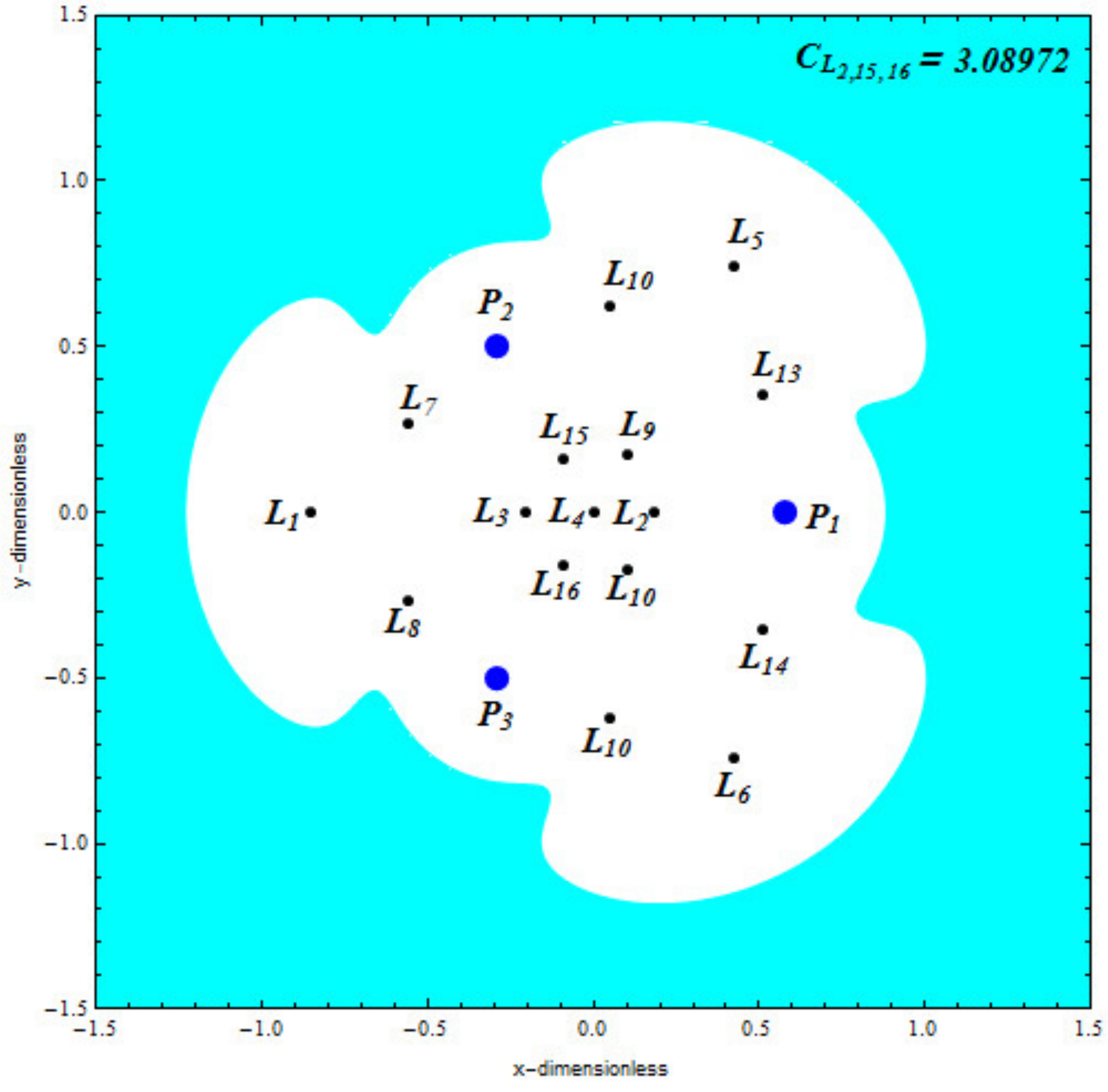}}
\end{center}
\caption{Evolution of the structure of the Hill's regions configuration for $\epsilon=0.69$, when 16 libration points
exist for the pseudo-Newtonian restricted four-body problem. The description of this figure is the same as
 in Fig. \ref{fig:7}. (Color figure online).} 
\label{fig:8}
\end{figure*}

In the case when sixteen libration points exist, we have   $C_{L_1}=C_{L_5}=C_{L_6}=2.83414$,
$C_1=C_{L_7}=C_{L_8}=C_{L_{11}}=C_{L_{12}}=C_{L_{13}}=C_{L_{14}}=3.01689$, $C_2=C_{L_4}=3.06573$, $C_3=C_{L_{3}}=C_{L_{9}}=C_{L_{10}}=3.07859$, and $C_4=C_{L_{2}}=C_{L_{15}}=C_{L_{16}}=3.08972$. From Fig. \ref{fig:8}, it can be observed that the forbidden region shown in white color increases with the increase of the Jacobian constant. From panel-(a), it is clear that the test particle is free to move everywhere except the three circular shaped islands around the primaries. For $C=C_1$, the forbidden region increases and six white coloured islands appear in panel-(b). The possible regions of motion split into the interior and the exterior regions for $C=C_2$ (panel-c). However, the test particle cannot move from the interior to the exterior region. Moreover, the interior region of possible motion constitutes three branches containing the libration points $L_{2, 15, 16}$  (panel-d).  For the Jacobian constant $C=C_4$, the interior possible region of motion disappears completely and the test particle is allowed to move only in the exterior cyan color region (Fig. \ref{fig:8}e).

\begin{figure*}[!t]
\begin{center}
(a)\resizebox{0.3\hsize}{!}{\includegraphics*{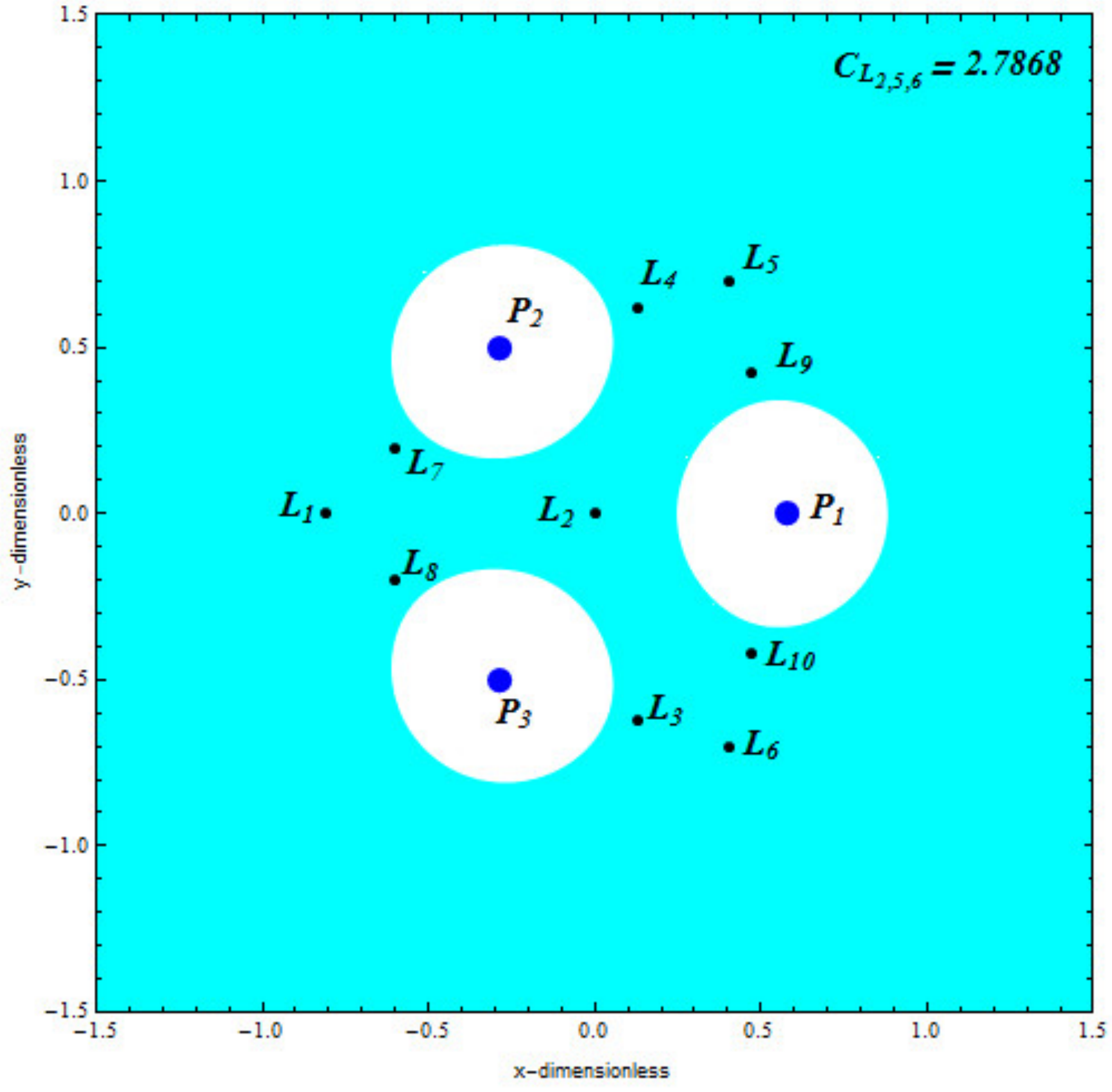}}
(a)\resizebox{0.3\hsize}{!}{\includegraphics*{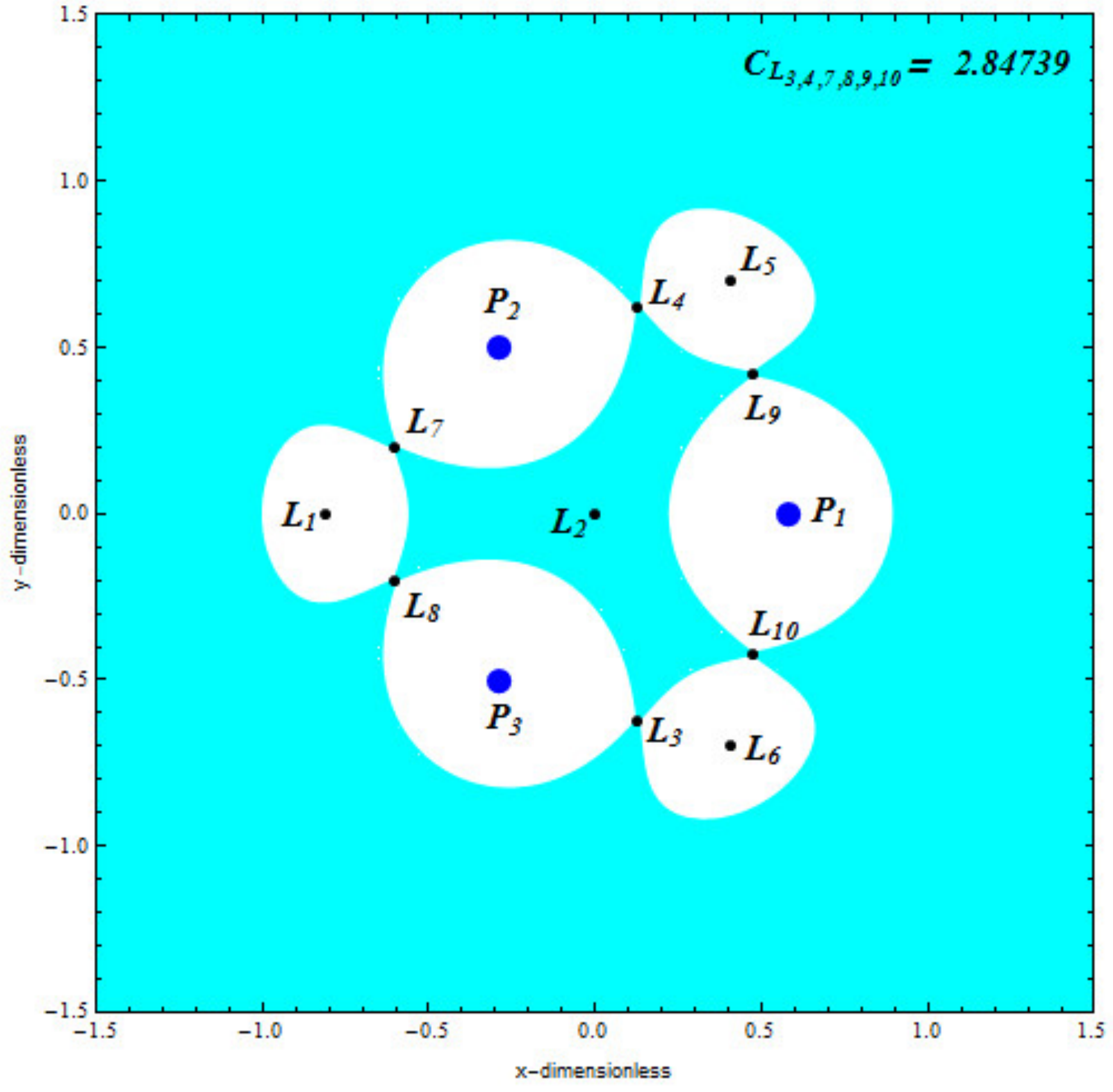}}
(c)\resizebox{0.3\hsize}{!}{\includegraphics*{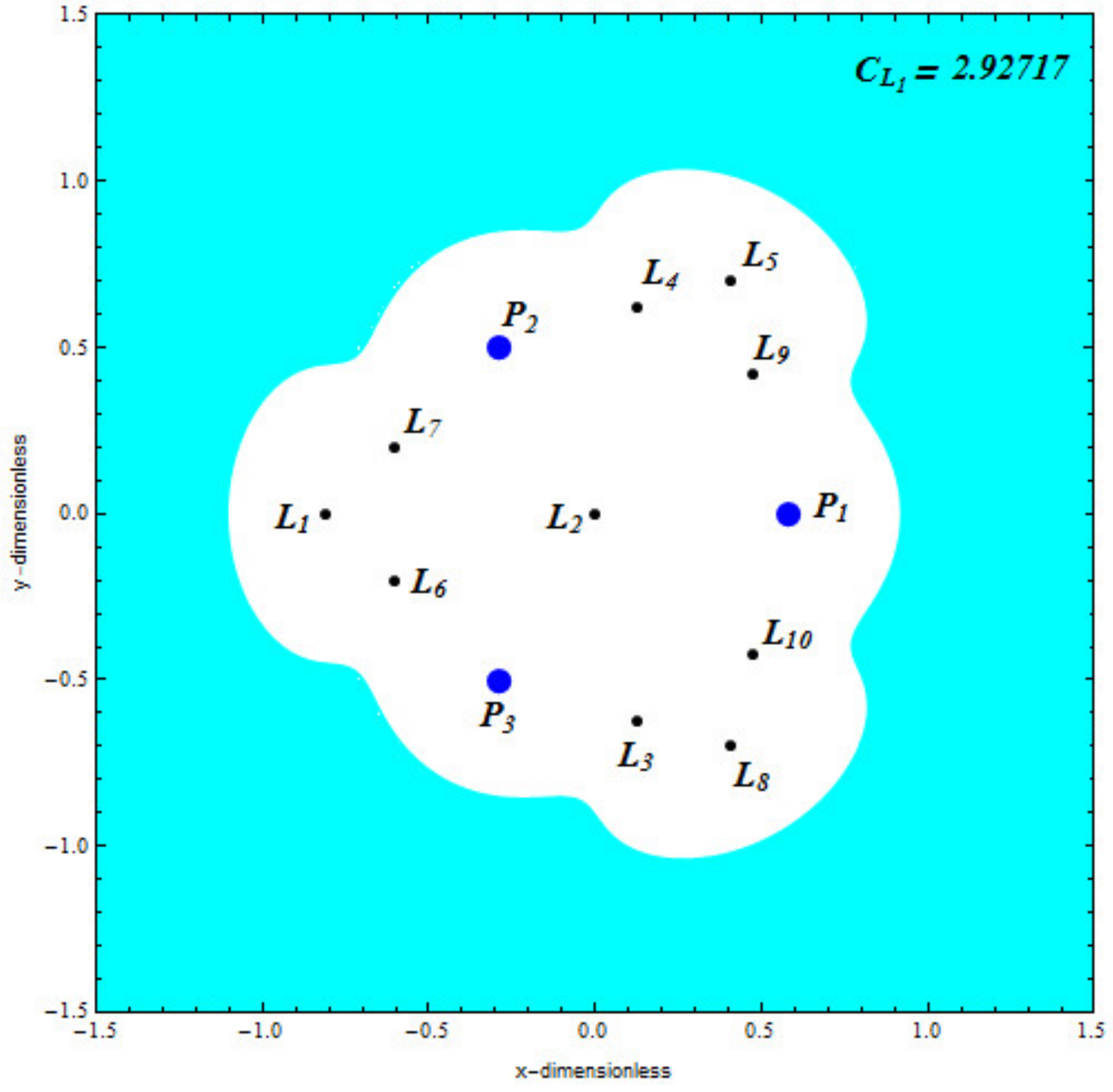}}
\end{center}
\caption{Evolution of the structure of the Hill's regions configuration for $\epsilon=0.93$, when 10
 libration points exist for the pseudo-Newtonian restricted four-body problem. The description of this figure
is the same as in Fig. \ref{fig:7}. (Color figure online).} 
\label{fig:9}
\end{figure*}

Figure \ref{fig:9} describes the ZVCs for $\epsilon=0.93$, when ten libration points exist.
\begin{figure*}[ht]
\begin{center}
(a)\resizebox{0.3\hsize}{!}{\includegraphics*{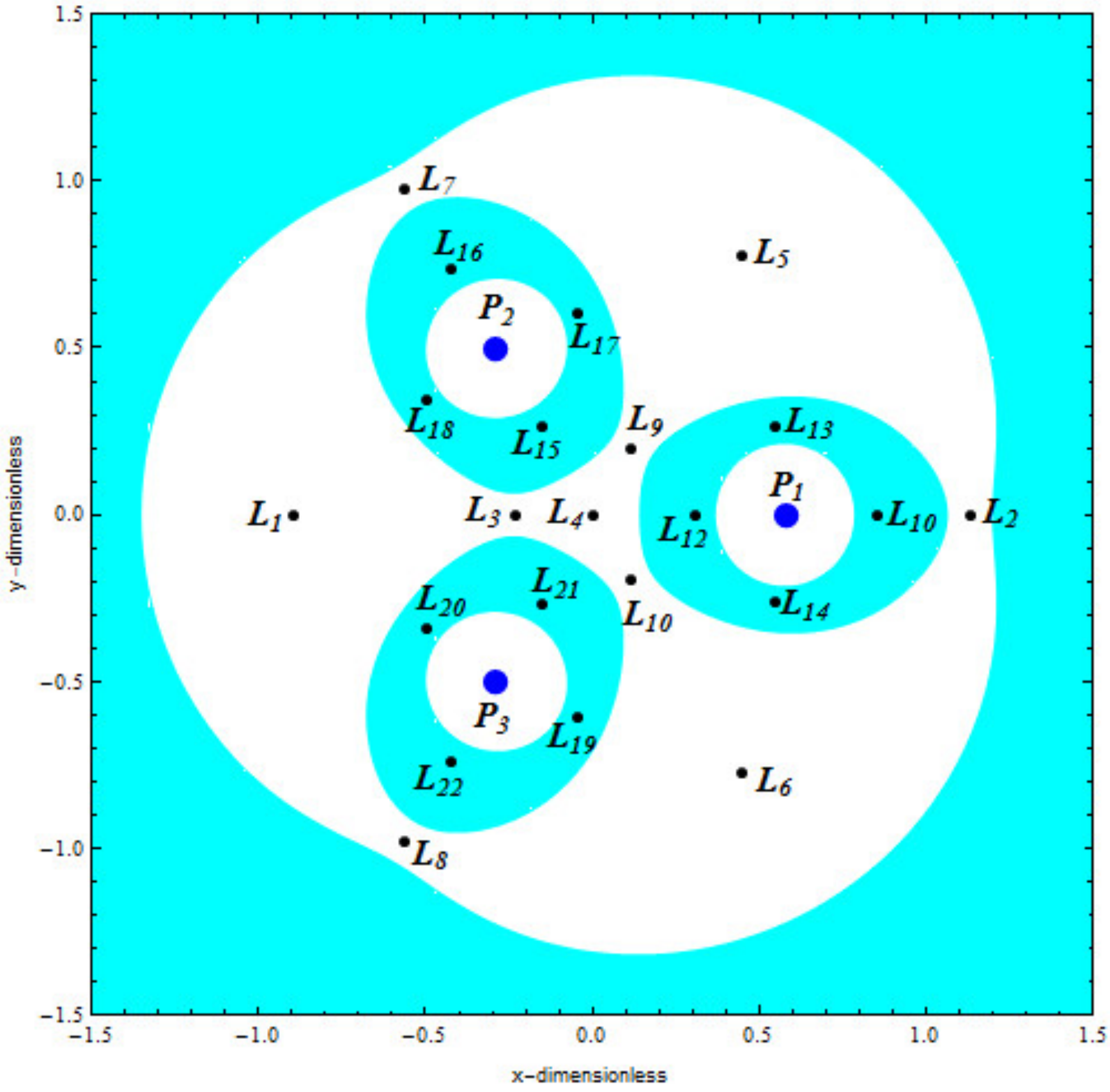}}
(b)\resizebox{0.3\hsize}{!}{\includegraphics*{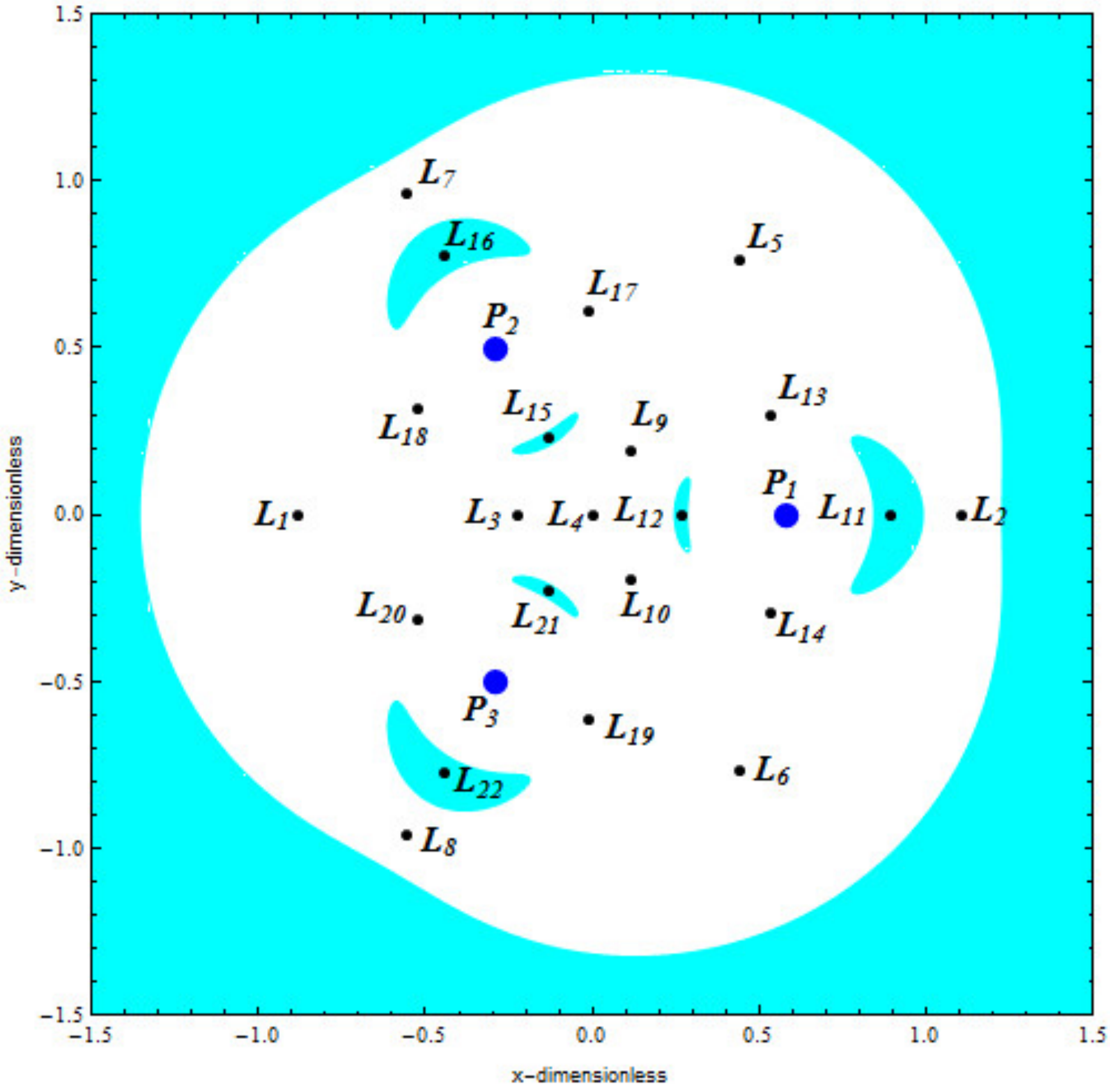}}
(c)\resizebox{0.3\hsize}{!}{\includegraphics*{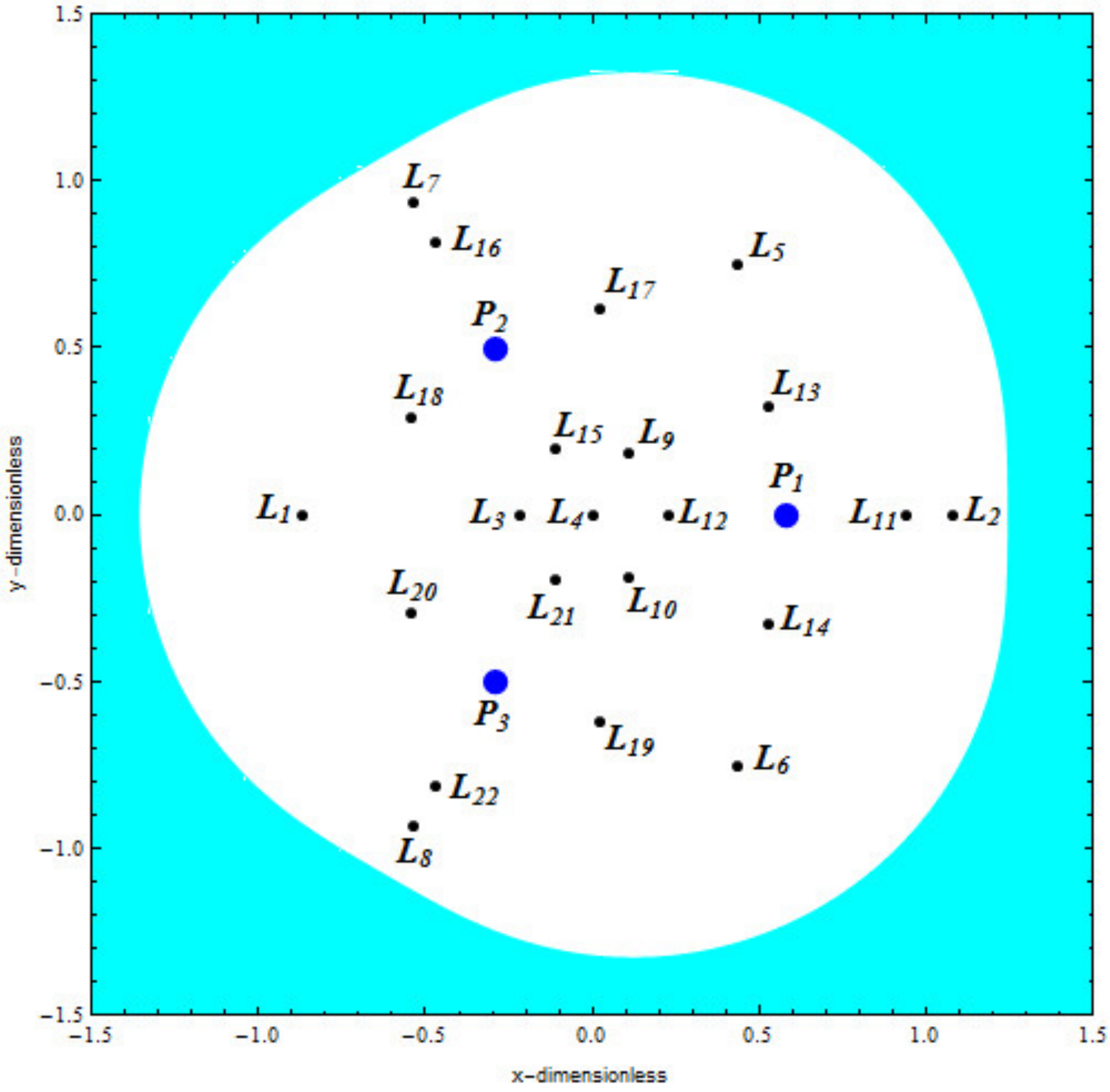}}
\end{center}
\caption{Evolution of the structure of the Hill's regions configuration, when 22 libration points exist
for the pseudo-Newtonian restricted four-body problem for $C=3.29$ and (a) $\epsilon=0.4$; (b) $\epsilon=0.5$; (c) $\epsilon=0.6$. The description of this figures is the same as in Fig. \ref{fig:7}. (Color figure online).} 
\label{fig:9a}
\end{figure*}
In Fig. \ref{fig:9a}, we depict the ZVCs for a fixed value of the Jacobian constant $C$ and for various values of the transition parameter $\epsilon$. It is observed when the transition parameter $\epsilon=0$, i.e., the classical restricted four-body problem, the infinitesimal mass is free to move from one primary to the other and vice-versa. On the other hand, there exist circular islands of white regions around each of the primary in the presence of transition parameter $\epsilon>0$, therefore the infinitesimal mass is prohibited to move from one primary body to the other and vice-versa. When $\epsilon=0.4$ the regions of possible motion, shown in cyan color, constitute three inner circle annuli containing the primaries in the interior region. However, the infinitesimal mass cannot move from the inner region to the outer region of possible motion. When $\epsilon=0.5$, the possible regions of motion occur in six tadpole shaped regions containing the libration points $L_{11, 12, 15, 16, 21, 22}$, respectively in the interior area, while for $\epsilon=0.6$ these tadpole shaped regions disappear completely and the infinitesimal mass cannot move inside the interior region. Moreover, our analysis suggests that as the transition parameter increases, the regions of possible motion decrease and the infinitesimal mass is confined to move in the outer region only.

\section{Newton-Raphson Basins of attractions}\label{Newton-Raphson Basin of attractions}
The location of the positions of the libration points, in any dynamical system, is a crucial issue. Unfortunately, for restricted problem of $N>3-$bodies, there are no analytical formulae to obtain the exact locations of the libration point. Therefore, to overcome this problem we can use one of the multivariate iterative scheme of numerical methods to solve the system of non-linear equations.
Moreover, the well known multivariate iterative scheme is the Newton-Raphson method. The iterative scheme is applicable to system of multivariate functions $f(\textbf{X})=0$ where as the associated iterative scheme is represented by
\begin{eqnarray}\label{Eq:18}
\mathbf{X}_{n+1}=\mathbf{X}_n-J^{-1}f(\mathbf{X}_n)
\end{eqnarray}
where $J^{-1}$  corresponds to the inverse Jacobian matrix of $f(\mathbf{X}_n)$. In the present model, the system of differential equations
 is described by the Eqs. (\ref{Eq:10}).
For the $(x, y)$  plane the iterative formulae for each co-ordinate are given by
\begin{eqnarray}
\label{Eq:19}
x_{n+1}&=&x_n-\frac{\Omega_{x_n}\Omega_{y_ny_n}-\Omega_{y_n}\Omega_{x_ny_n}}{\Omega_{x_nx_n}\Omega_{y_ny_n}-\Omega_{x_ny_n}\Omega_{y_nx_n}},\nonumber\\
y_{n+1}&=&y_n+\frac{\Omega_{x_n}\Omega_{y_nx_n}-\Omega_{y_n}\Omega_{x_nx_n}}{\Omega_{x_nx_n}\Omega_{y_ny_n}-\Omega_{x_ny_n}\Omega_{y_nx_n}},
\end{eqnarray}
where the values of $x$ and $y$ coordinates at the $n$-th step of the Newton-Raphson iterative scheme are represented by $x_n$ and $y_n$. In addition, the subscripts of $\Omega(x, y)$ correspond to the respective partial derivatives of the effective potential function.

The Newton-Raphson iterative scheme works under the prescribed numerical algorithm and the code is activated with an initial condition $(x_0, y_0)$ on the configuration plane, while the iterative scheme terminates when the attractor of the described system is reached, with predefined accuracy. The method converges only if the particular initial condition leads to one of the attractor of the system. It is necessary to note that not every initial condition on the configuration $(x,y)$ plane converge to one of the attractors of the system. If the fixed initial condition converges to one of the libration point which acts as attractor, we claim that the iterative scheme converges for that fixed initial condition.  The set of initial conditions which converge to same libration points compose the domain of the basins of convergence.
Finally, the Newton-Raphson basins of attraction, also called as basins of convergence, are composed of the initial conditions which lead to an attractor of the system.

One can notice that the iterative formulae of Eqs. (\ref{Eq:19}) consist the first and the second order partial derivatives of the effective potential function $\Omega(x,y)$, which reflect some of the most intrinsic qualitative properties of the dynamical  system. This makes the study of the Newton-Raphson basins of attraction very crucial. The domain of the basins of convergence is obtained by a double scan of the configuration $(x, y)$ plane by defining $(x_0, y_0)$ nodes as initial conditions used for the multivariate iterative scheme. Further, for the iterative scheme, the allowed maximum number of iterations are five hundred whereas the multivariate iterative scheme terminates when the coveted accuracy is reached.

\begin{figure}[!t]
\begin{center}
\resizebox{0.6\hsize}{!}{\includegraphics*{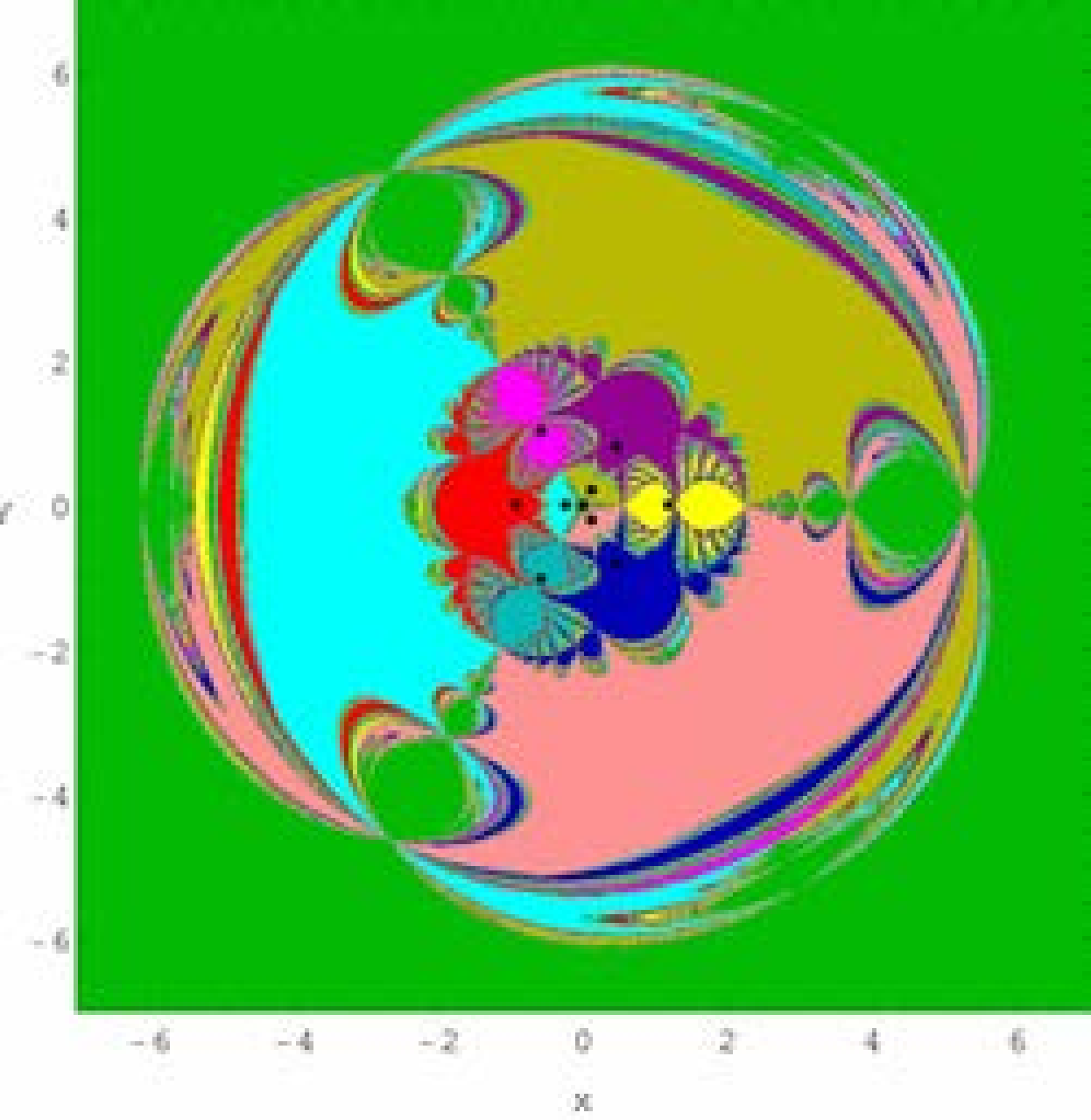}}
\end{center}
\caption{The Newton-Raphson basins of attraction on the configuration
$(x, y) $ plane  for $\epsilon=0$.  The color code denoting the ten attractors is as follows: $L_1$ (\emph{red});
 $L_2$ (\emph{yellow}); $L_3$ (\emph{cyan}); $L_4$ (\emph{green}); $L_5$ (\emph{purple}); $L_6$ (\emph{blue});
 $L_7$ (\emph{magenta}); $L_8$ (\emph{persian green});  $L_9$ (\emph{olive}); $L_{10}$ (\emph{pink});
 non-converging points (\emph{white}).  The \emph{black} dots show the positions of the libration points.
 (Color figure online).} 
\label{fig:11n}
\end{figure}

The Newton-Raphson basins of convergence when $\epsilon=0$  corresponds to the classical restricted four-body problem are presented  using different color codes to denote the different basins of attraction corresponding to each libration point. The positions of  ten liberation points are pinpoint by black dots (see Fig. \ref{fig:11n}). It is clear that in the restricted four-body problem when primaries have equal masses, the problem admits four collinear libration points on the $x-$axis and six non-collinear libration points on the $(x, y)$ plane (off the $x-$axis). The classical restricted four-body problem admits a symmetry and all the ten libration points lie on the $(x, y)$-plane symmetrical to the axes of symmetry $y=0$, and $y=\pm\sqrt{3}$.

\begin{figure*}[!t]
\begin{center}
(a)\resizebox{0.4\hsize}{!}{\includegraphics*{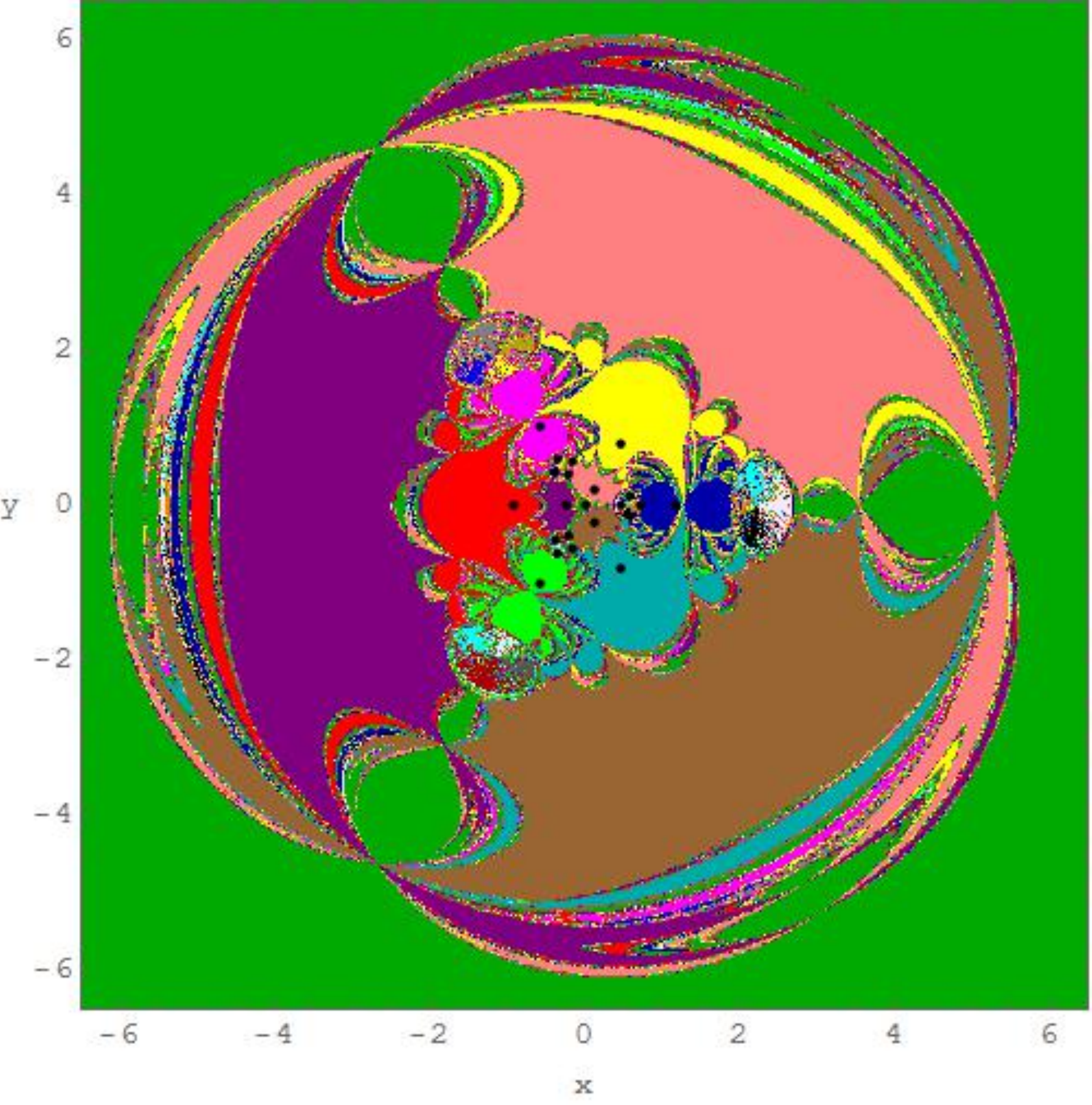}}
(b)\resizebox{0.4\hsize}{!}{\includegraphics*{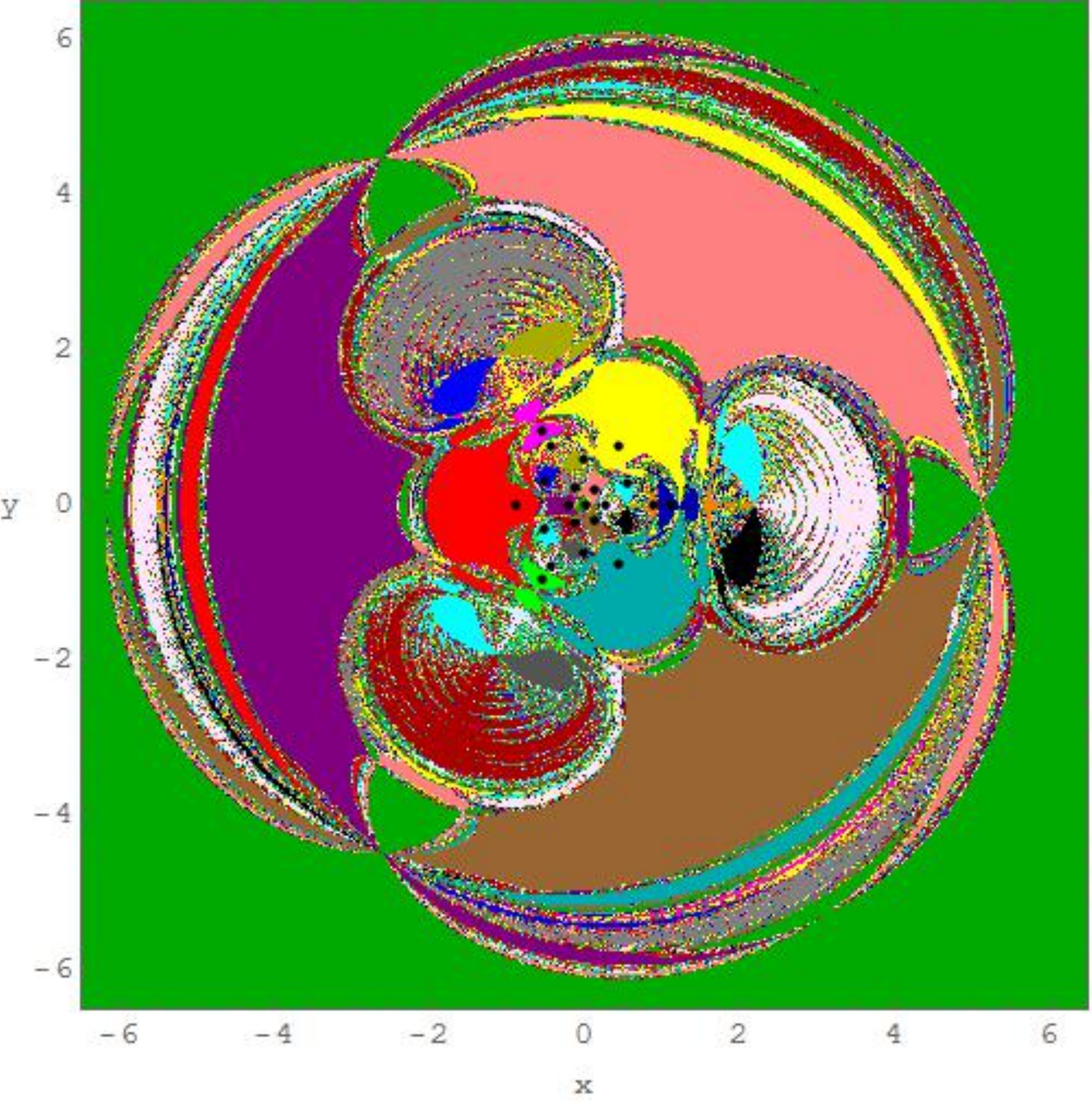}}
(c)\resizebox{0.4\hsize}{!}{\includegraphics*{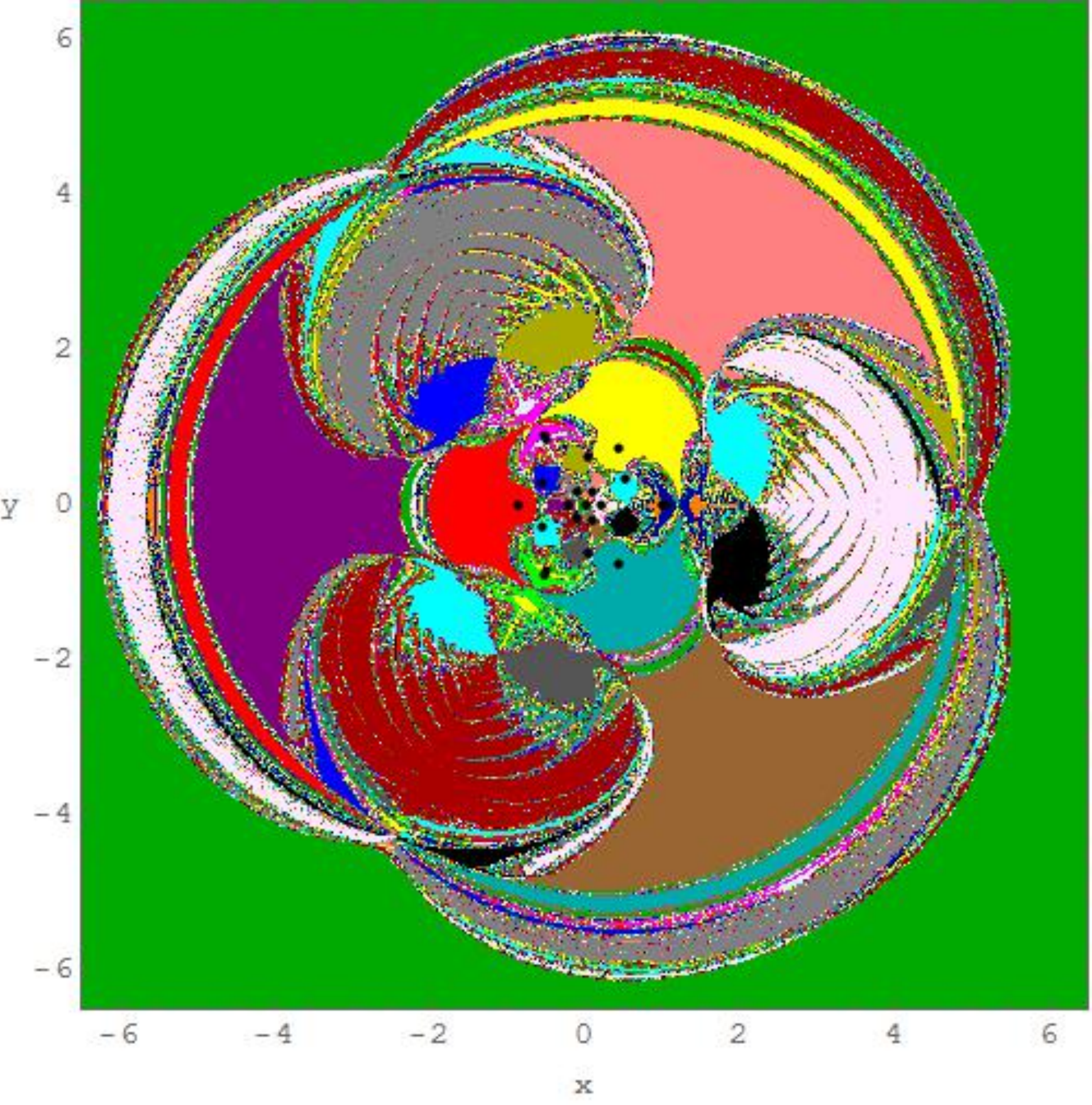}}
(d)\resizebox{0.4\hsize}{!}{\includegraphics*{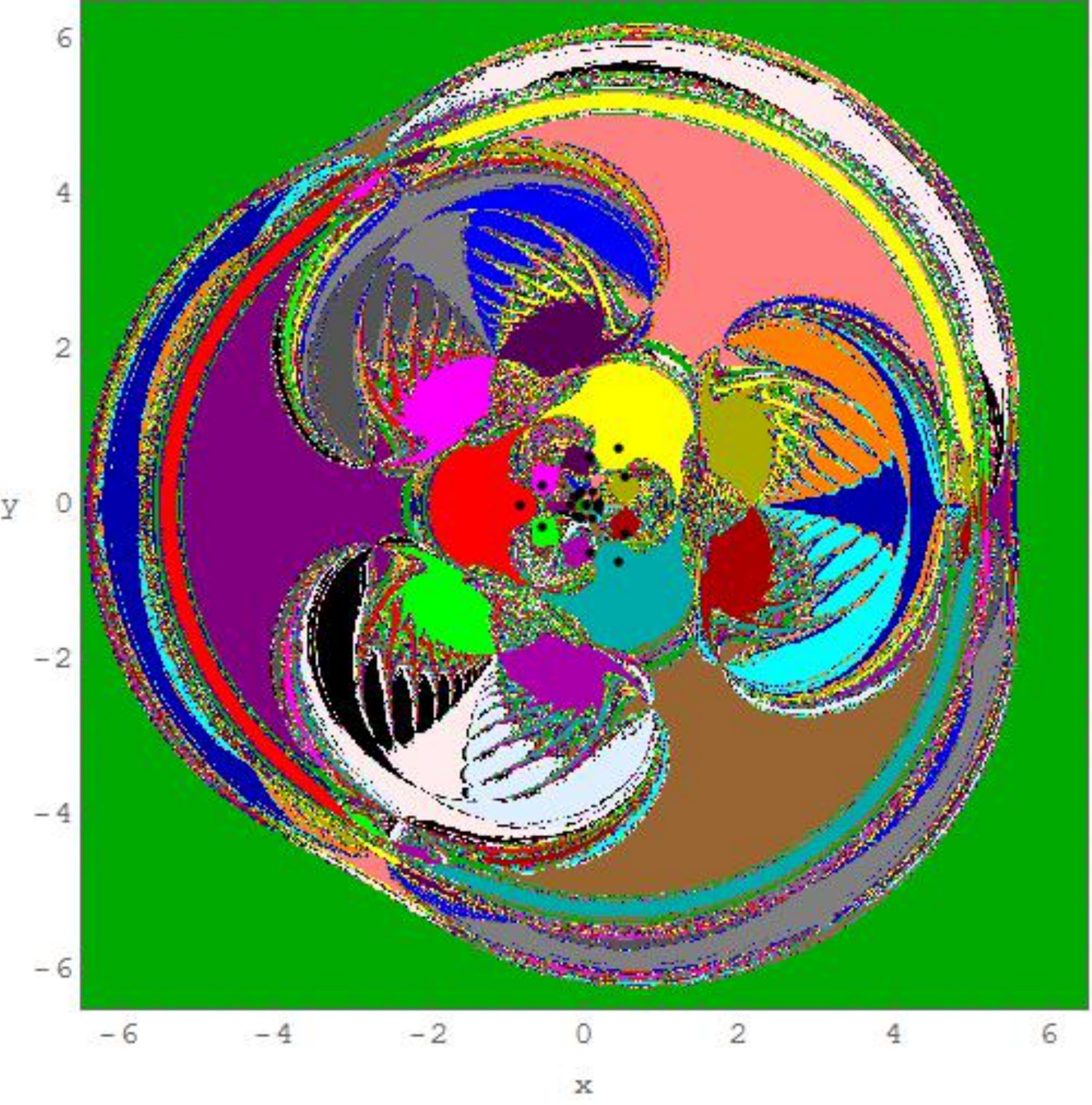}}
(e)\resizebox{0.4\hsize}{!}{\includegraphics*{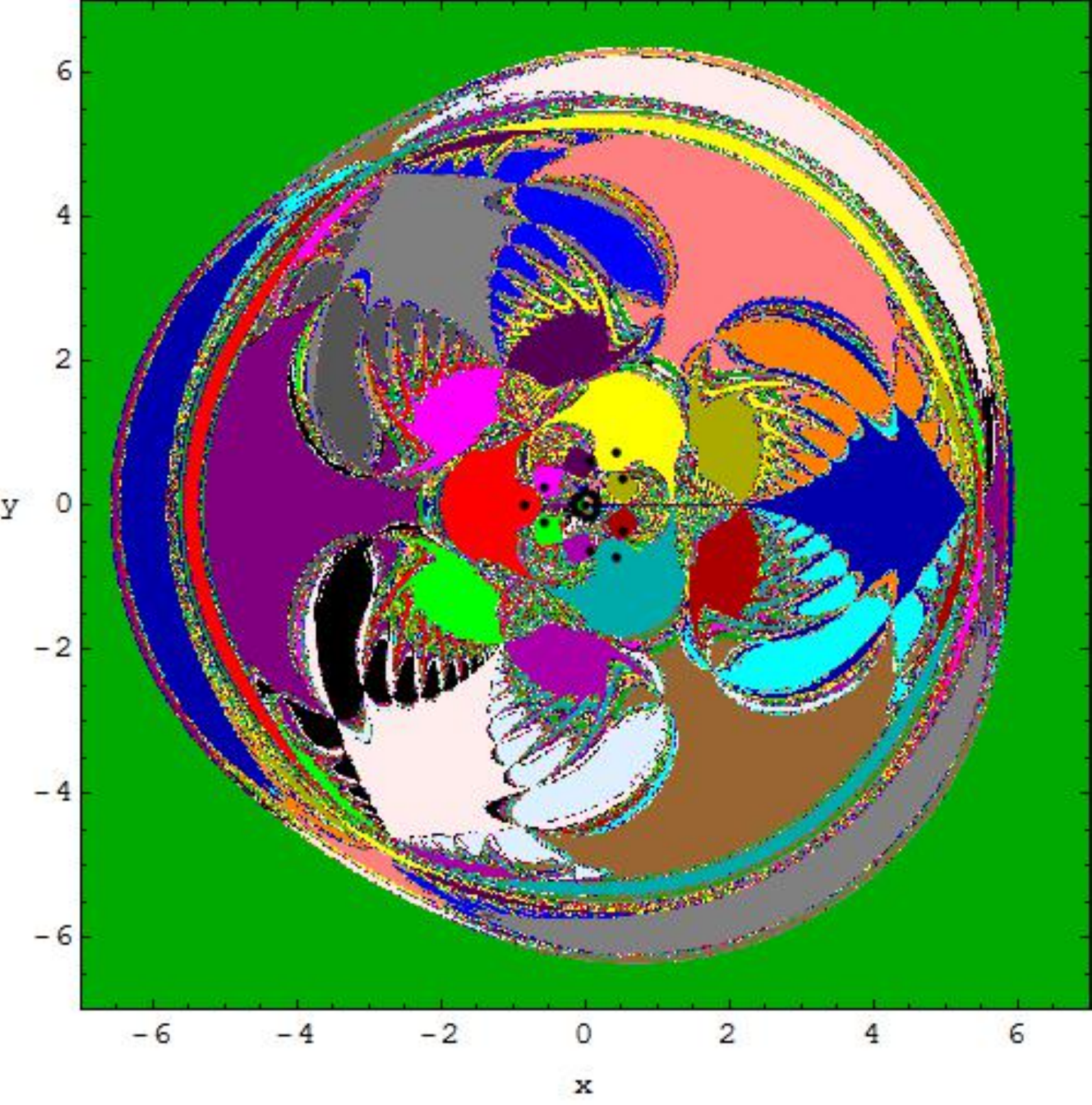}}
(f)\resizebox{0.4\hsize}{!}{\includegraphics*{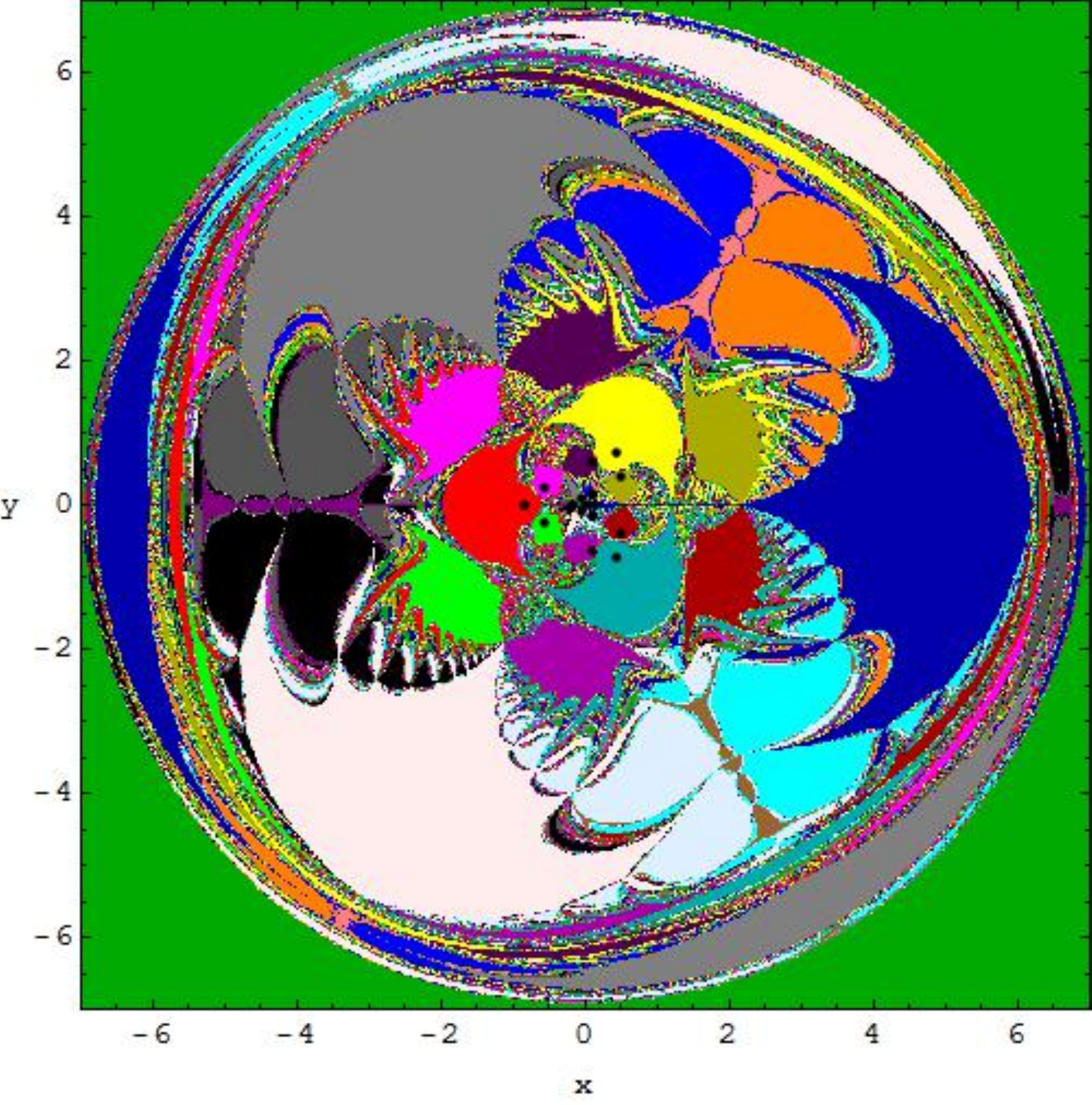}}
\end{center}
\caption{The Newton-Raphson basins of attraction on the
configuration $(x, y)$ plane for the case, where twenty two
libration points are present.  (a) $\epsilon=0.1$, (b) $\epsilon=0.5$, (c) $\epsilon=0.66$, (d) $\epsilon=0.714768$,
 (e) $\epsilon=0.75$,  and (f) $\epsilon=0.8125$. The color code denoting 22 attractors is as follows: $L_1$ (\emph{red});
 $L_2$ (\emph{duke blue}); $L_3$ (\emph{purple}); $L_4$ (\emph{darker green}); $L_5$ (\emph{yellow}); $L_6$ (\emph{persian cyan});
 $L_7$ (\emph{magenta}); $L_8$ (\emph{green});  $L_9$ (\emph{pink}); $L_{10}$ (\emph{brown}); $L_{11}$ (\emph{orange}); $L_{12}$
 (\emph{cyan}); $L_{13}$ (\emph{olive}); $L_{14}$ (\emph{maroon}); $L_{15}$ (\emph{gray}); $L_{16}$ (\emph{blue}); $L_{17}$
  (\emph{darker magenta}); $L_{18}$ (\emph{darker purple}); $L_{19}$ (\emph{darker gray}); $L_{20}$ (\emph{black}); $L_{21}$
   (\emph{light pink}); $L_{22}$ (\emph{light blue});
non-converging points (\emph{white}).  The \emph{black} dots show the positions of the libration points. (Color figure online).} 
\label{fig:11}
\end{figure*}
\begin{figure*}[!t]
\begin{center}
(a)\resizebox{0.45\hsize}{!}{\includegraphics*{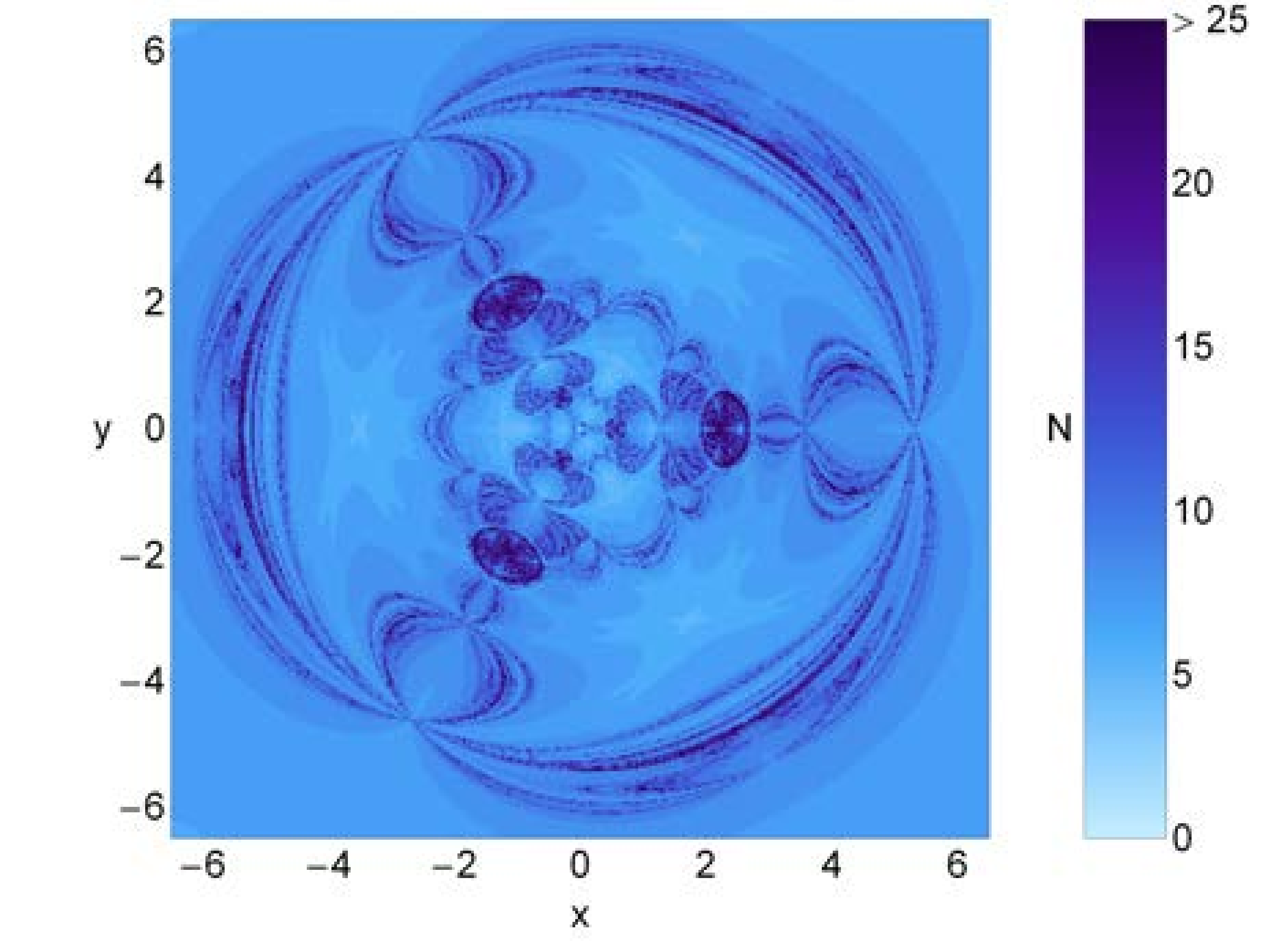}}
(b)\resizebox{0.45\hsize}{!}{\includegraphics*{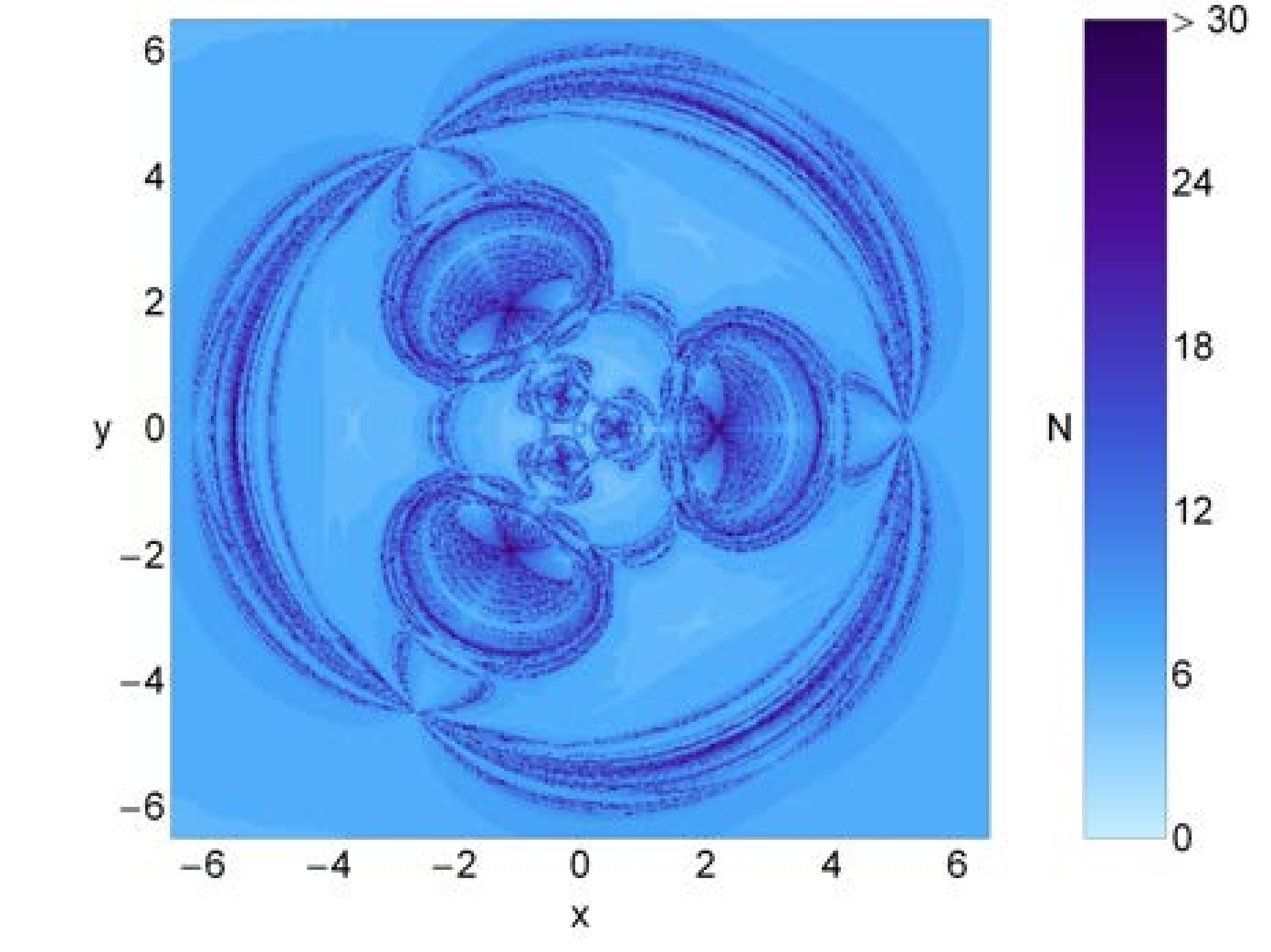}}
(c)\resizebox{0.45\hsize}{!}{\includegraphics*{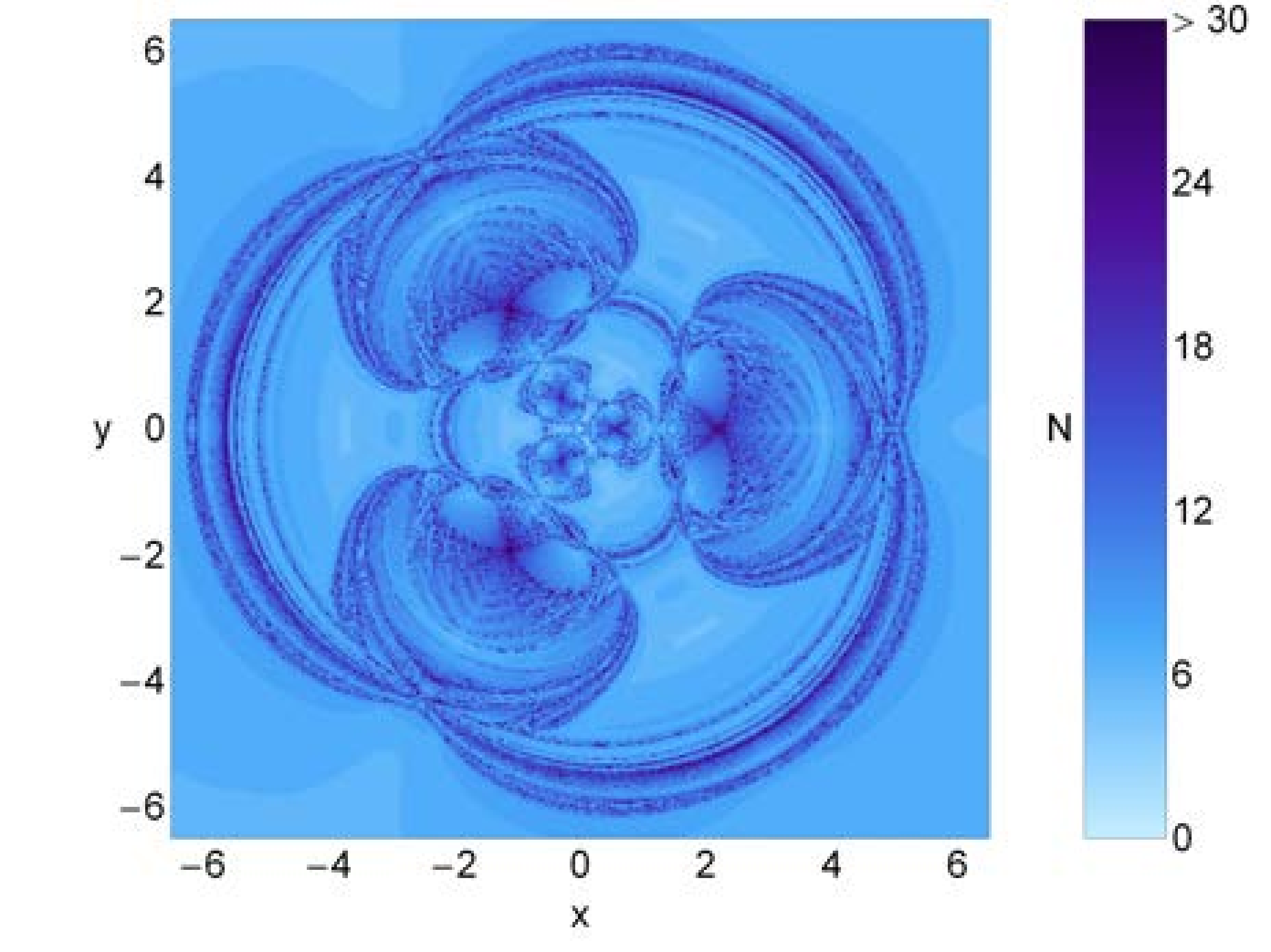}}
(d)\resizebox{0.45\hsize}{!}{\includegraphics*{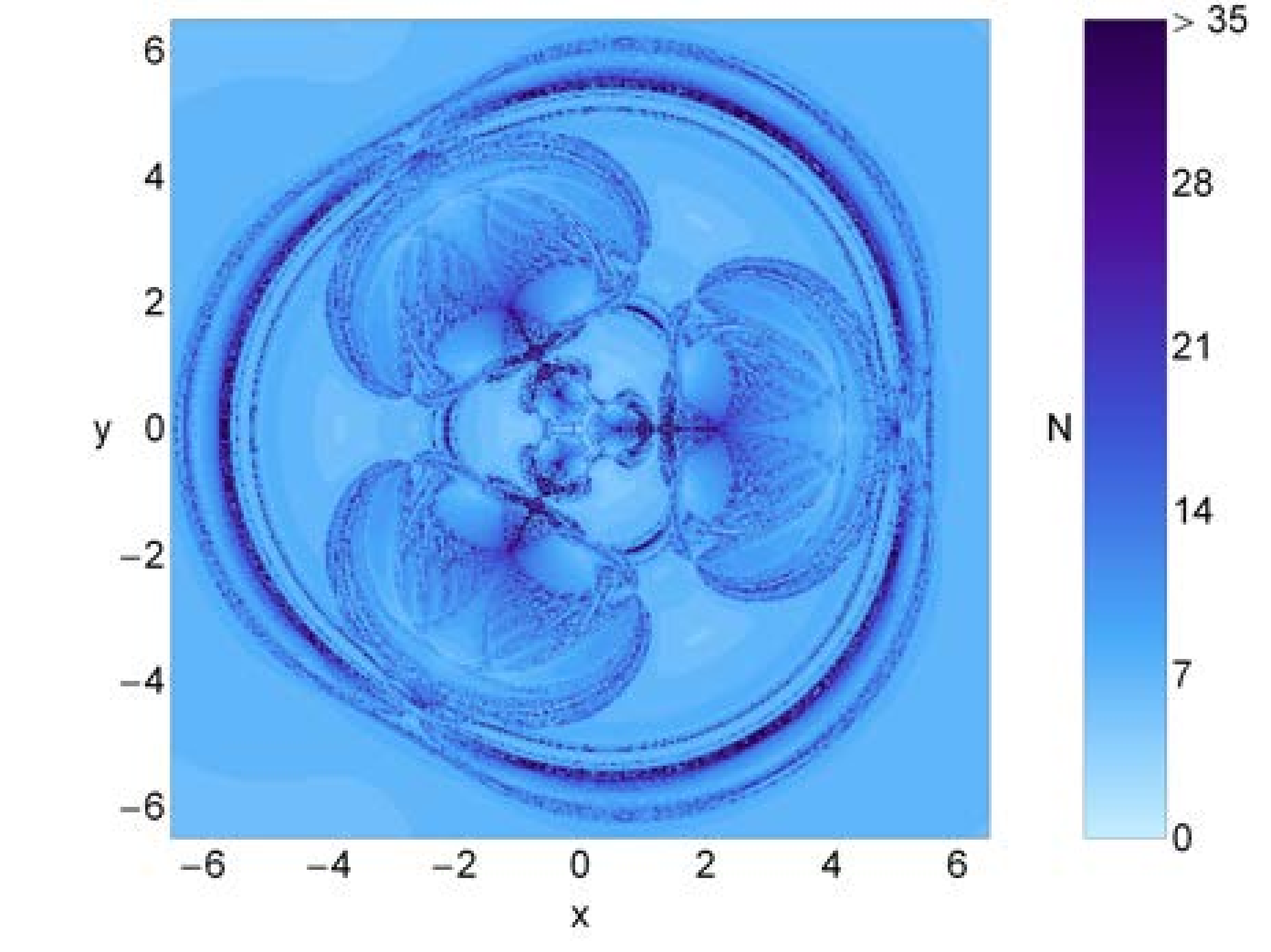}}
(e)\resizebox{0.45\hsize}{!}{\includegraphics*{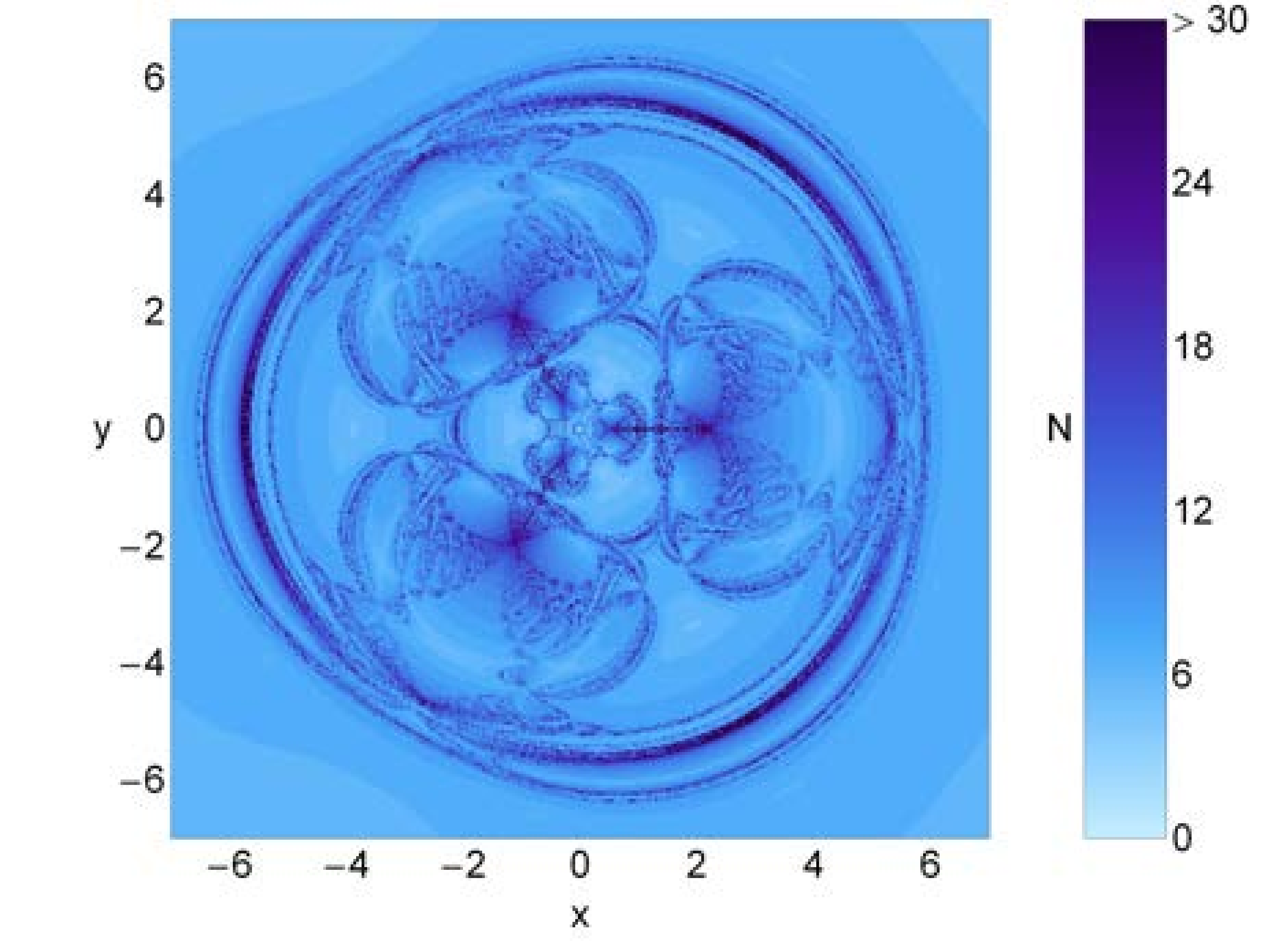}}
(f)\resizebox{0.45\hsize}{!}{\includegraphics*{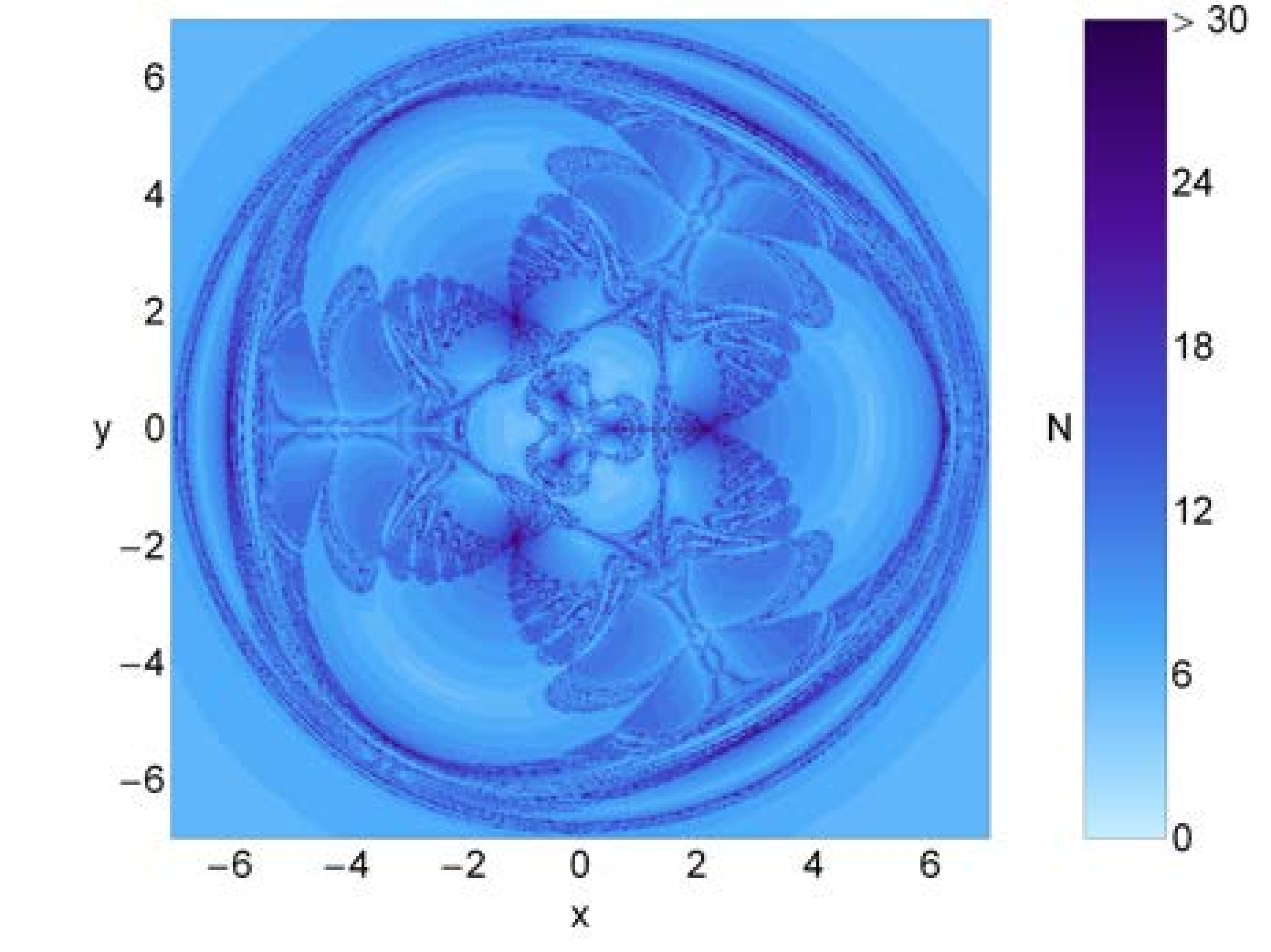}}
\end{center}
\caption{The distribution of the
corresponding number $N$ of
required iterations for obtaining
the Newton-Raphson basins of
attraction shown in Fig. \ref{fig:11}(a--f). (Color figure online).} 
\label{fig:11a}
\end{figure*}
\begin{figure*}[!t]
\begin{center}
(a)\resizebox{0.4\hsize}{!}{\includegraphics*{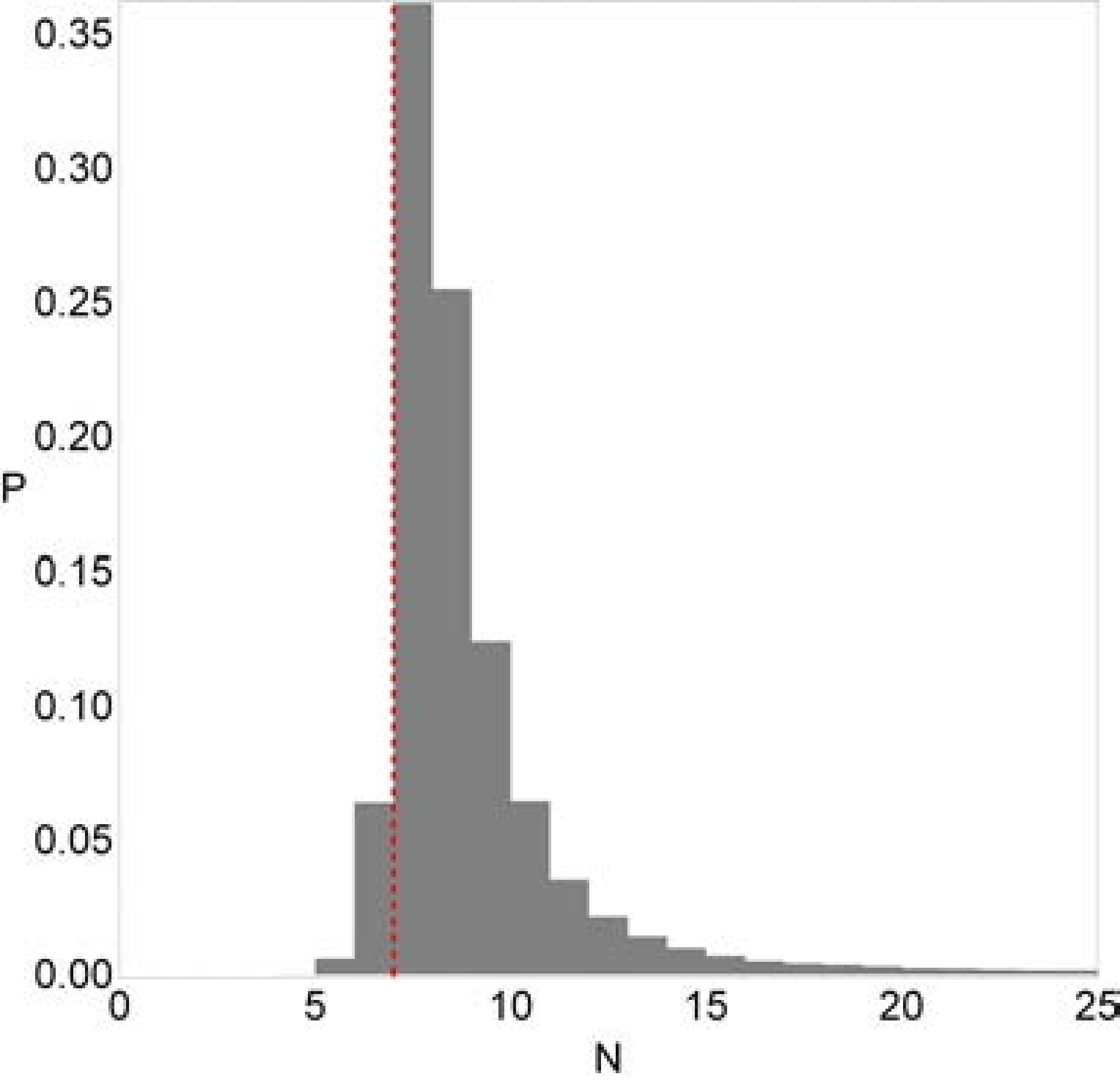}}
(b)\resizebox{0.4\hsize}{!}{\includegraphics*{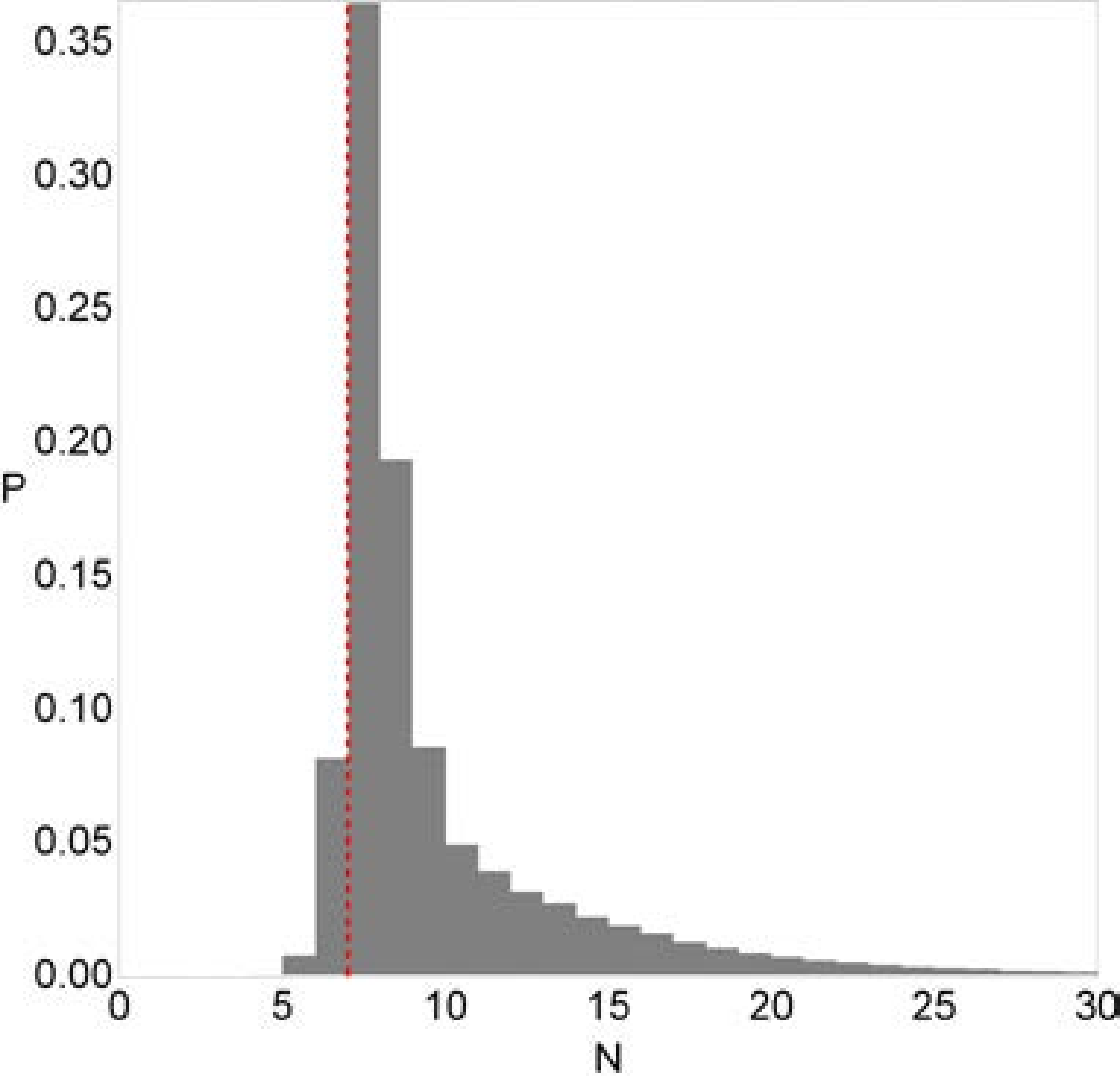}}
(c)\resizebox{0.4\hsize}{!}{\includegraphics*{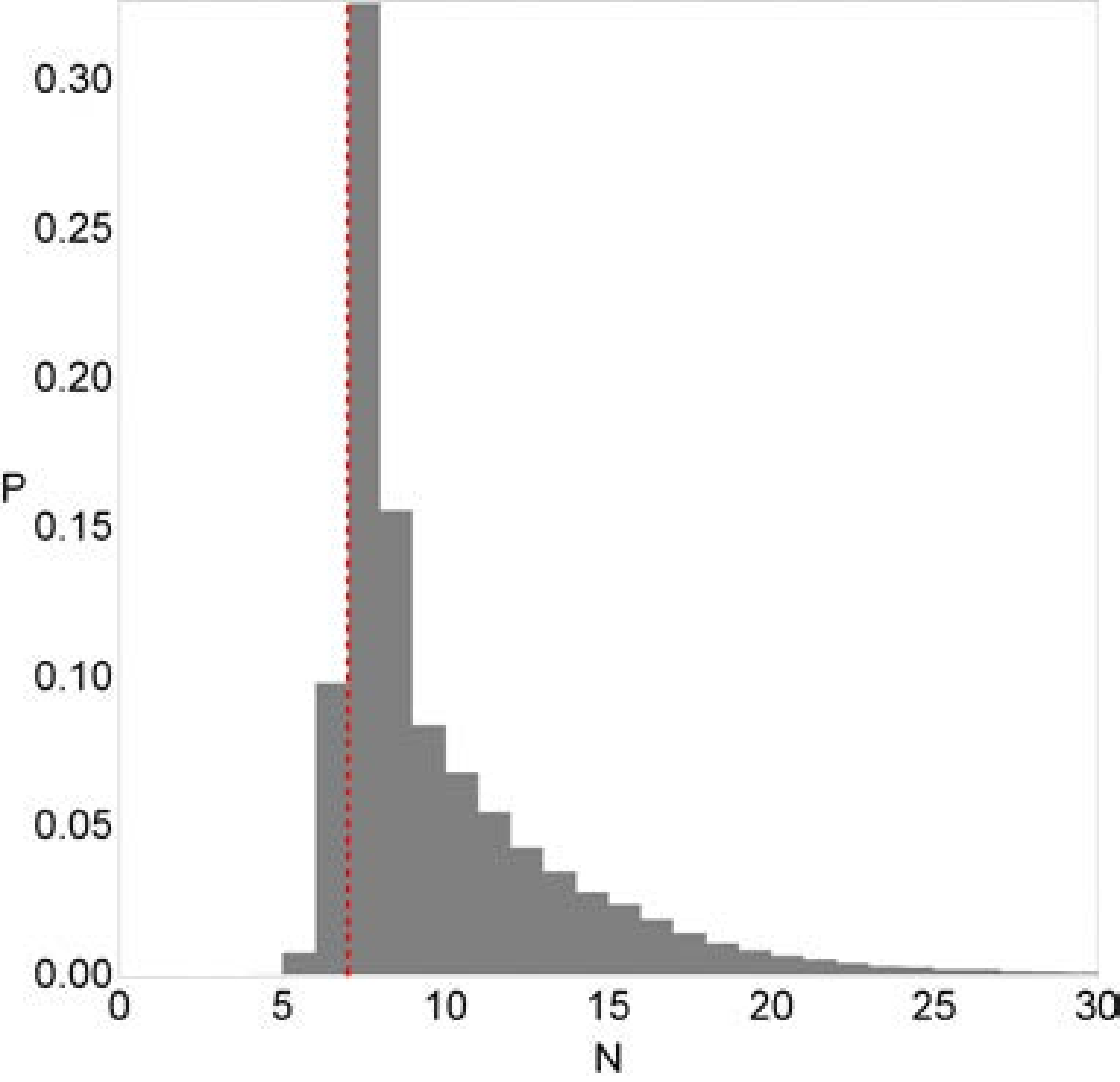}}
(d)\resizebox{0.4\hsize}{!}{\includegraphics*{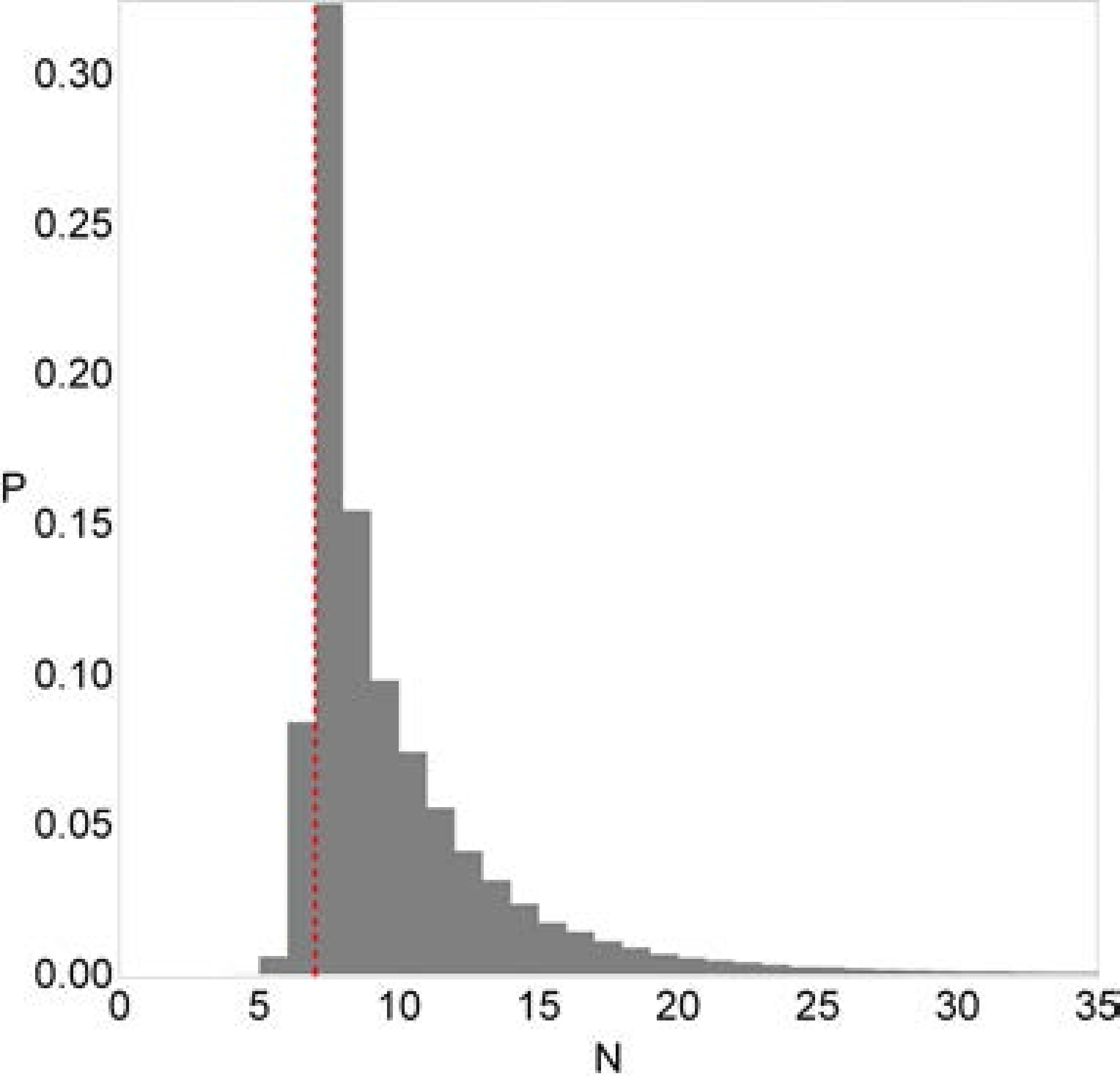}}
(e)\resizebox{0.4\hsize}{!}{\includegraphics*{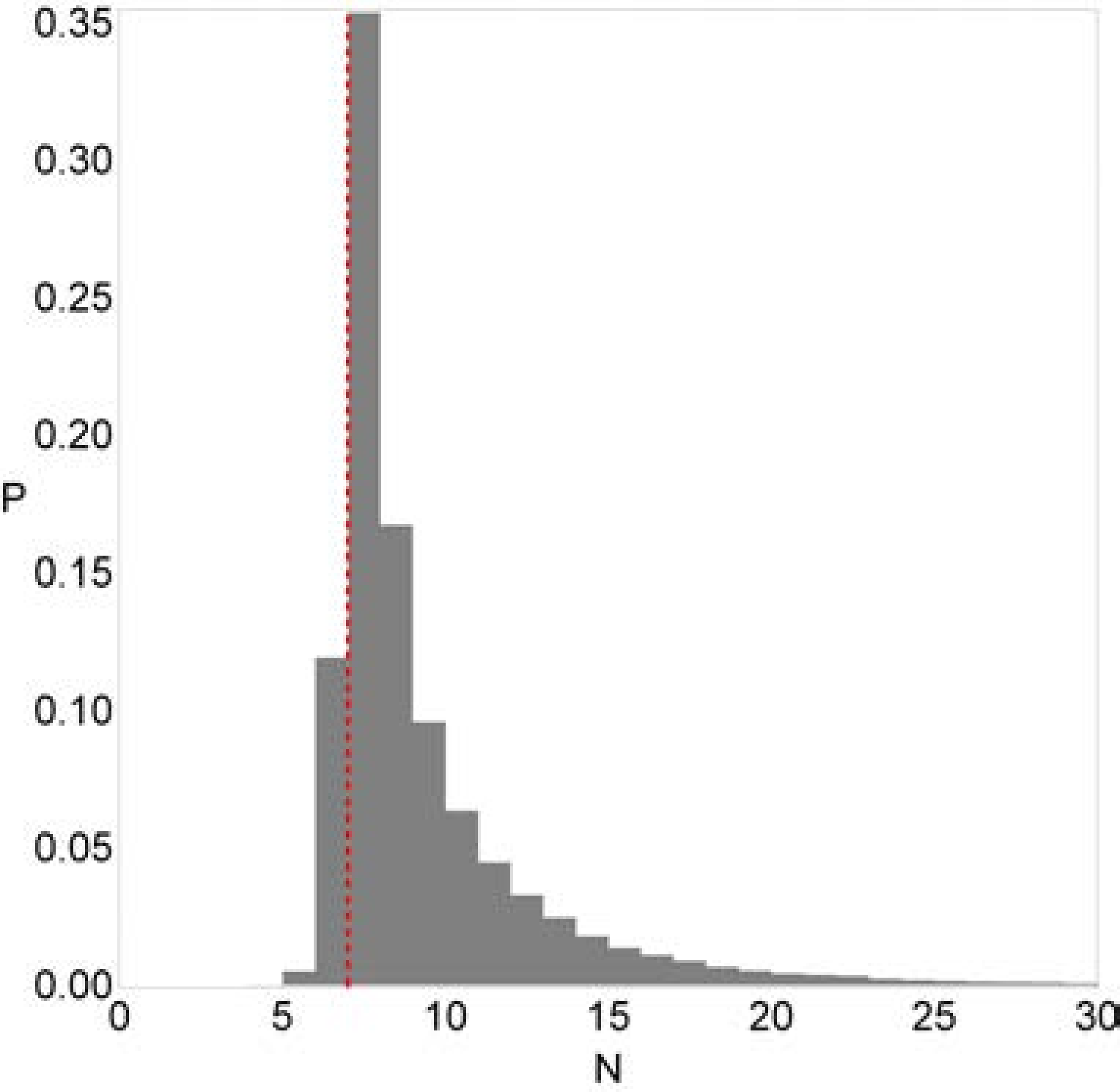}}
(f)\resizebox{0.4\hsize}{!}{\includegraphics*{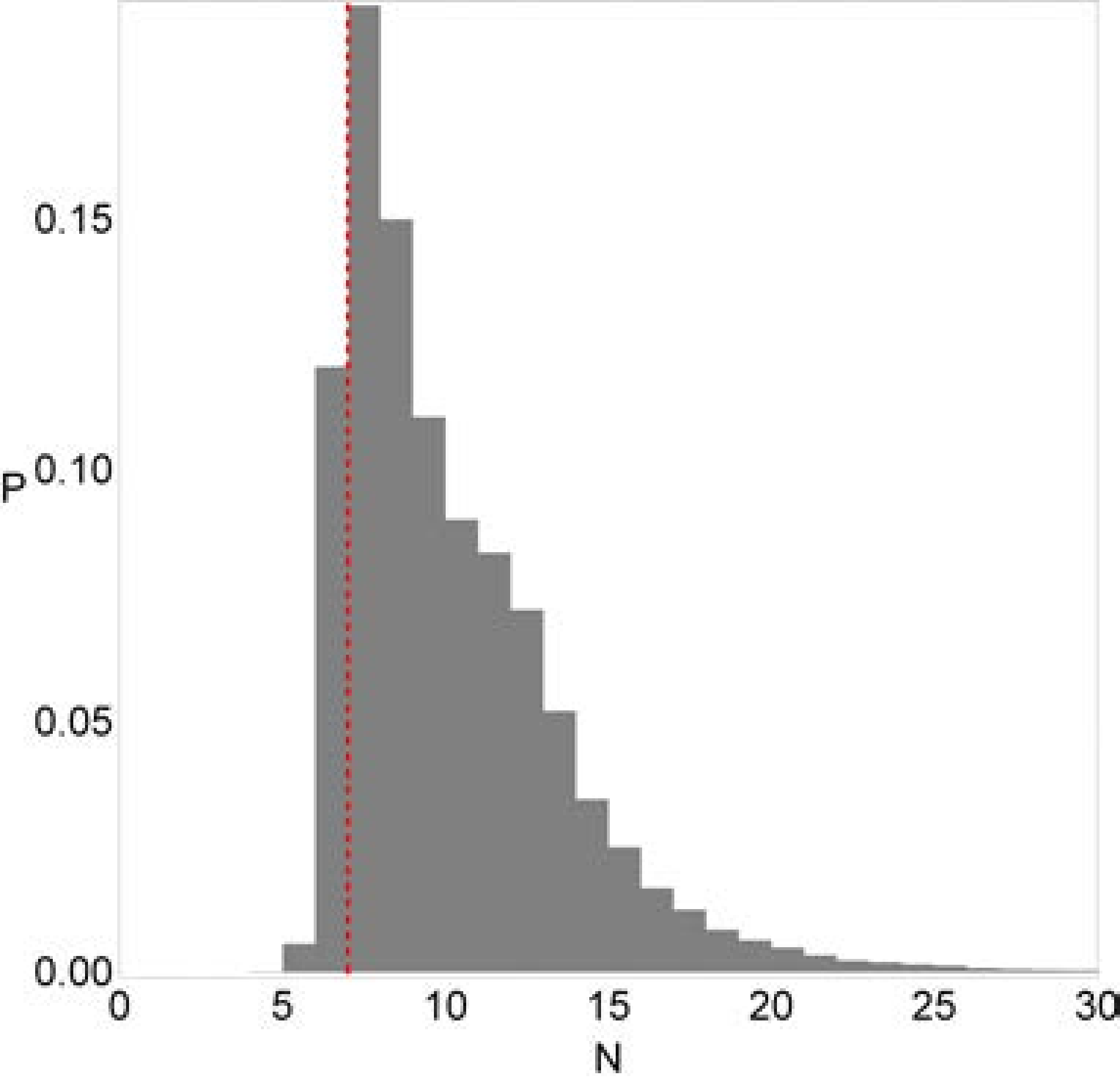}}
\end{center}
\caption{The corresponding
probability distribution of
required iterations for obtaining
the Newton-Raphson basins of
attraction shown in Fig. \ref{fig:11}(a-f).
The vertical dashed red line
indicates, in each case, the most
probable number $N^{*}$ of
iterations. (Colour figure online).} 
\label{fig:11b}
\end{figure*}
\subsection{Case 1: Twenty-two libration points} \label{Case 1: Twenty-two libration points}
We begin with the first case where twenty-two libration points exist, i.e., when $0<\epsilon \leq 0.67752839$ and  $0.704528<\epsilon \leq0.812528$ . In Fig. \ref{fig:11}, we present the evolution of the Newton-Raphson basins of attraction for six different values of the transition parameter. To separate the different initial conditions on the configuration $(x, y) $ plane, the color-coded diagrams are used, where each pixel is assigned a color, according
to the attractor of the initial conditions. It is observed that most of the configuration $(x, y)$ plane is covered
by well formed Newton-Raphson basins of attraction, while all basins boundaries
are highly chaotic.
From Fig. \ref{fig:11}, it is observed that in all the panels, the extent of the basins of convergence associated with the libration  point $L_4$ is infinite. Furthermore, in Fig.\ref{fig:11}a, we observe that the shape of the basins of convergence, associated with the libration points $L_2$, $L_7$, and $L_8$ look like exotic bugs with many legs and many antennas. It is also observed that the shape of the basins of convergence, associated with the libration points $L_1$, $L_5$, and $L_6$ are identical in all the panels. Moreover, due to the symmetry, the shape of the basins of convergence associated with the non-collinear libration points are identical and exist in pairs. It is interesting to note that the geometry of the basins of convergence and their topology,
when the transition parameter $\epsilon \in (0,  0.67752839)$  are completely different in comparison of the geometry of the basins of convergence when the transition parameter $\epsilon \in (0.704528, 0.812528)$. However, in both the cases there exist twenty-two libration points. \\
The geometry of the Newton-Raphson basins of convergence of the configuration $(x, y)$ plane changes
substantially as the value of transition parameter $\epsilon$ increases. Following are the significant changes that take place, when the transition parameter $\epsilon \in (0,  0.67752839)$:
\begin{description}
  \item[--] The extent of the basins of attraction, associated with the libration points $L_1, L_5$, and $L_6$, remains almost unchanged.
  \item[--] The extent of the basins of attraction, associated with the libration points $L_2, L_7$, and $L_{8}$,  decreases. The exotic bugs-like regions, corresponding to these libration points, are distorted and disappear completely with the increase in transition parameter.
  \item[--] The basins of attraction, associated with the libration points $L_{3}, L_{9}$, and $L_{10}$, decrease.
\end{description}
Following are the most significant changes when the transition parameter $\epsilon \in (0.704528, 0.812528)$:
\begin{description}
  \item[--] The extent of the basins of attraction associated with the libration points $L_1, L_5$, $L_6$,  $L_{7}, L_{8}$, $L_{13}, L_{14}$,  $L_{17}$, and $L_{19}$  remains almost unchanged.
  \item[--] The extent of the basins of attraction associated with the libration points $L_3, L_9$, and $L_{10}$  decreases.
  \item[--] The extent of the basins of attraction associated with the libration points $L_{2},  L_{15},$ and $L_{21}$ increases.
\end{description}
In Fig. \ref{fig:11a} (a-f), the distribution of the corresponding number $N$ of the iterations are shown, using tons of blue. Our analysis reveals that the initial conditions inside the attracting regions converge relatively fast $(N <10)$. On the other hand, the initial conditions in the neighbourhood of the basins boundaries are the slowest converging points $(N>30)$. The corresponding probability distributions of iterations are presented in Fig. (\ref{fig:11b}). If $N_t$ is the total number of initial conditions in every color-coded diagram and $N_0$ represents the initial condition $(x_0, y_0)$  converges to one of the attractors after $N$ iterations, then the probability $P$ is defined by $P=\frac{N_0}{N_t}$. It is also observed that the most probable number $N^{*}$ of iterations is equal to 7 (the red vertical dashed line in Fig.\ref{fig:11b}). The value remains unfazed throughout this region for values of the transition parameter.

\subsection{Case II: Sixteen libration points} \label{Case II: Sixteen libration points}
\begin{figure*}[!t]
\begin{center}
(a)\resizebox{0.25\hsize}{!}{\includegraphics*{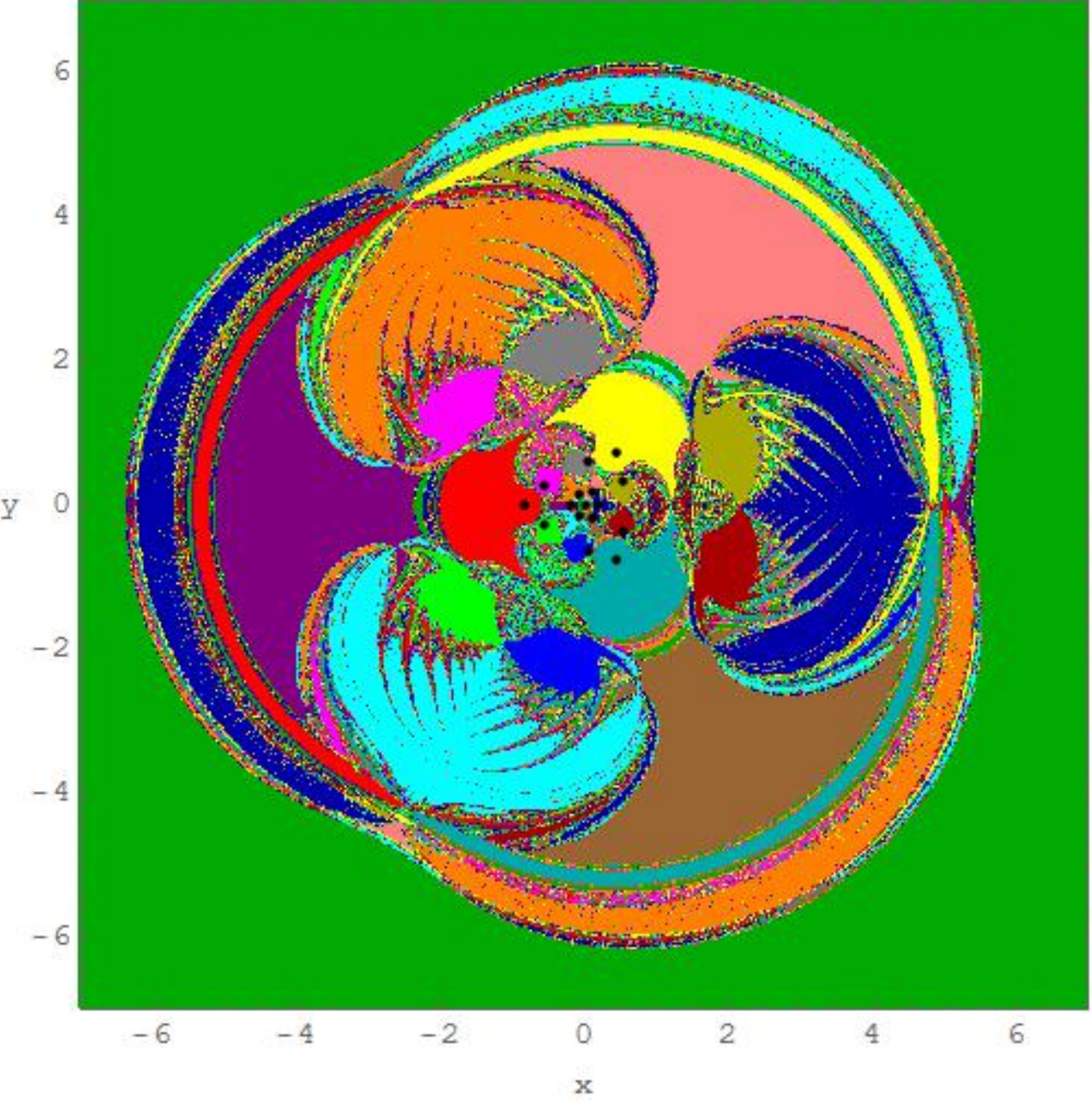}}
(b)\resizebox{0.33\hsize}{!}{\includegraphics*{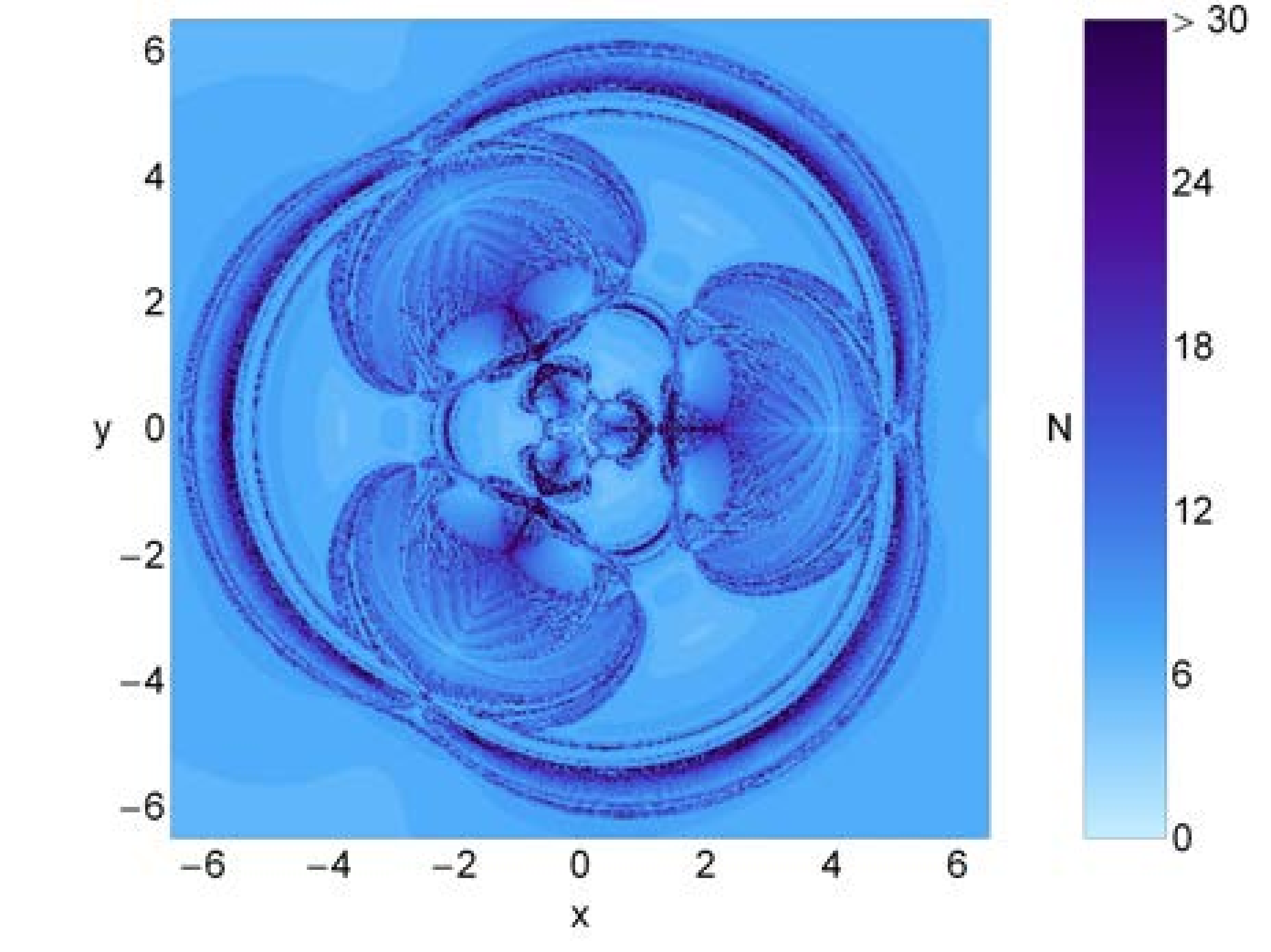}}
(c)\resizebox{0.25\hsize}{!}{\includegraphics*{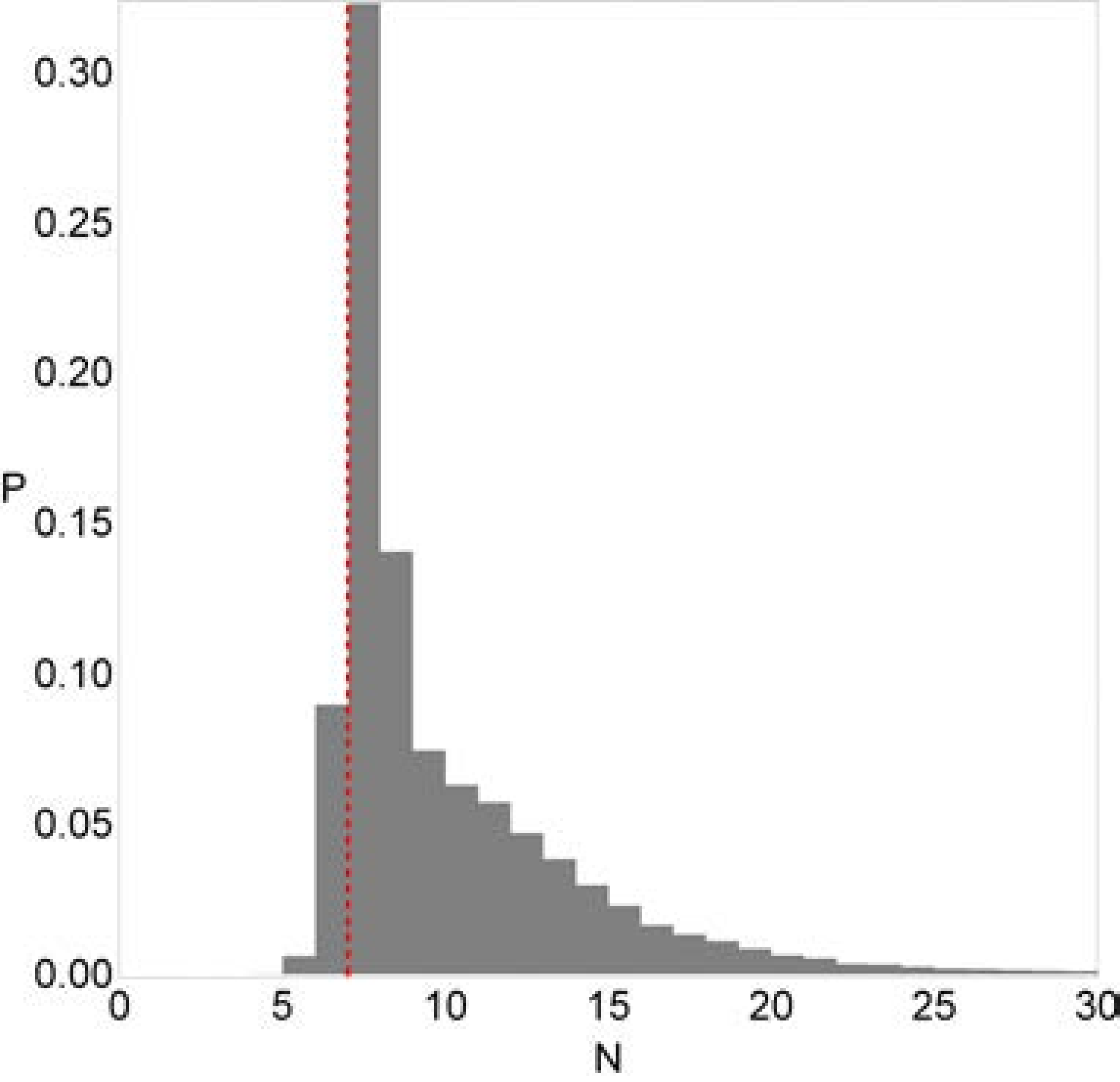}}\\
(d)\resizebox{0.25\hsize}{!}{\includegraphics*{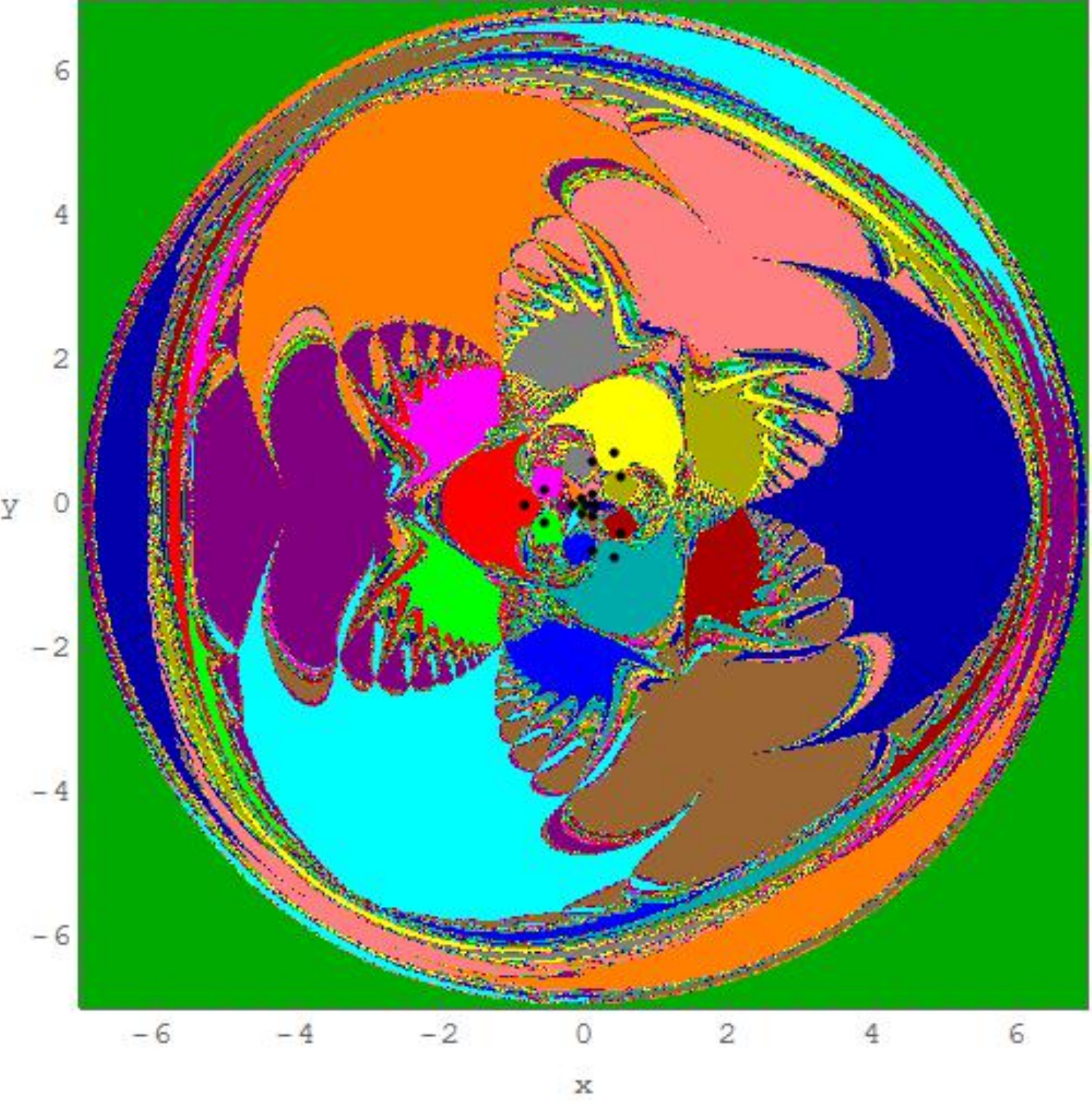}}
(e)\resizebox{0.33\hsize}{!}{\includegraphics*{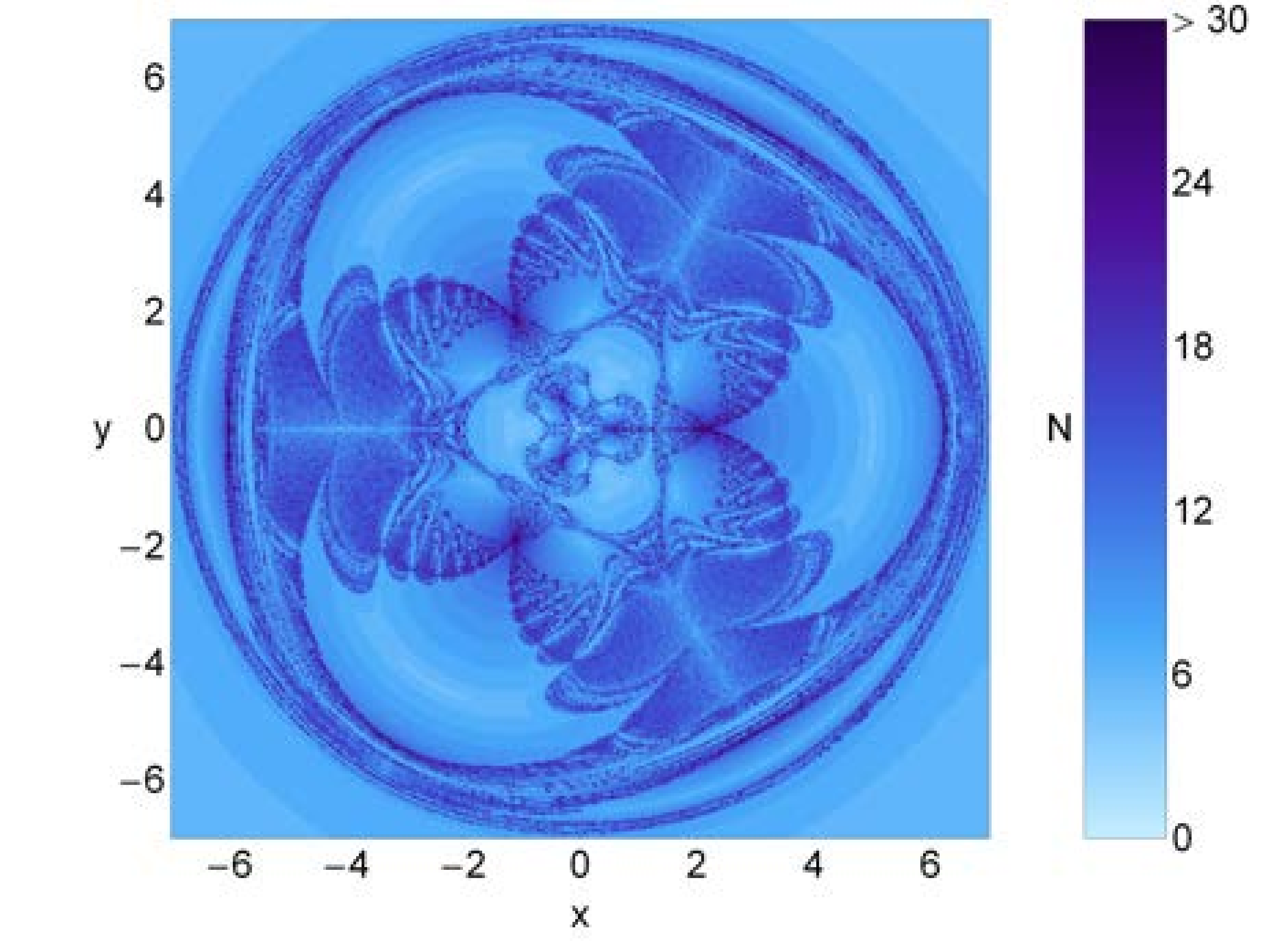}}
(f)\resizebox{0.25\hsize}{!}{\includegraphics*{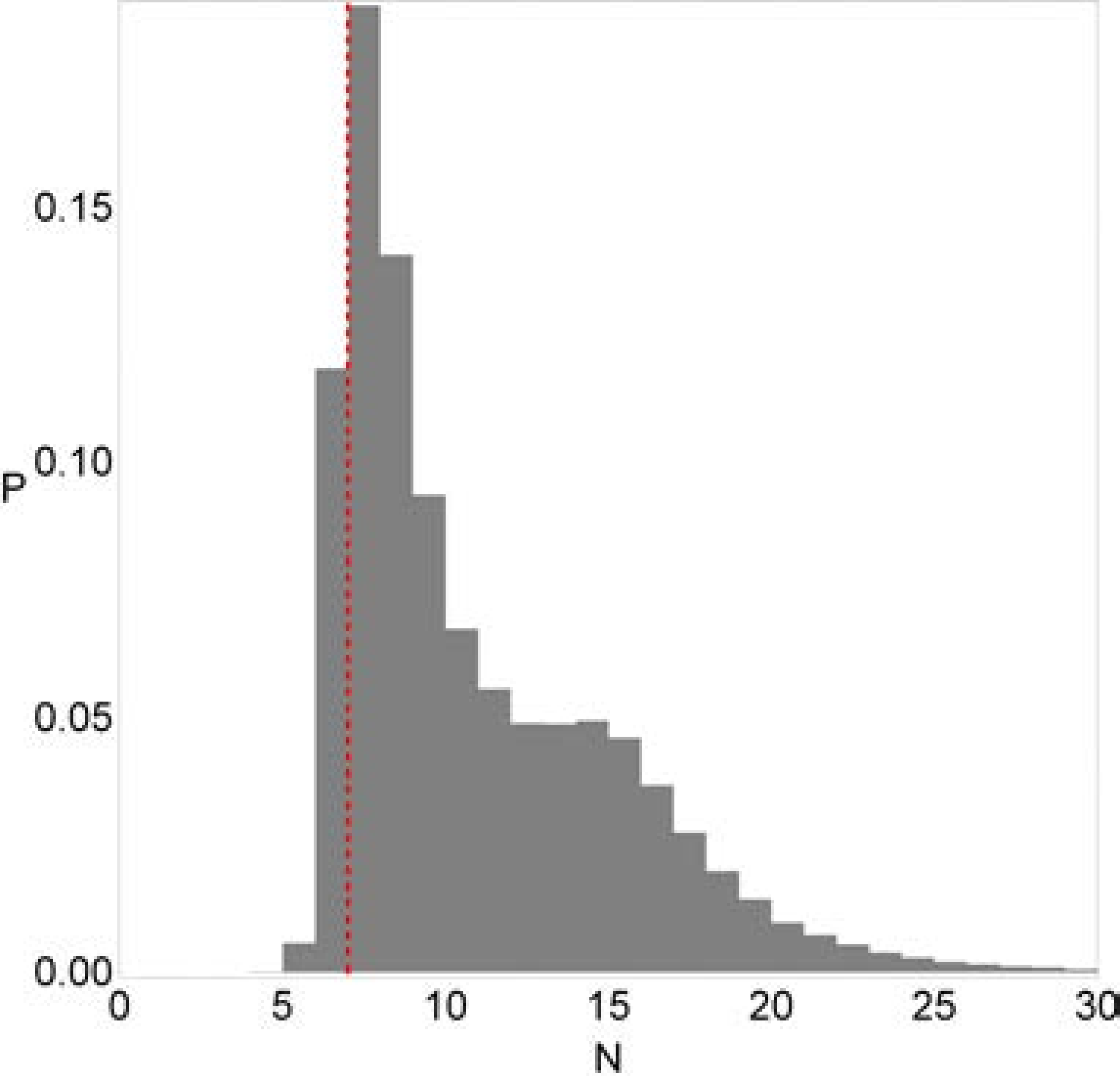}}\\
(g)\resizebox{0.25\hsize}{!}{\includegraphics*{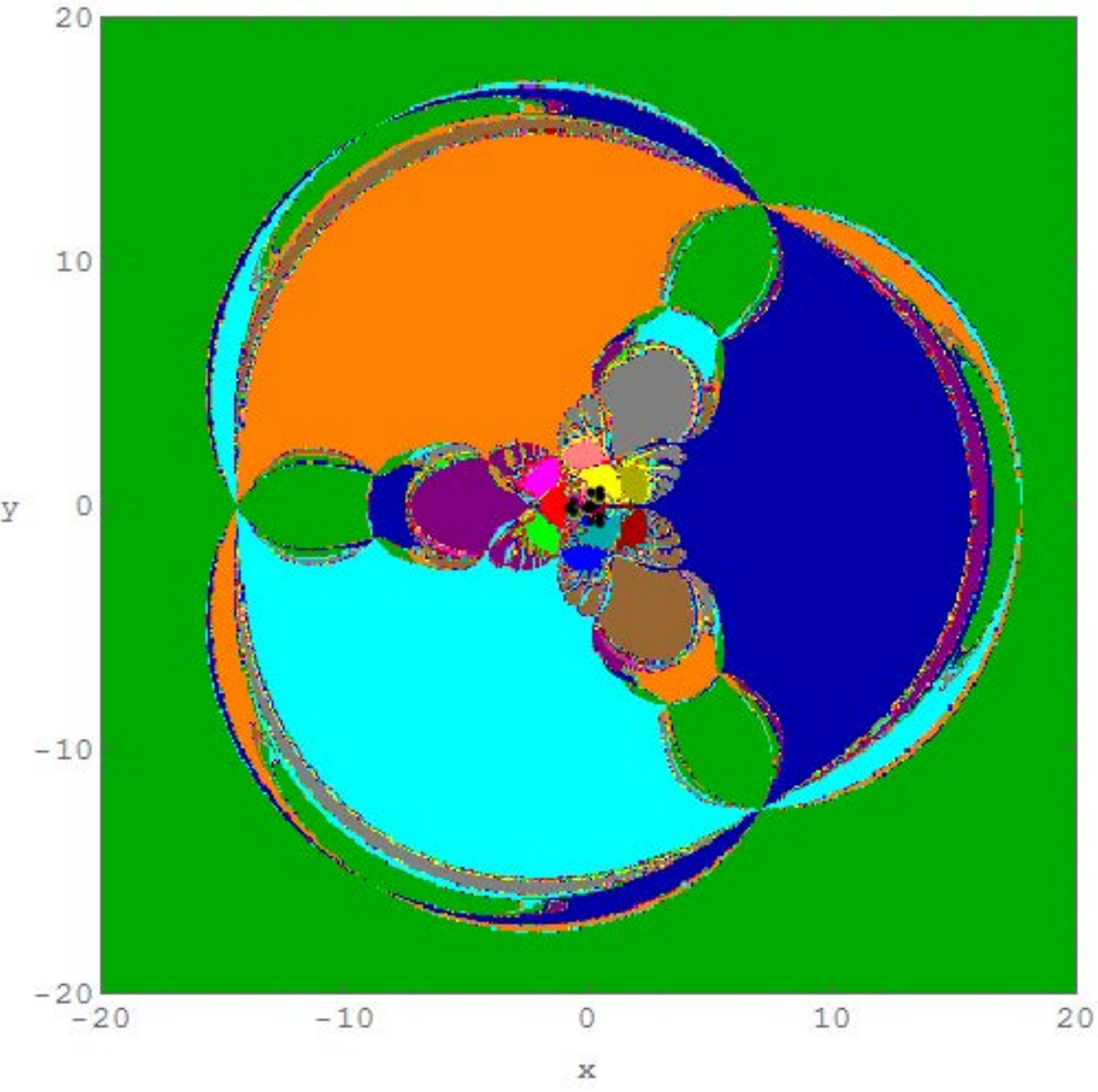}}
(h)\resizebox{0.33\hsize}{!}{\includegraphics*{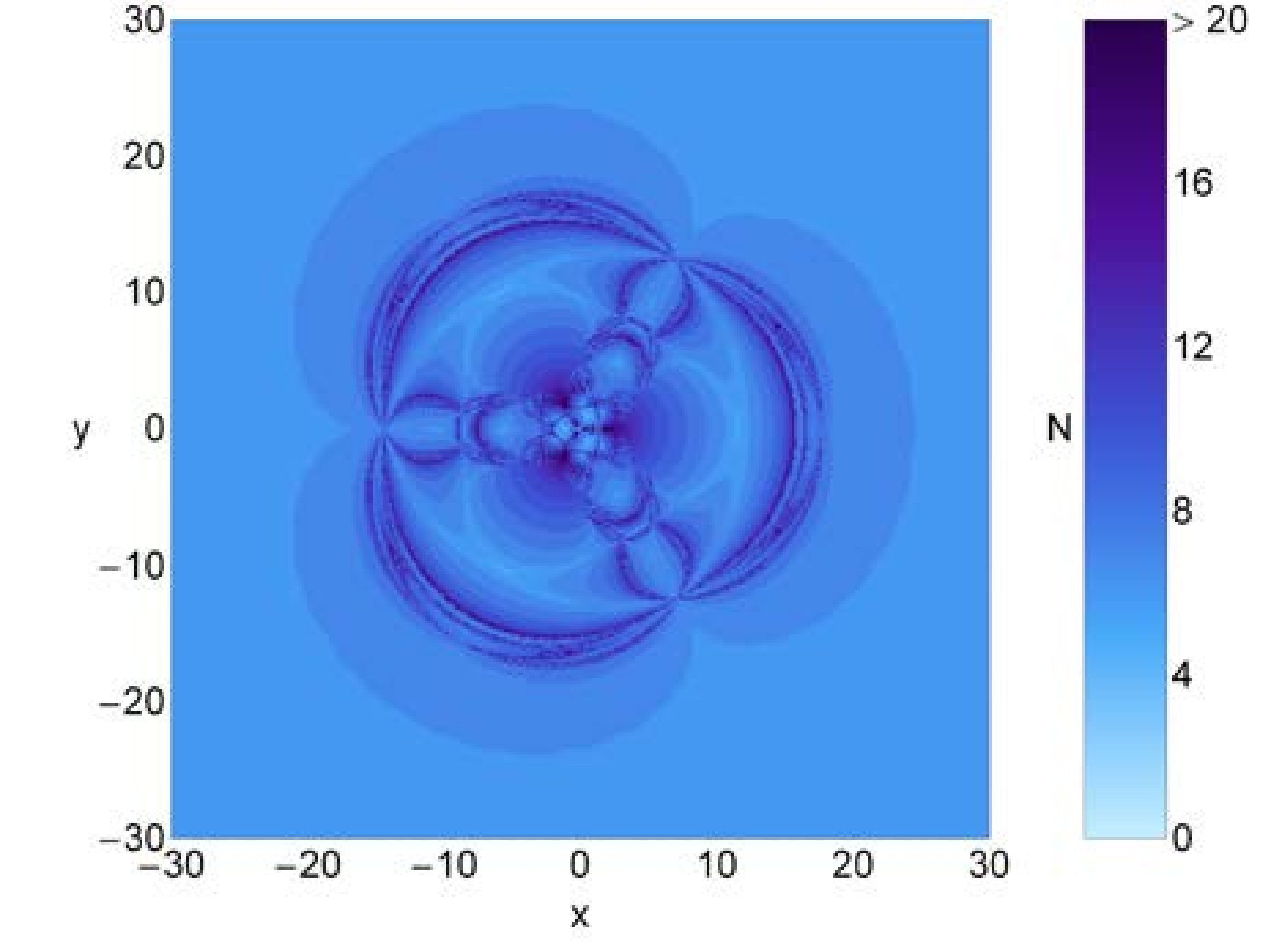}}
(i)\resizebox{0.25\hsize}{!}{\includegraphics*{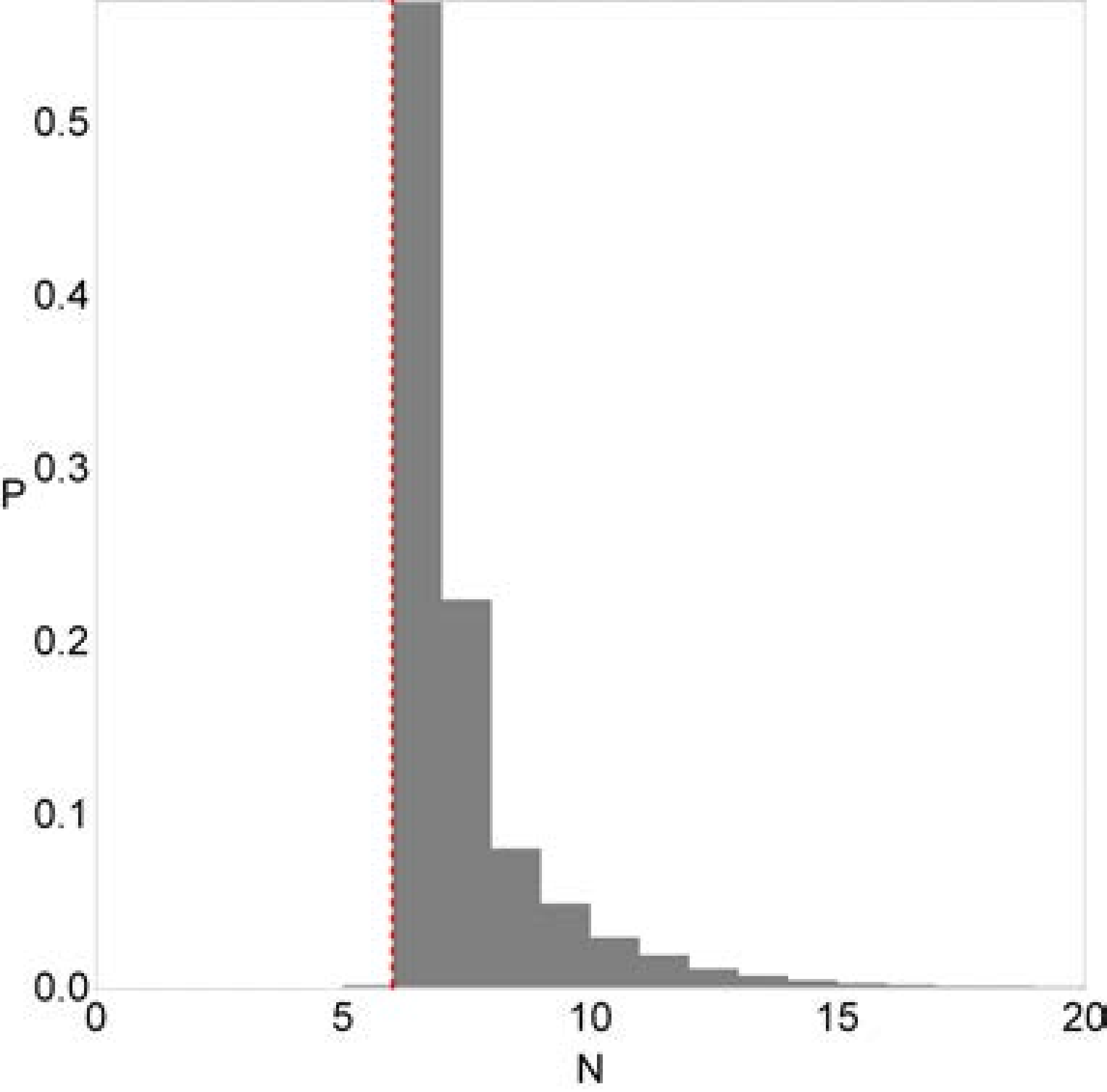}}
(j)\resizebox{0.25\hsize}{!}{\includegraphics*{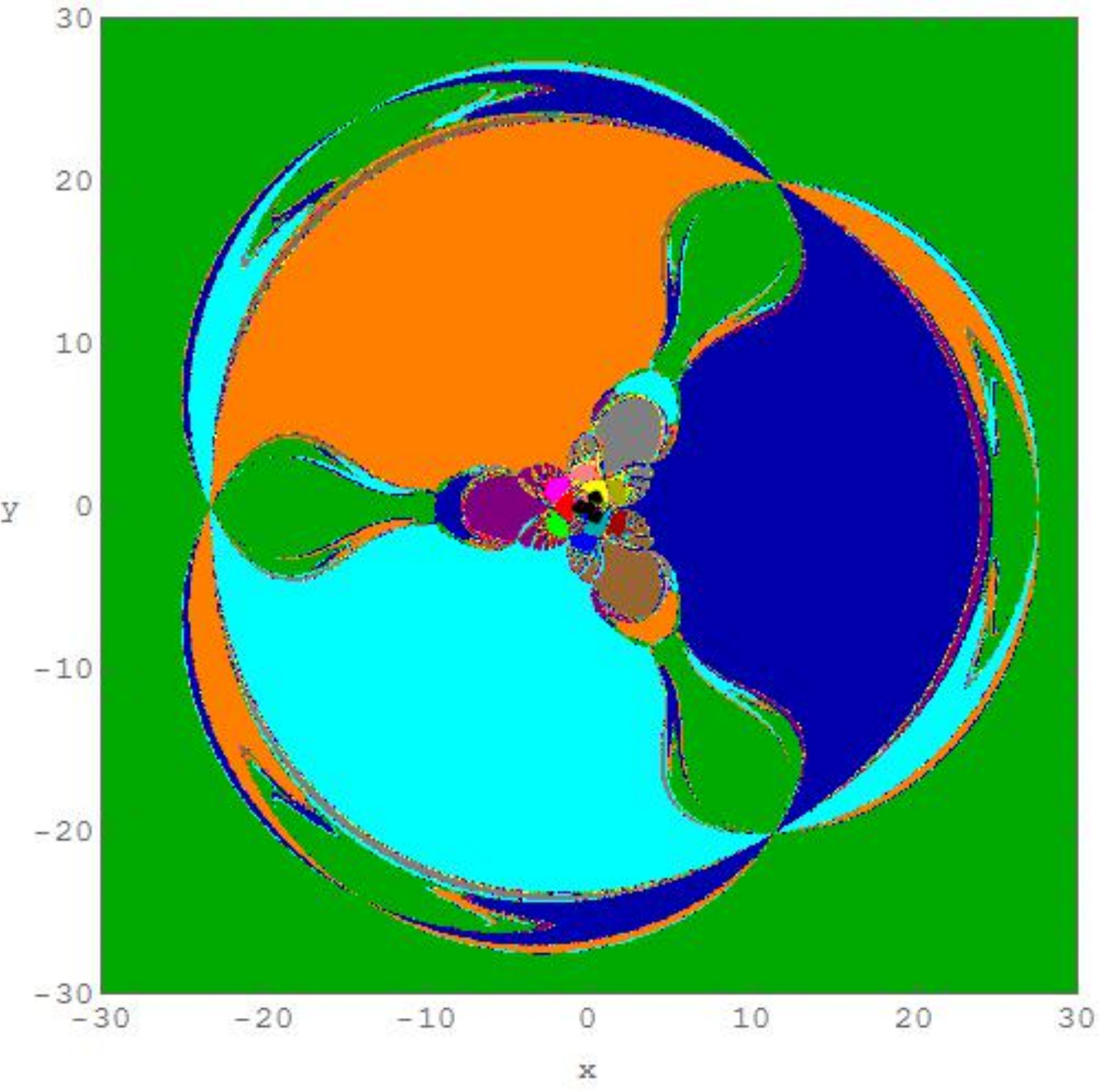}}
(k)\resizebox{0.33\hsize}{!}{\includegraphics*{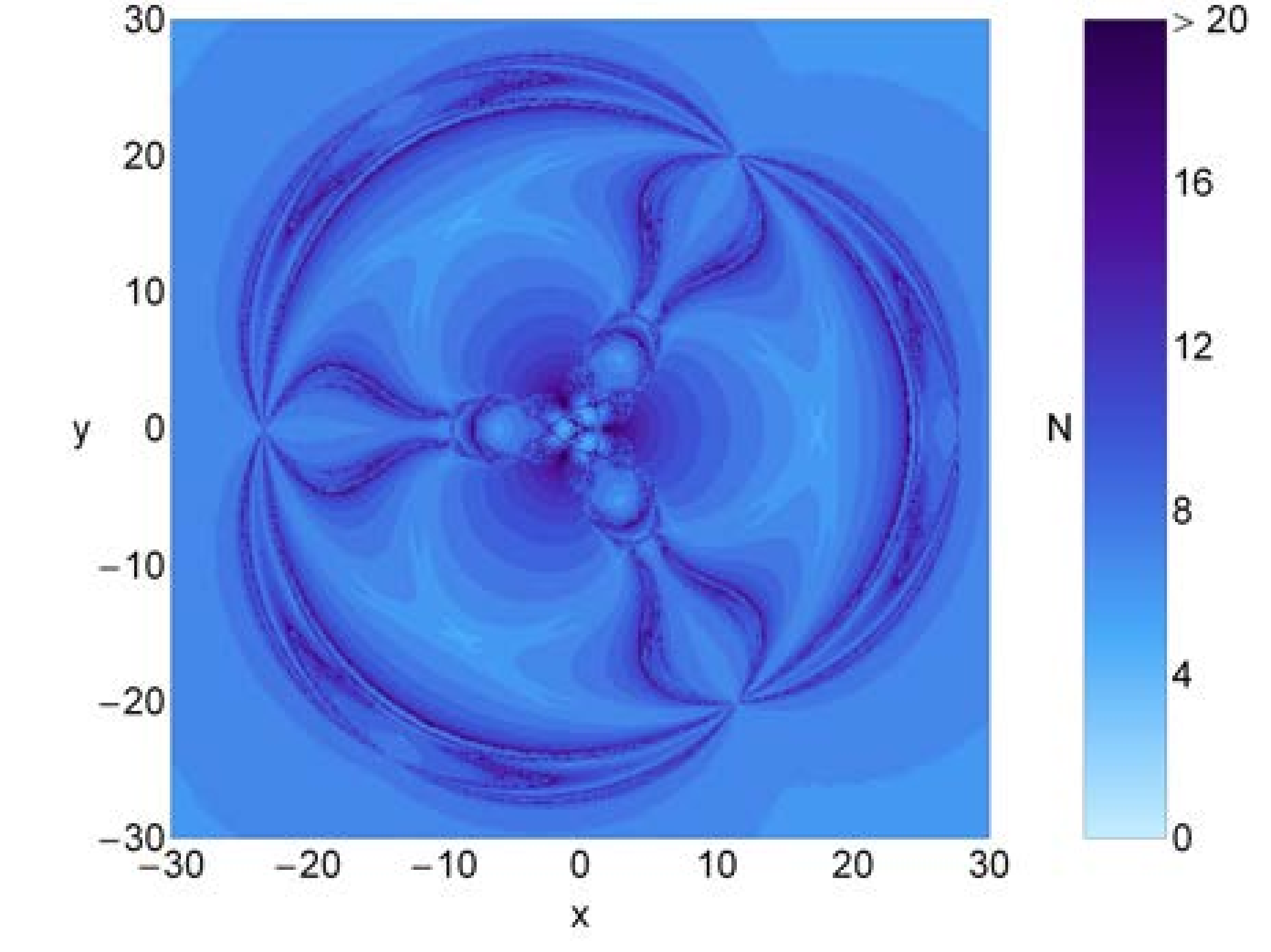}}
(l)\resizebox{0.25\hsize}{!}{\includegraphics*{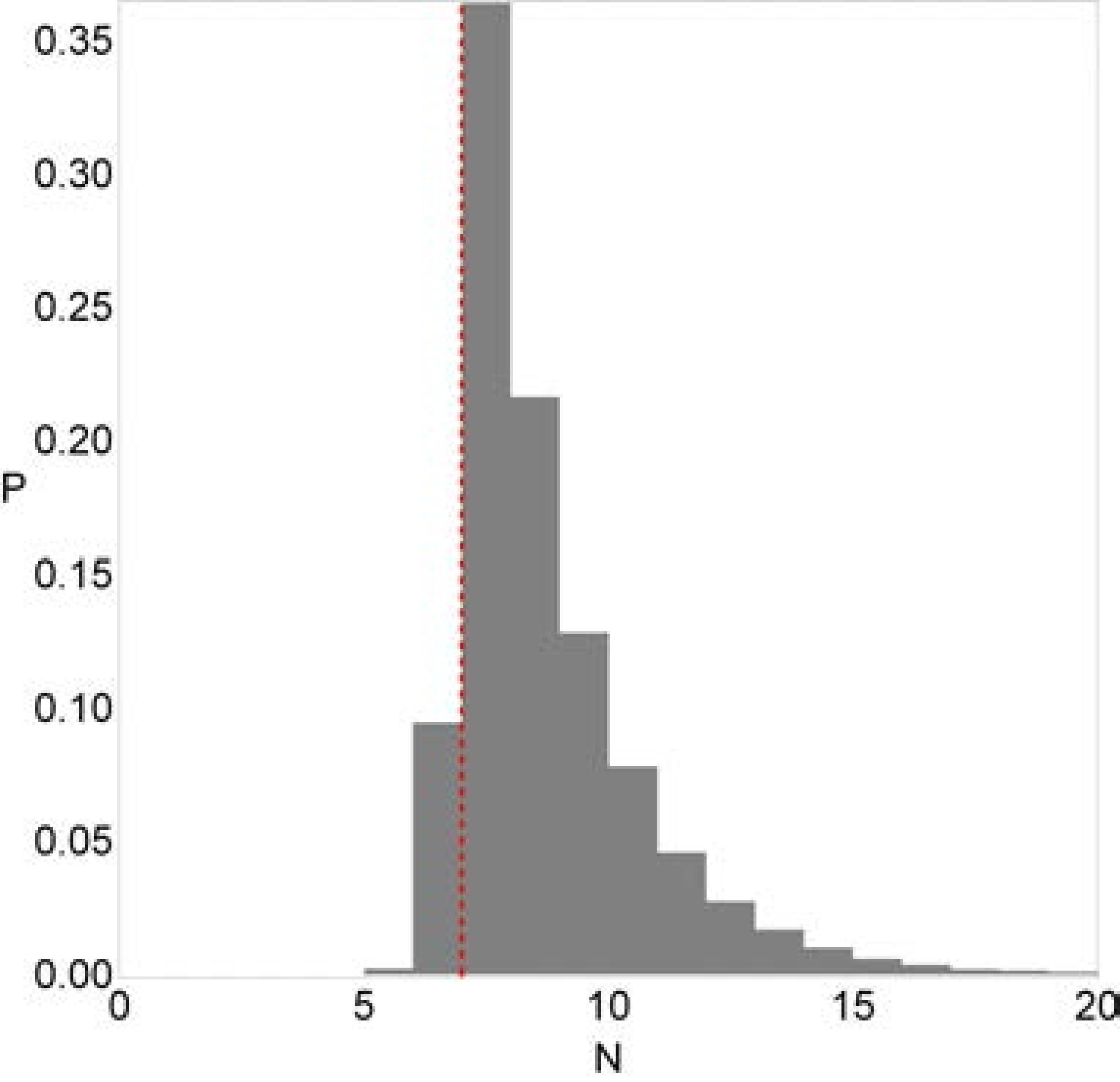}}
(m)\resizebox{0.25\hsize}{!}{\includegraphics*{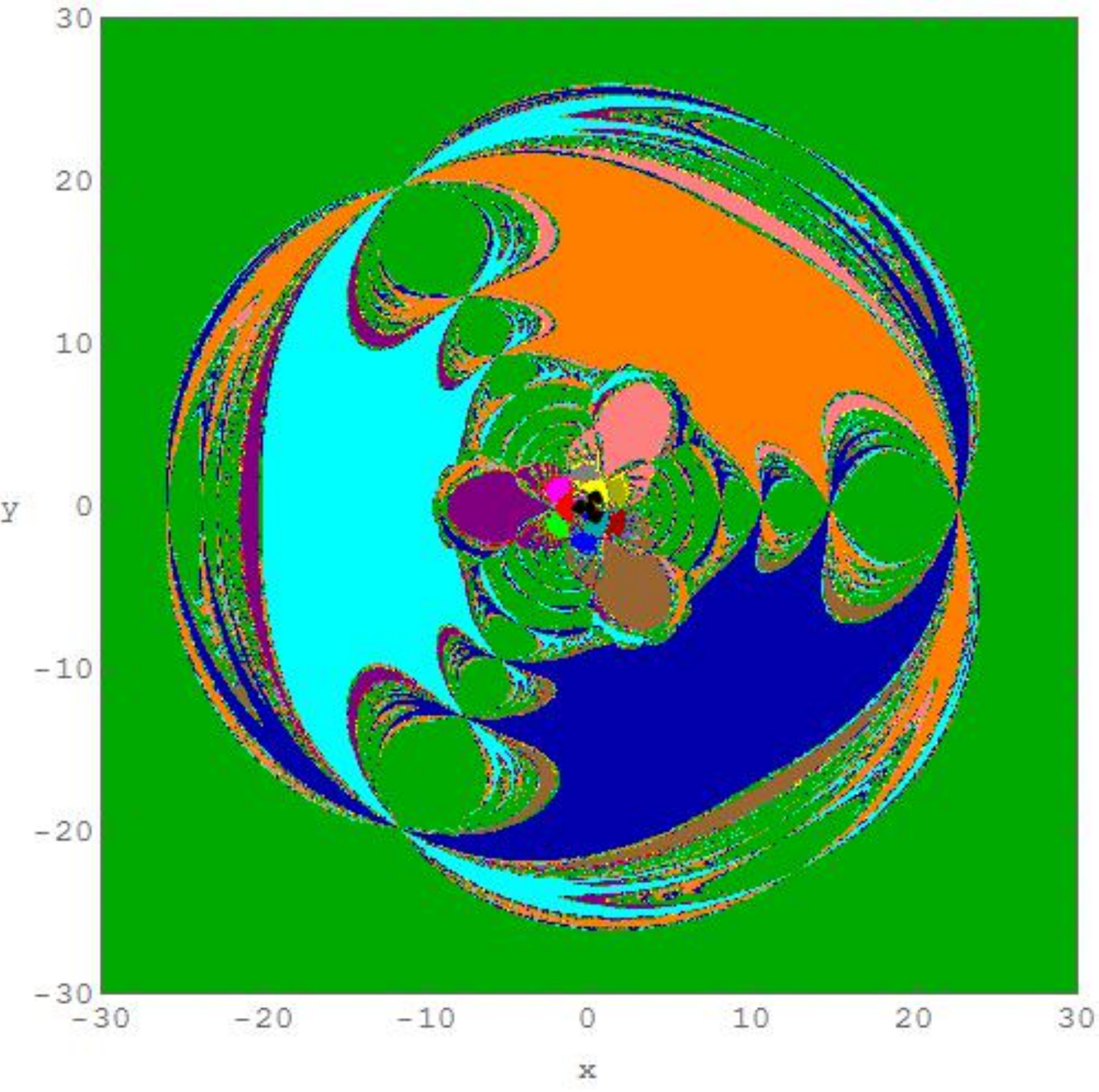}}
(n)\resizebox{0.33\hsize}{!}{\includegraphics*{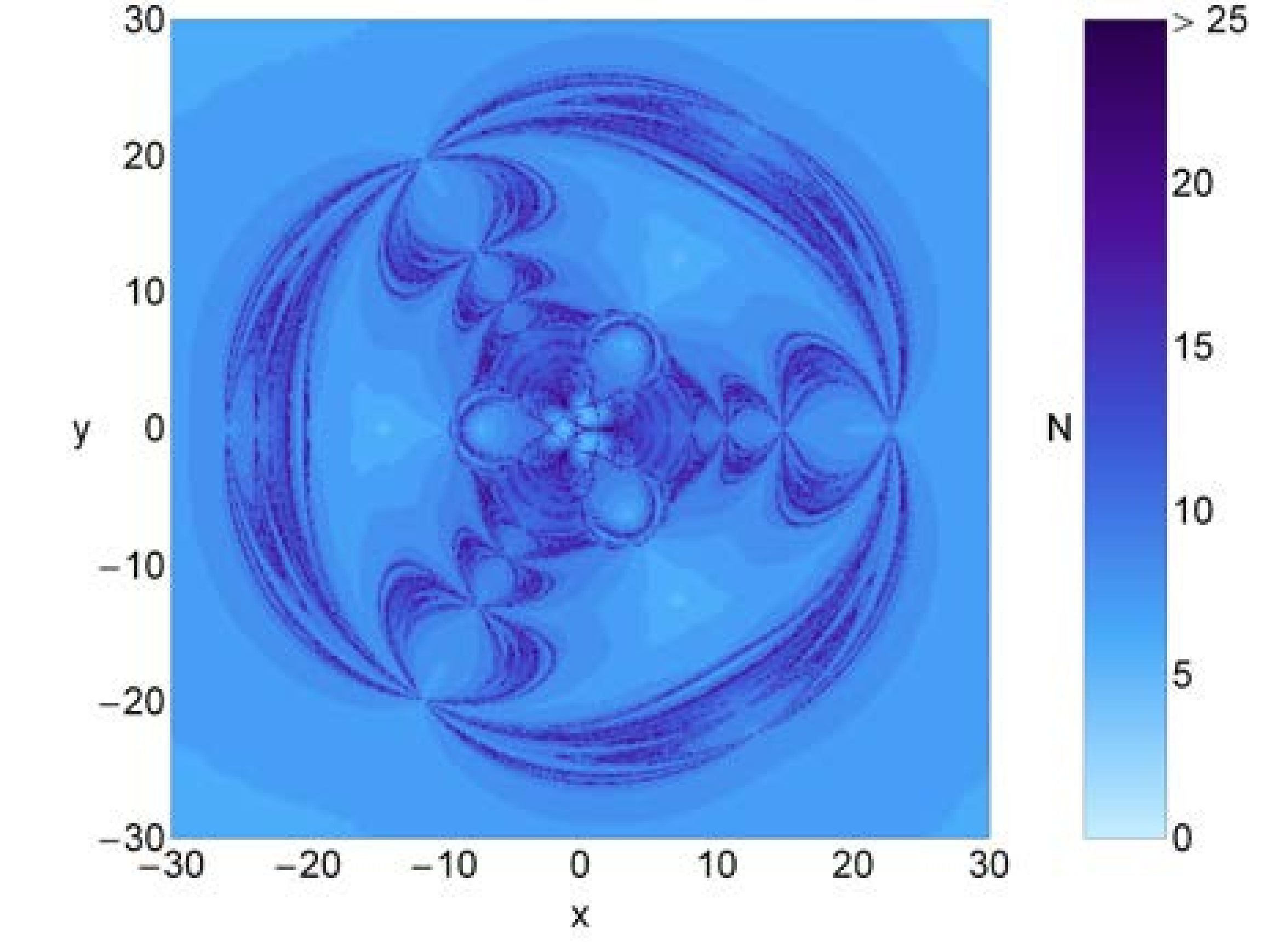}}
(o)\resizebox{0.25\hsize}{!}{\includegraphics*{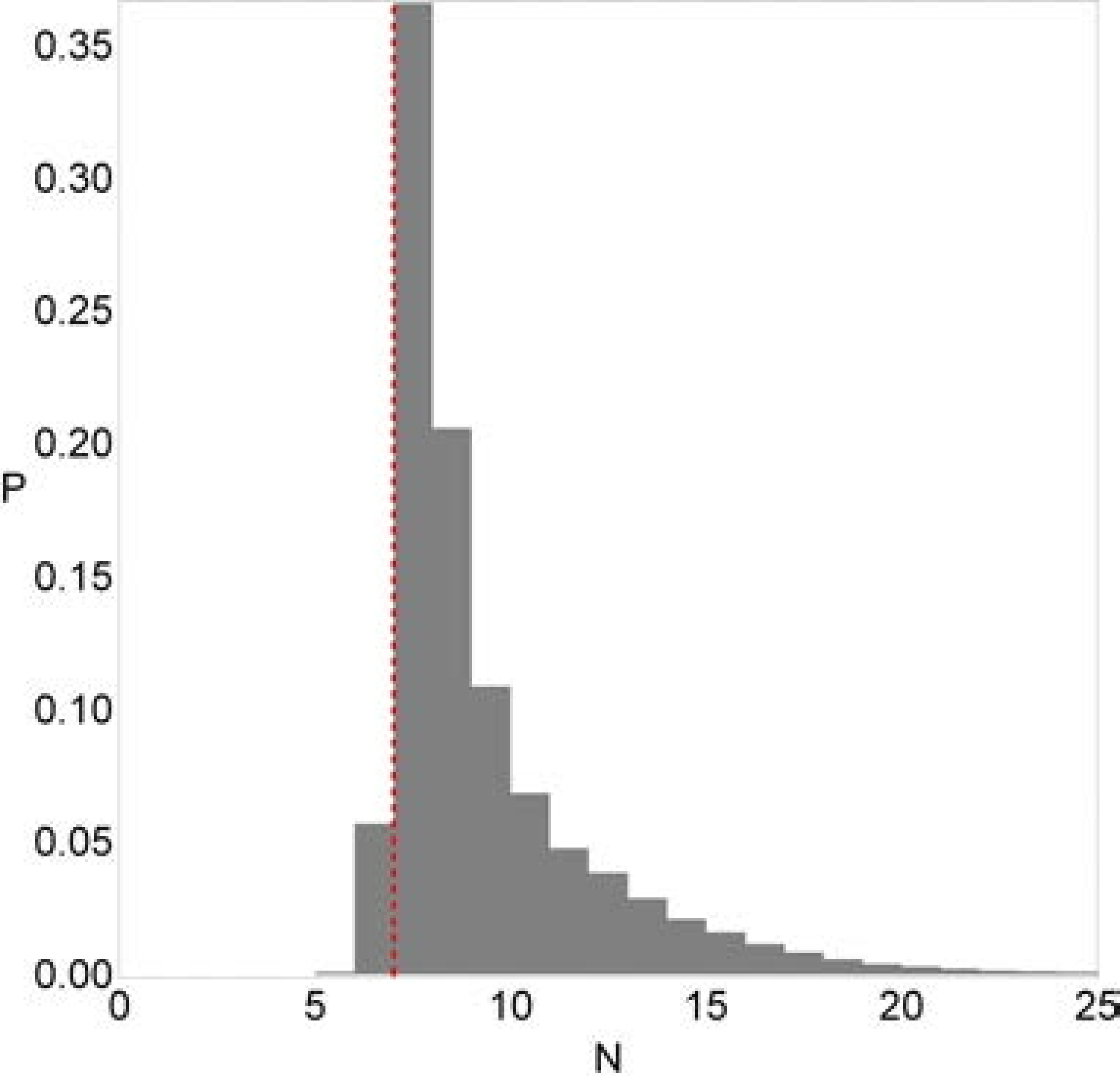}}
\end{center}
\caption{(First column): The Newton-Raphson basins of attraction on the configuration $(x,y)$ plane for the case-II, where sixteen
libration points are present. (a) $\epsilon=0.694528$, (d) $\epsilon=0.813528$,  (g) $\epsilon=0.913528$, (j) $\epsilon=0.919528$, (m) $\epsilon=0.927125$. The positions of the libration points are denoted by black dots. The color code, denoting the 16 attractors ($L_{1}- L_{16}$) is the same as in Fig. (\ref{fig:11}). (Second column): The distribution of the corresponding number $N$ of required iterations for obtaining the Newton-Raphson basins of convergence. The non-converging points are shown in white.  (Third column): The corresponding
probability distribution of required iterations for obtaining the Newton-Raphson basins of attraction.
The vertical dashed red line indicates, in each case, the most probable number $N^{*}$ of iterations. (Color figure online).}
\label{fig:12}
\end{figure*}

The case when $\epsilon \in ( 0.67752839, 0.704528]$ and $\epsilon \in (0.812528, 0.929528]$, there are sixteen libration points: four lie on $x-$axis and remaining twelve libration points on the $(x, y)$-plane. In Fig. \ref{fig:12}(a-d), we present the Newton-Raphson basins of convergence for different values of the transition parameter $\epsilon$. Moreover, the size of the color-coded diagrams (i.e., the minimum and the maximum values of the coordinates $x$ and $y$) is taken differently in different panels in order to have a complete view regarding the topology of the basins of attraction acquired by the different attracting domains. For $\epsilon=0.694528$, panel (a) of Fig.\ref{fig:12} reveals that the geometry of the Newton-Raphson basins of convergence is well formed and finite, corresponding to all the libration points except $L_4$. We further observed that the extent of all the attracting domains changes with the change in the transition parameter. Therefore, we may claim as the value of the transition  parameter varies in described intervals, the topology of the Newton-Raphson basins of convergence changes significantly. It is also noticed that the geometry of the Newton-Raphson basins of convergence also admits three axes of symmetry $y=0, y = \sqrt{3}$ and $y = -\sqrt{3}$ for all the values  of transition parameter. \\
Following are the most crucial changes observed:
\begin{description}
  \item[--] The extent of the basins of attraction associated with the collinear libration point $L_4$ is infinite.
  \item[--] The extent of the attracting domains associated with the libration points $L_2, L_{11},$ and $L_{12}$ increases.
  \item[--] The extent of the basins of attraction associated with the libration points $L_3, L_9$, and $L_{10}$ decreases.
\end{description}
The basins of convergence in all the panels may be divided into exterior and interior regions. The interior regions are the regions enclosed by the basins of convergence corresponding to the libration points $L_1$(red), $L_5$(yellow), and $L_6$(persian cyan), while the exterior region is the remaining area of the basins of convergence including boundary of the ball shaped finite region. We further observed as the transition parameter increases, the outer region constitutes three branches of basins of convergence and  for $\epsilon= 0.927125$ these branches (the major parts are shown by cyan, blue, and orange colours) separate from the interior region, and the boundary between them are just the chaotic mixture of the initial conditions.
\begin{figure*}[!t]
\begin{center}
(a)\resizebox{0.25\hsize}{!}{\includegraphics*{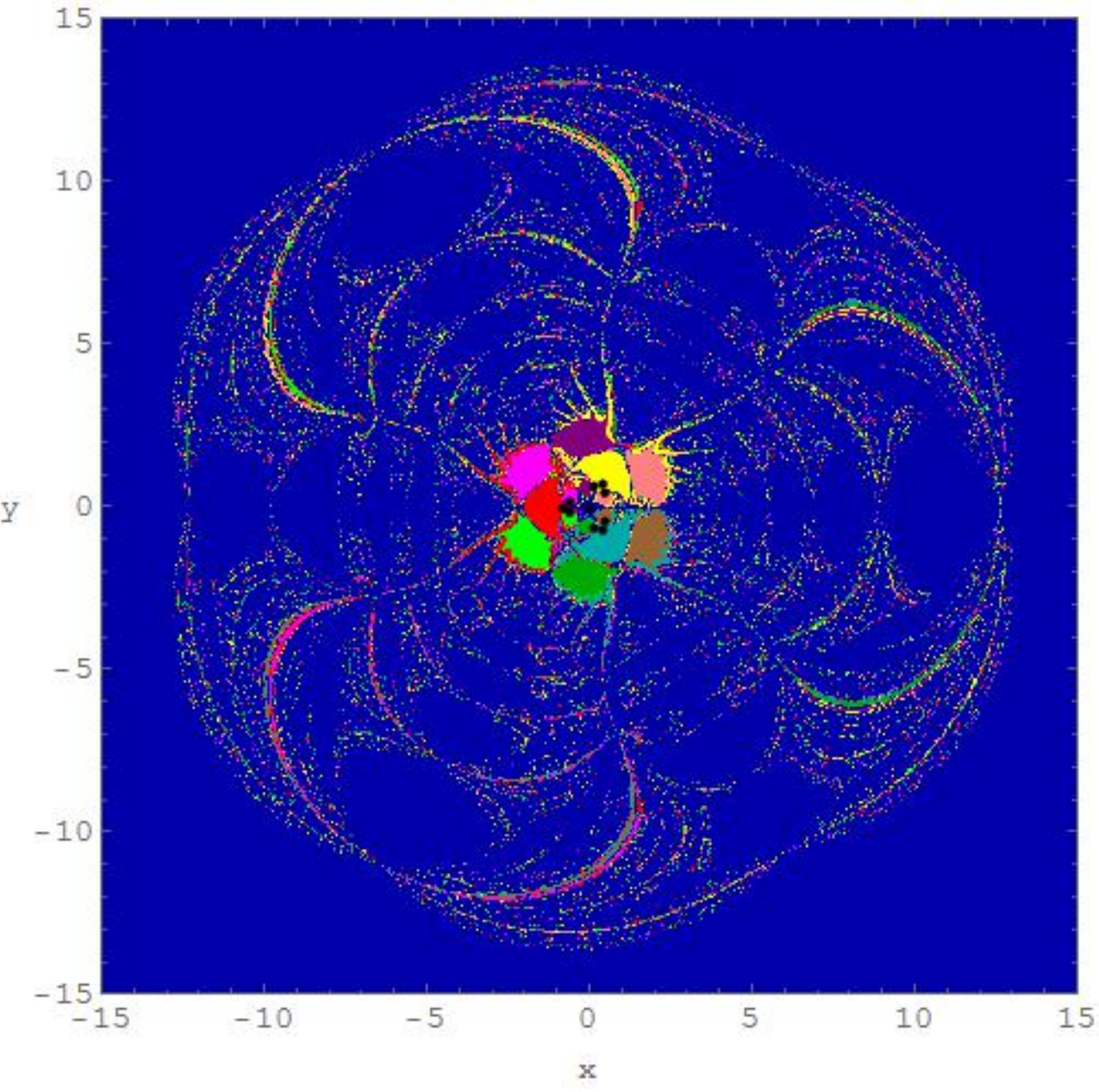}}
(b)\resizebox{0.35\hsize}{!}{\includegraphics*{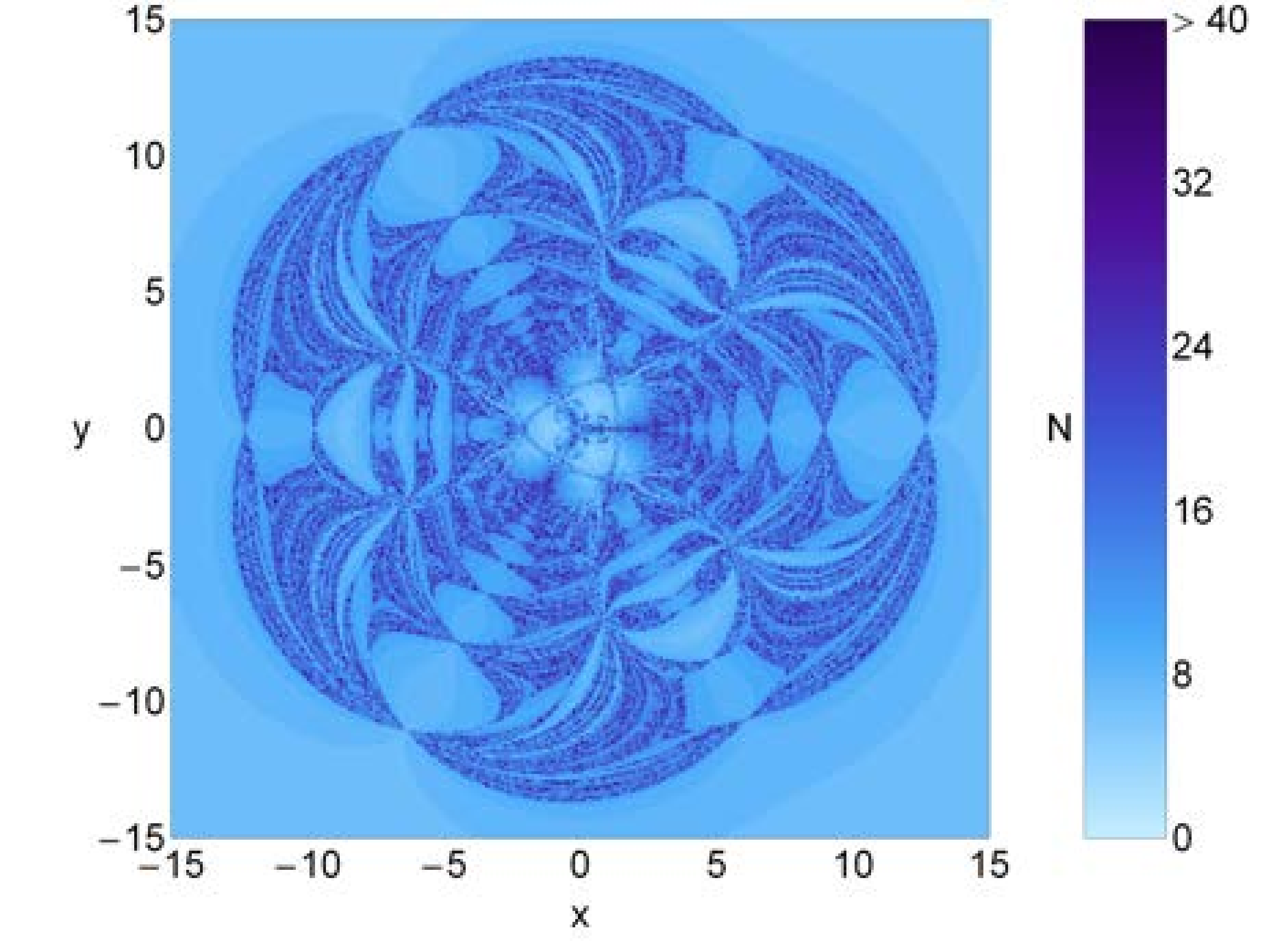}}
(c)\resizebox{0.25\hsize}{!}{\includegraphics*{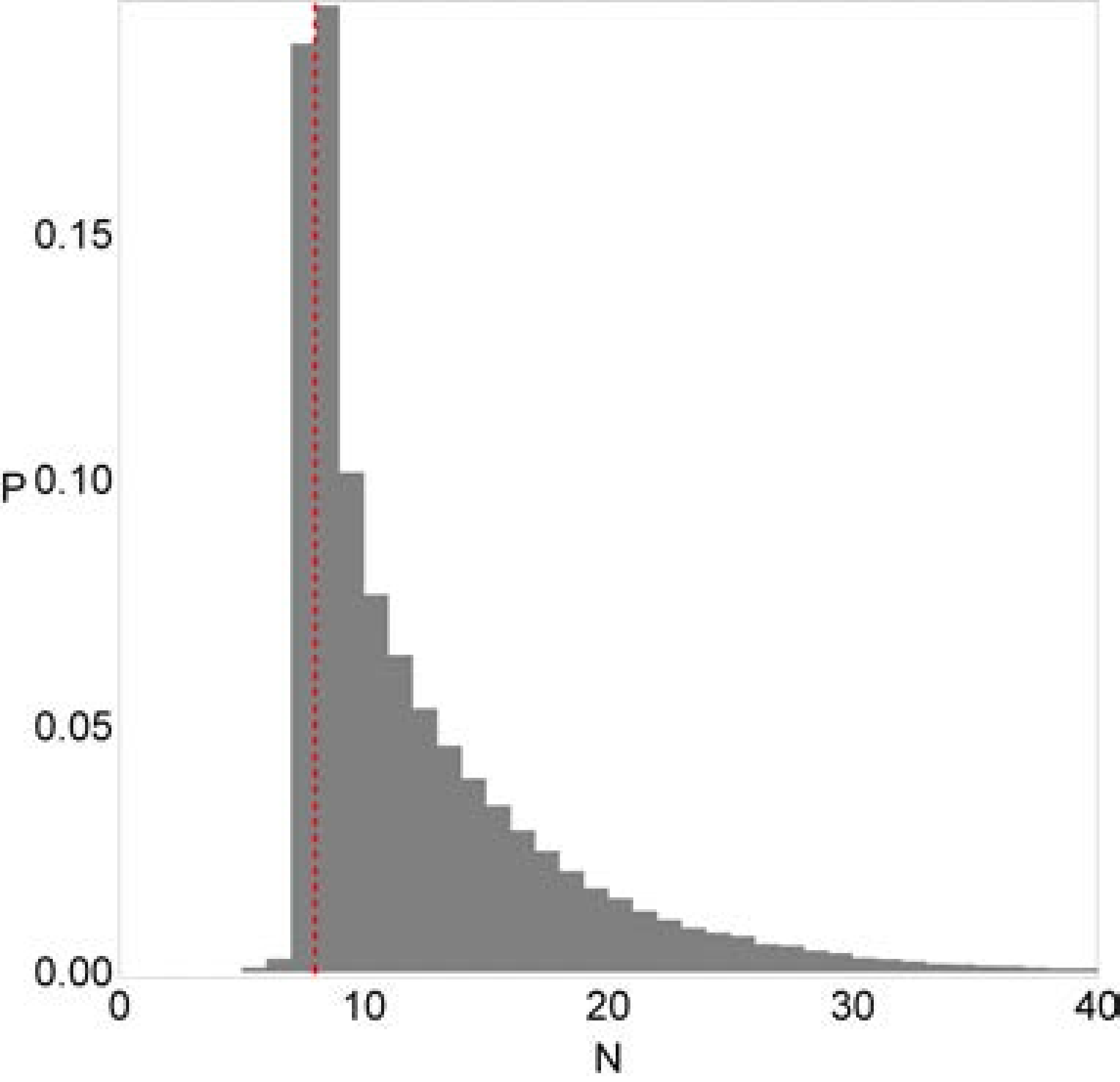}}\\
(d)\resizebox{0.25\hsize}{!}{\includegraphics*{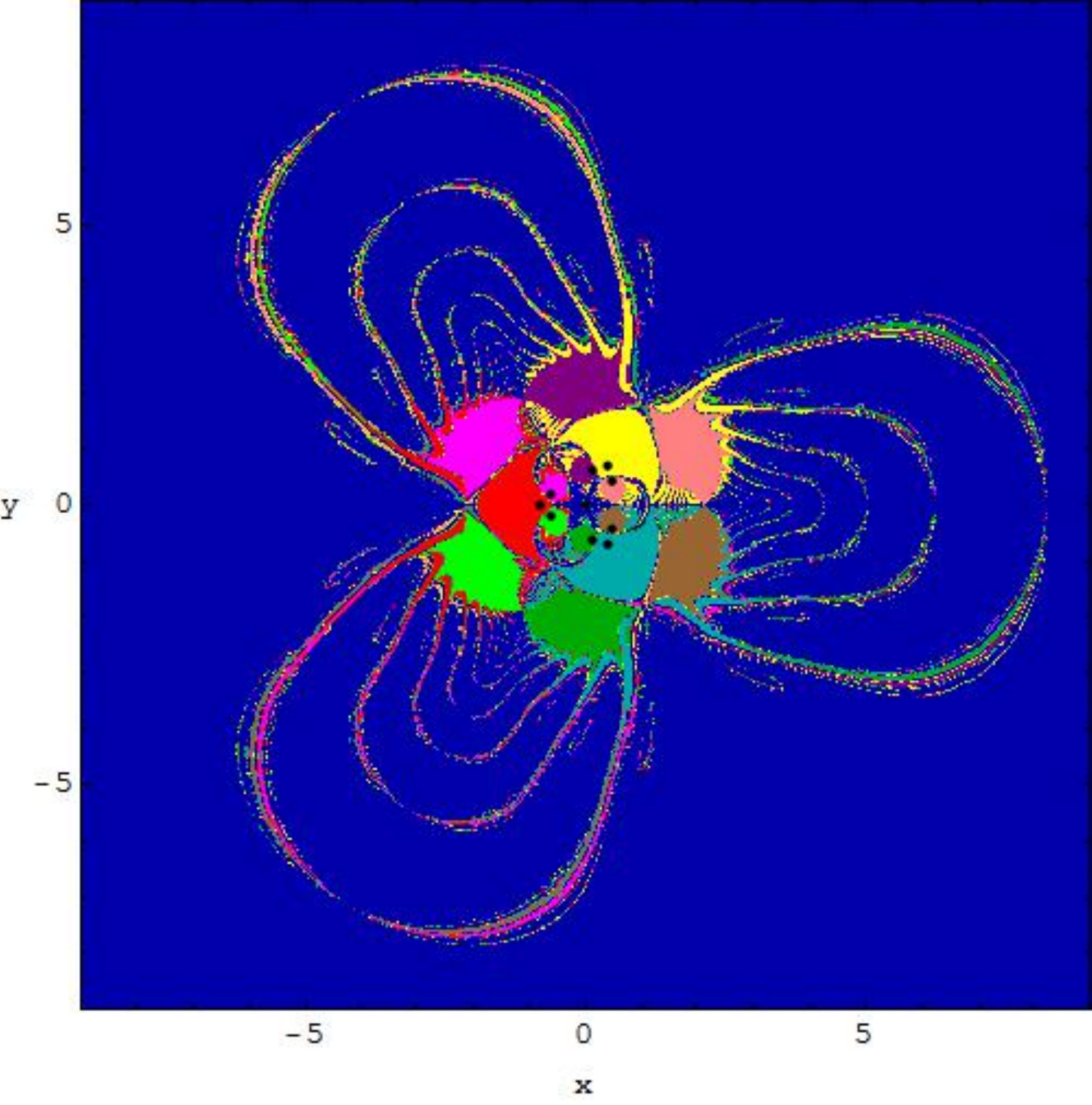}}
(e)\resizebox{0.35\hsize}{!}{\includegraphics*{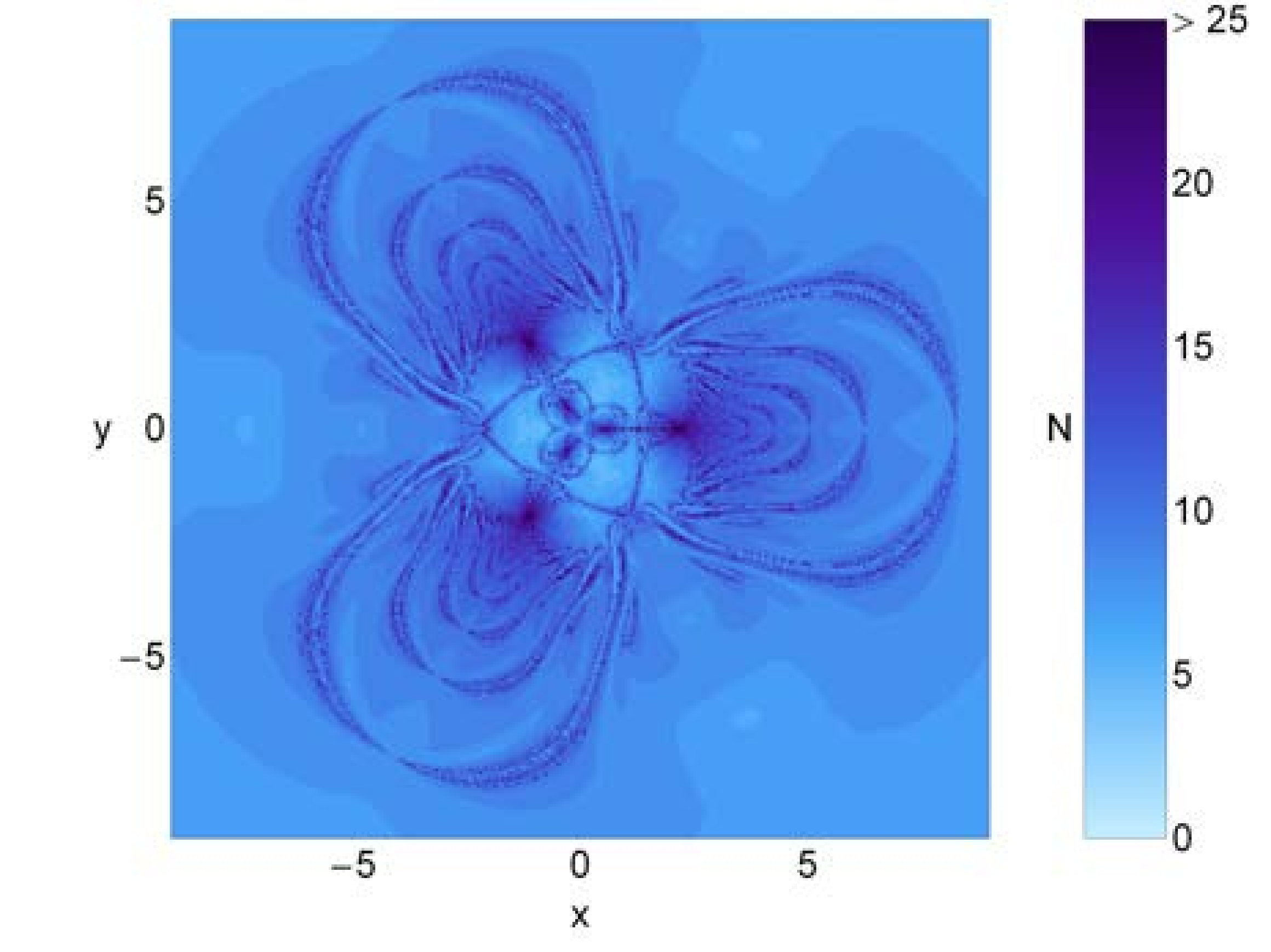}}
(f)\resizebox{0.25\hsize}{!}{\includegraphics*{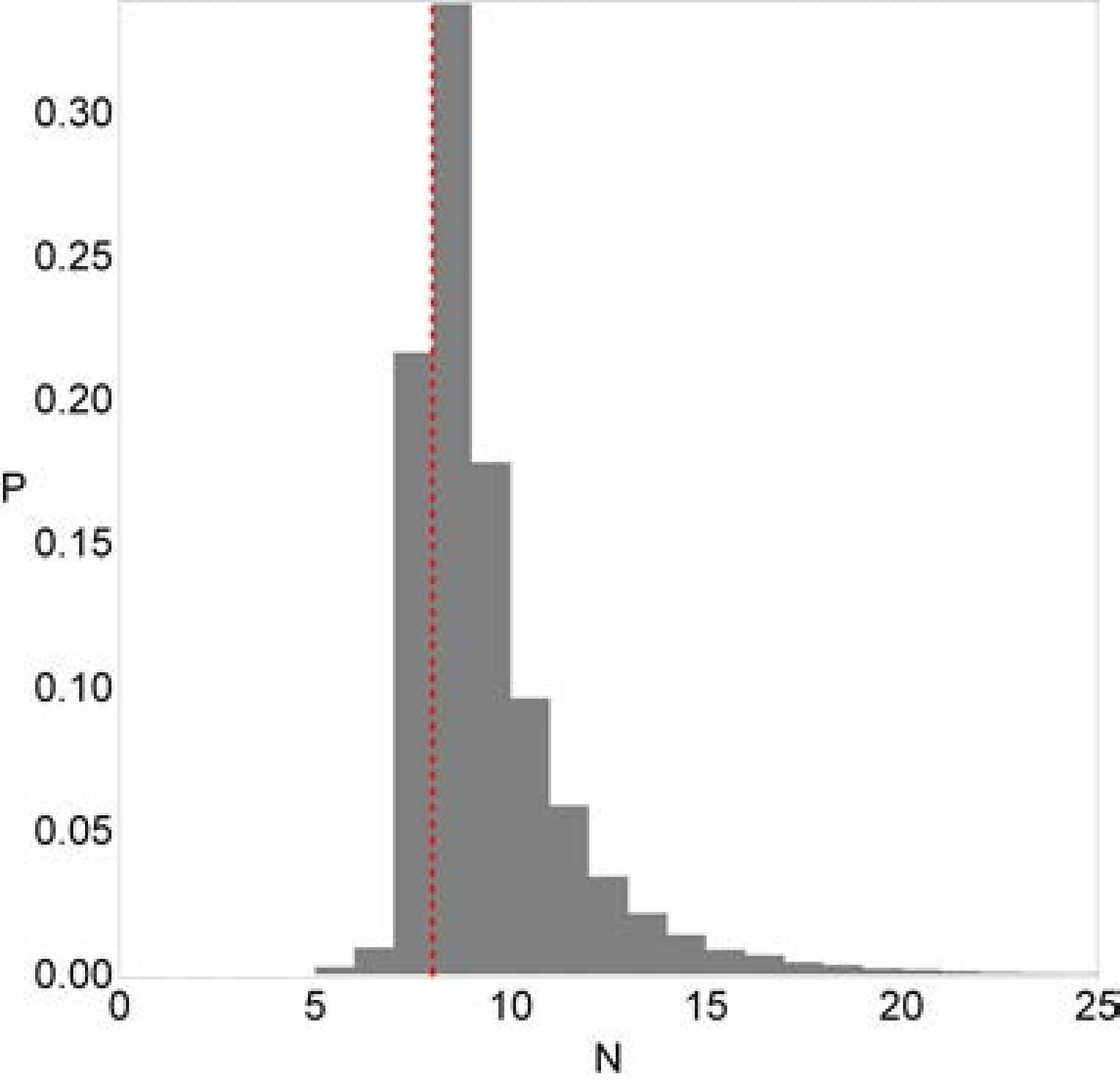}}
(g)\resizebox{0.25\hsize}{!}{\includegraphics*{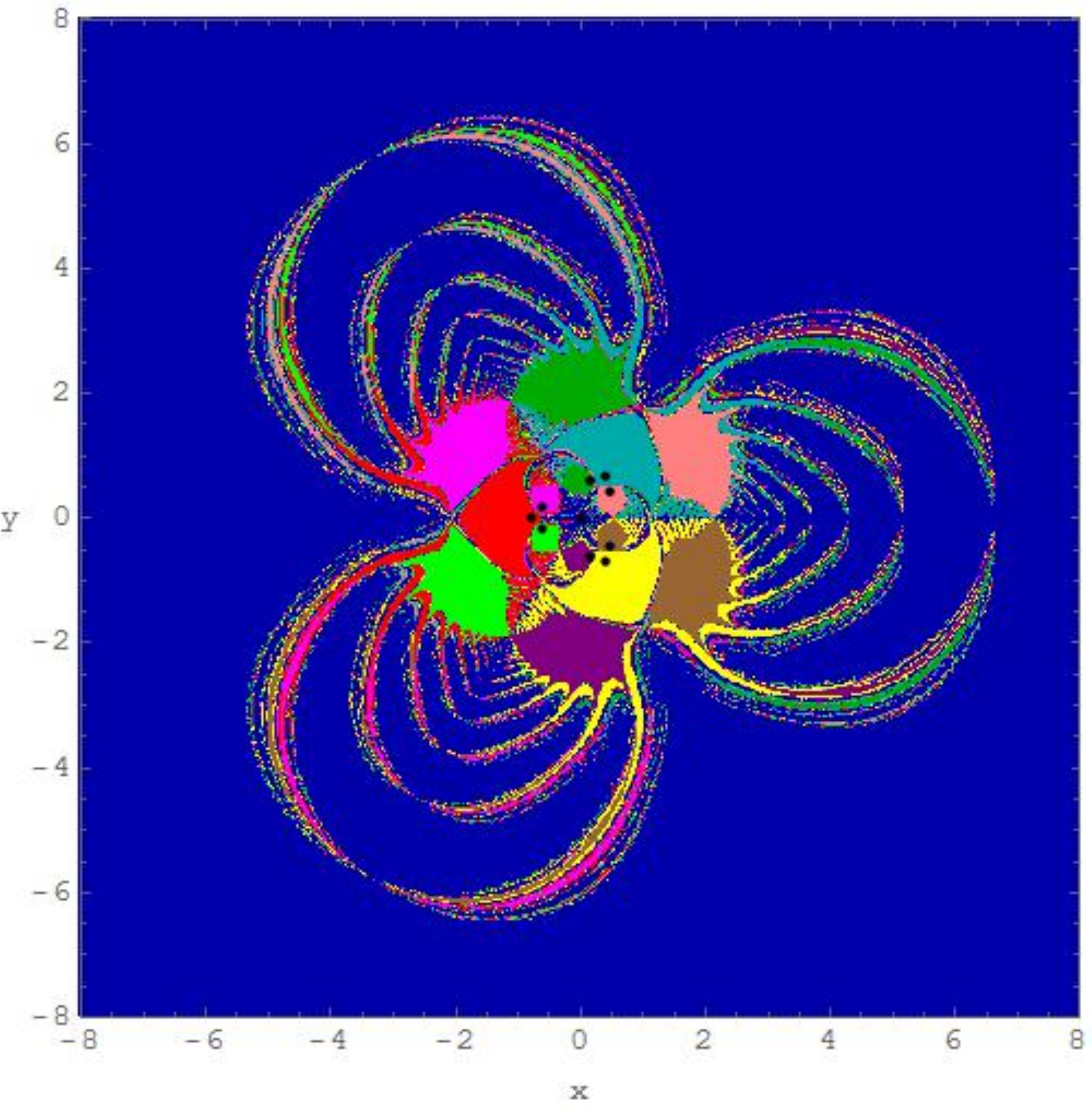}}
(h)\resizebox{0.35\hsize}{!}{\includegraphics*{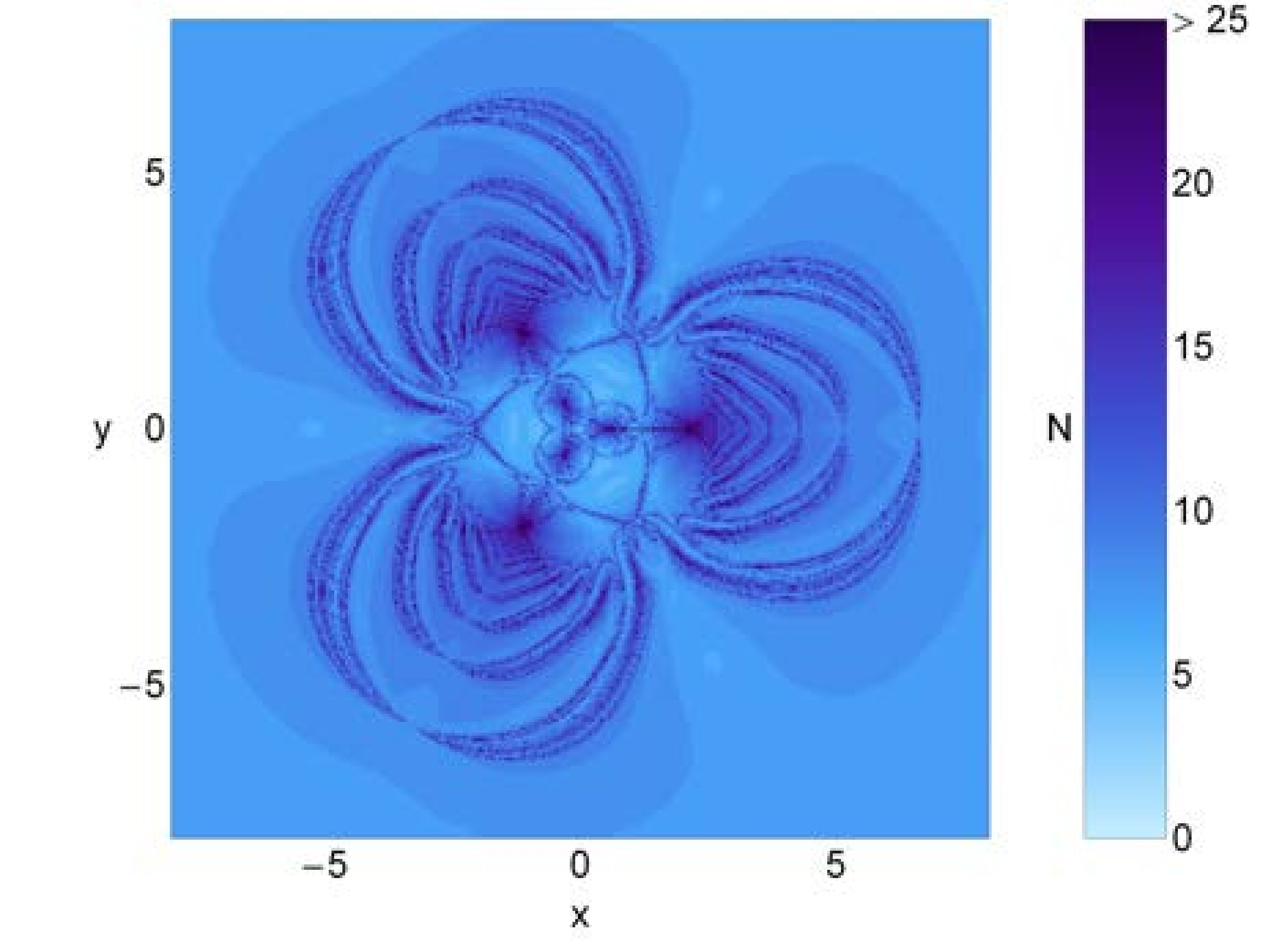}}
(i)\resizebox{0.25\hsize}{!}{\includegraphics*{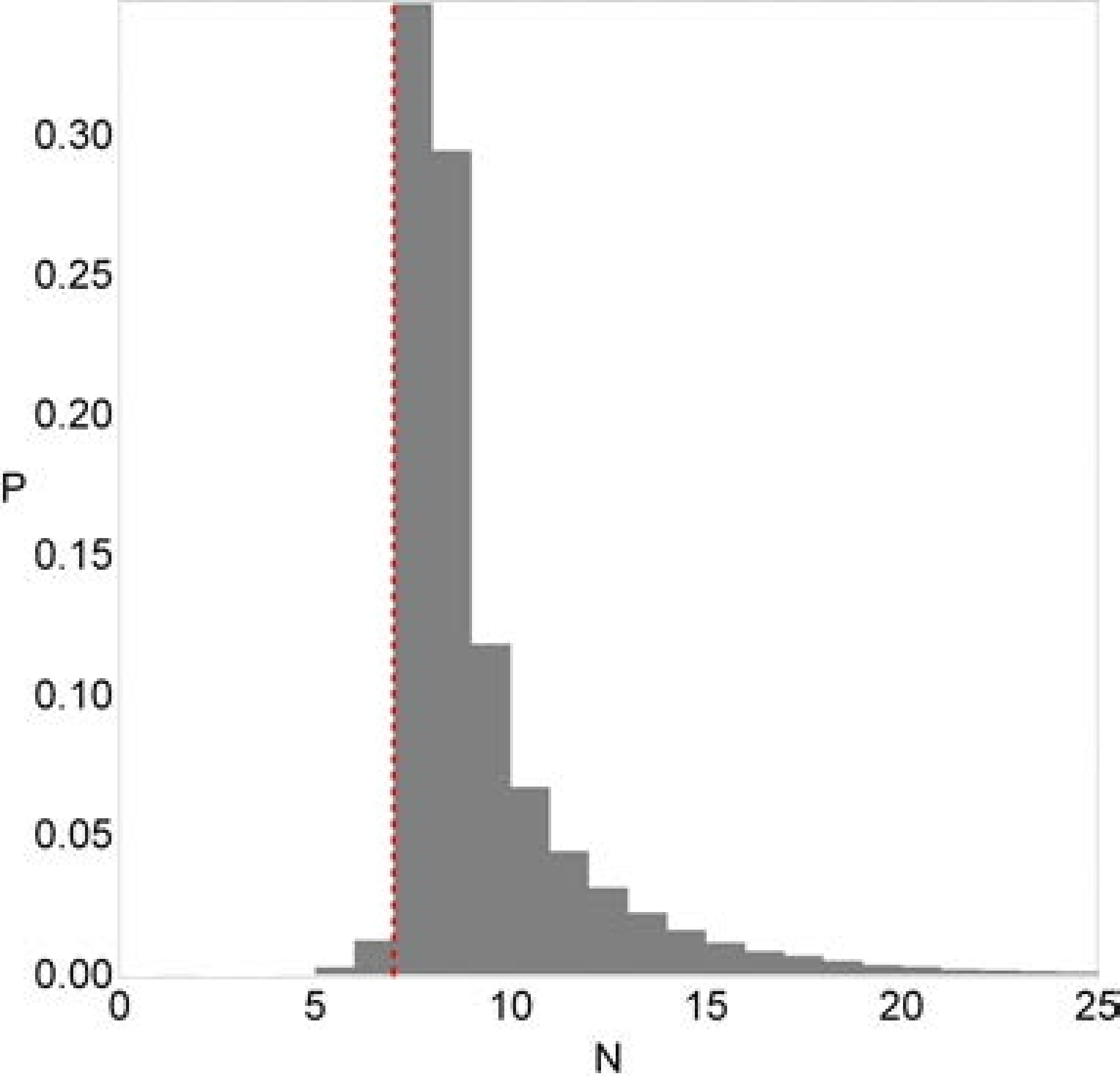}}
\end{center}
\caption{The Newton-Raphson
basins of attraction on the
configuration (x, y) plane for
the case-III, where ten
libration points are present. (a) $\epsilon=0.935967$, (d) $\epsilon=0.957285$ , (g) $\epsilon=0.999555$. The positions of the libration
points are denoted by black
dots. The color code, denoting
the 10 attractors ($L_{1}- L_{10}$) is the same
as in Fig. (\ref{fig:11}). (Second column): The distribution of the corresponding number $N$ of required iterations for obtaining the Newton-Raphson basins of convergence. The non-converging points are shown in white.  (Third column): The corresponding
probability distribution of required iterations for obtaining the Newton-Raphson basins of attraction.
The vertical dashed red line indicates, in each case, the most probable number $N^{*}$ of iterations. (Color figure online).}
\label{Fig:16}
\end{figure*}
\subsection{Case III: Ten libration points} \label{Case III: Ten libration points}
The last case corresponds to  $0.929528<\epsilon\leq1$, when there exist only ten libration points. In Fig. (\ref{Fig:16}), we have presented the Newton-Raphson basins of convergence for three values of the transition parameter. In panel (a), we have observed a very interesting and unexpected phenomenon. It is observed that most of the configuration plane is occupied by the initial conditions corresponding to collinear libration point $L_2$ which has infinite extent. On the other hand, the basins of convergence corresponding to all other libration points are finite. There exists a chaotic sea of initial conditions in the vicinity of the boundaries of the finite region of basins of convergence. From panels (d, g) of Fig.  (\ref{Fig:16}), we may observe as the value of transition parameter increases, the chaotic sea composed of the initial conditions shrinks and turn into a shape which looks like 3-leaves of a flower. For further increase in transition parameter, these leaves shrink further and become almost circular in shape.
In Fig. \ref{Fig:16} (b, e, h), the distribution of the corresponding number $N$ of the iterations are illustrated. The analysis unveils that the initial conditions inside the attracting regions converge relatively fast $(N <8)$ whereas the initial conditions inside the chaotic sea needs more iterations to converge one of the attractors. . The corresponding probability distributions of iterations are presented in Fig. \ref{Fig:16}(c, f, i).

\section{Discussion and conclusions}\label{Discussion}
The aim of this study was to unveil the effect of transition parameter $\epsilon$  on the position as well as on the linear stability of the libration points. We determined the regions of possible motion and computed numerically the basins of convergence,  associated with the libration points in the pseudo-Newtonian planar circular restricted four-body problem, when the primaries have equal masses.  The multivariate version of the Newton-Raphson iterative scheme was used to reveal the topology of the basins of convergence on various two-dimensional planes. We successfully completed the task by monitoring how the Newton-Raphson basins of
convergence  changes with respect to the transition parameter.

Following are the most important conclusions of our analysis:
\begin{enumerate}
  \item The existence as well as the number of the libration points strongly depend upon the value of the transition parameter. There exist either 22 or 16 or 10 libration points for different intervals of the transition parameter.
  \item The linear stability analysis reveals that most of the libration points of the dynamical system are always linearly unstable. The libration point $L_4$ is always linearly stable, while the libration points $L_{11}, $ $L_{12}, L_{15},$ $L_{16}, L_{21}$, and $L_{22}$ are linearly stable for those values of transition parameter for which twenty-two libration points exist. Moreover, $L_{2, 3, 9, 10}$ are linearly stable for those values of transition parameter for which  sixteen libration points exist while on the other hand none of the libration point is stable except the libration point $L_2$  for the interval of transition parameter for which  ten libration points exist.
  \item The regions of possible motion decrease with the increase in the value of the transition parameter as well as Jacobian constant.
  \item The Newton-Raphson basins of convergence revealed that the configuration $(x, y)$ plane is composed of a complicated mixture of attracting domains with highly fractal basins boundaries. It is almost impossible to predict in advance the final state of an initial condition in the neighborhood of the basins boundaries, where the degree of fractality is high.
  \item In all the examined cases, the basins of attraction, corresponding to central collinear libration points i.e. ($L_4$ for those values of $\epsilon$ for which $22$ or $16$ and $L_2$ for those values of $\epsilon$ for which $10$ libration points exist), extend to infinity while on the other hand, the domain of the basins of convergence associated to the remaining libration points are always finite.
  \item The most probable number of iterations $N^*$  in the configuration $(x, y)$  plane was found 7, mainly for those values of the transition parameter for which 22 or 16 libration points exist, while the number of iterations vary between 9 to 7 for those values of transition parameter for which 10 libration points exist.
\end{enumerate}
It should be noted that all the calculations and graphics have been created using the latest version 11 of Mathematica$^\circledR$ (\cite{wol03}). Our Mathematica code requires approximately 2hrs of CPU time on an Intel$^\circledR$ Core(TM) i5 2.67 GHz PC, for the numerical calculations of the sets of initial conditions on the $(x, y)$ plane. This required time may increase or decrease, depending of course on the number of iterations required for predefined accuracy.
In future, we can use other types of iterative schemes to compare the similarities as well as the differences on the corresponding basins of convergence. We believe that it would certainly reveal some very useful and unexpected results corresponding to the attracting domains of the libration points in the restricted four-body problem.
\section*{Acknowledgments}
\footnotesize
The authors are thankful to Center for Fundamental Research in Space dynamics and Celestial mechanics(CFRSC), New Delhi, India for providing research facilities.\par

The authors would like to express their warmest thanks to the anonymous referee for the careful reading of the manuscript and for all the apt suggestions and comments which permitted us to improve both the quality and the clarity of the paper.\\
\textbf{Compliance with Ethical Standards}
\begin{description}
  \item[-] Funding: The authors state that they have not received any research
grants.
  \item[-] Conflict of interest: The authors declare that they have no conflict of
interest.
\end{description}


\begin{thebibliography}{}

\bibitem[Abouelmagd (2012)]{abo12} Abouelmagd, E.I.: Existence and stability of triangular points in the restricted three-body problem with numerical applications. Astrophys. Space Sci. \textbf{342}, 45--53 (2012)

\bibitem[Abouelmagd \& El-Shaboury (2012)]{aboandsha12} Abouelmagd, E.I. El-Shaboury, S.M.: Periodic orbits under combined effects of oblateness and radiation in the restricted problem of three bodies. Astrophys. Space Sci. \textbf{341}, 331--341 (2012)

\bibitem[Aggarwal et al. (2018)]{agg18} Aggarwal, R.,  Mittal, A.,   Suraj, M. S.,   Bisht, V., The effect of small perturbations in the Coriolis and centrifugal forces on the existence of libration points in the restricted four‐body problem with variable mass. Astronomical notes. (online published)
    
\bibitem[Asique et al.(2015a)]{cha15a} Asique, M.C., Umakant, P., Hassan, M., Suraj, M.S.: On the R4BP when third primary is an oblate spheroid. Astrophys. Space Sci. \textbf{357}, 87 (2015a)
    
\bibitem[Asique et al.(2015b)]{cha15b} Asique, M.C., Umakant, P., Hassan, M.R., Suraj, M.S.: On the photogravitational R4BP when third primary is an oblate/prolate spheroid. Astrophys. Space Sci. \textbf{360}, 313 (2015b)

\bibitem[Asique et al. (2016)]{cha16} Asique, M.C., Umakant, P., Hassan, M.R., Suraj, M.S.: On the R4BP when third primary is a triaxial rigid body. Astrophys. Space Sci. \textbf{361}, 1 (2016)
    
\bibitem[Asique et al. (2017)]{cha17} Asique, M.C., Umakant, P.,Hassan, M.R., Suraj, M.S.: On the R4BP when third primary is an ellipsoid. J of Astronaut. Sci. \textbf{64}, 231 (2017) 
    
\bibitem[Baltagiannis \& Papadakis (2011)]{bal11a} Baltagiannis, A.N., Papadakis, K.E.: Equilibrium points and their stability in the restricted four-body poblem. Int. J. Bifurc. Chaos. \textbf{21}, 2179-2193 (2011)
    
\bibitem[Bhatnagar \& Hallan(1978)]{bha78} Bhatnagar, K.B., Hallan, P.P.: Effect of perturbations in Coriolis and centrifugal forces on the stability of libration points in the restricted problem. Celest. Mech. \textbf{18}, 105--112 (1978)

\bibitem[Bhatnagar \& Chawla(1979)]{bha79} Bhatnagar, K.B., Chawla, J.M: A study of Lagrangian points in the photogravitational restricted three body problem. Indian J. Appl. Math. \textbf{10} (11), 1443--1451 (1979)

\bibitem[Bhatnagar and Hallan (1983)]{bha83} Bhatnagar, K.B., Hallan, P.P.: The effect of perturbations in Coriolis and centrifugal forces on the nonlinear stability of equilibrium points in the restricted problem of three bodies. Celestial Mechanics, \textbf{30}: 97 (1983)
    
\bibitem[Brumberg(1972)]{bru72} Brumberg, V.A.: Relativistic Celestial Mechanics. Nauka, Moscow (1972)

\bibitem[Cheng and She(2017)]{che17}Cheng, X., She, Z.: Study on chaotic dynamics of the restricted four-body problem with an equilateral triangle configuration.  Int. J. Bifurcation Chaos 27, 1750026 (2017).

\bibitem[Contopoulos(1976)]{con76} Contopoulos, G., Kotsakis, D.: In Memoriam D. Eginitis, Athens, 159 (1976)

\bibitem[Croustalloudi \& Kalvouridis (2007)]{cro07} Croustalloudi, M., Kalvouridis, T.J.: Attracting domains in ring-type N-body formations. Planetary and Space Science \textbf{55}, 53--69 (2007)
    
\bibitem[Croustalloudi \& Kalvouridis (2013)]{cro13}Croustalloudi, M., Kalvouridis, T.J.: The restricted 2+2 body problem: Parametric variation of the equilibrium states of the minor bodies and their attracting regions. ISRN Astronomy and Astrophysics, Volume 2013, Article ID 281849 (2013)

\bibitem[Dubeibe et al.(2017a)]{dub17a} Dubeibe, F.L., Lora-Clavijo, F.D., Guillermo, A.G.: Pseudo-Newtonian planar circular restricted 3-body problem. Phys. Lett. A \textbf{381}, 563--567 (2017a)

\bibitem[Dubeibe et al.(2017b)]{dub17b} Dubeibe, F.L., Lora-Clavijo, F.D., Guillermo, A.G.: On the conservation of the Jacobi integral in the post-Newtonian
    circular restricted three-body problem. Astrophys. Space Sci. \textbf{362}, 97 (2017b)

\bibitem[Einstein et al.(1938)]{ein38} Einstein, A., Infeld, L., Hoffmann, B.: The gravitational equations and the problem of motion. Ann. Math. \textbf{39}, 65--100 (1938)

\bibitem[Huang \& Wu(2014)]{hua14} Huang, G., Wu, X.: Dynamics of the post-Newtonian circular restricted three-body problem with compact objects. Phys. Rev. D, \textbf{89}, 124-134 (2014)
    
\bibitem[Idrisi \& Taqvi (2013)]{jav13}Idrisi, M.J.,  Taqvi, Z.A.: Restricted three-body problem when one of the primaries is an ellipsoid. Astrophys Space Sci, \textbf{348}: 41 (2013)
    
\bibitem[Kumar \& Choudhry(1986)]{kum86} Kumar, V., Choudhry, R.K.: Existence of libration points in the generalised photogravitational restricted problem of three bodies. Celest. Mech. \textbf{39}, 159--171 (1986)

\bibitem[Krefetz(1967)]{kre67} Krefetz, E.: Restricted three-body problem in the post-Newtonian approximation. Astron. J. \textbf{72}, 471 (1967)

\bibitem[Mittal et al. (2016)]{mit16} Mittal, A., Aggarwal, R., Suraj, M.S., Bisht, V.S.: Stability of libration points in the restricted four-body problem with variable mass. Astrophys. Space Sci. \textbf{361}, 329 (2016)
    
\bibitem[Mittal et al. (2018)]{mit18}Mittal, A., Aggarwal, R., Suraj, M.S.,  Arora, M., On the photo-gravitational restricted four-body problem with variable mass, Astrophys. Space Sci. \textbf{363} 109 (2018)
    
\bibitem[Papadouris \& Papadakis(2013)]{pap13} Papadouris, J.P., Papadakis,  K.E.: Equilibrium points in the photogravitational restricted four-body problem. Astrophys. Space Sci. \textbf{344}, 21--38 (2013)

\bibitem[Papadakis(2016)]{pap16} Papadakis,K.E.: Families of three dimensional periodic solutions in the circular restricted four-body problem. Astrophys. Space Sci. \textbf{361}, 129 (2016)

\bibitem[Sharma \& Subba Rao(1975)]{sha75} Sharma, R.K., Subba Rao, P.V.: Collinear equilibria and their characteristic exponents in the restricted three-body problem when the primaries are oblate spheroids. Celest. Mech. \textbf{12}(2), 189--201 (1975)
    
\bibitem[Shalini et al.(2017)]{sha17} Shalini, K., Suraj, M.S., Aggarwal, R.: The Nonlinear Stability of L4 in the R3BP when the Smaller Primary is a Heterogeneous Spheroid. J of Astronaut. Sci. \textbf{64}, 18(2017)
    
\bibitem[She et al.(2013)]{she13} She, Z., Cheng, X., Li, C.: The existence of transversal homoclinic orbits in a planar circular restricted four-body problem. Celest Mech Dyn Astr (2013) 115: 299
    
\bibitem[Suraj et al.(2014)]{Sur14} Suraj, M.S., Hassan, M.R., Asique, M.C.: The photo-gravitational R3BP when the primaries are heterogeneous spheroid with three layers. J. of Astronaut. Sci. \textbf{61}, 133 (2014)
    
\bibitem[Suraj and Hassan(2014)]{Sur14b} Suraj, M.S., Hassan, M.R.: Sitnikov  restricted four-body problem with radiation pressure. Astrophys. Space Sci. \textbf{349}(2):705-716 (2014)
    
\bibitem[Suraj et al.(2017a)]{Sur17a} Suraj, M.S., Aggarwal, R., Arora, M.: On the restricted four-body problem  with the effect of small perturbations in the Coriolis and centrifugal forces.  Astrophys. Space Sci. \textbf{362}:159 (2017a)
    
\bibitem[Suraj et al.(2017b)]{Sur17b} Suraj, M.S., Asique, M.C., Prasad, U. et al.: Fractal basins of attraction in the restricted four-body problem when the primaries are triaxial rigid bodies.  Astrophys. Space Sci. \textbf{362}:211 (2017b)

\bibitem[Suraj et al.(2018a)]{Sur18a}Suraj, M.S., Mittal, A., Arora, M., Aggarwal, R., Exploring the fractal basins of convergence in the restricted four-body problem with oblateness, Int. J. Nonlinear Mech. \textbf{102} (2018a) 62--71.

\bibitem{sur18b} Suraj, M.S., Zotos, E.E., Kaur, C., Aggarwal, R., et al.: Fractal basins of convergence of libration points in the planar Copenhagen problem with a repulsive quasi-homogeneous Manev--type potential.  Int. J. Non-Linear Mech. 103 (2018b) 113-127.

\bibitem[Wolfram (2003)]{wol03}  Wolfram, S.: The Mathematica Book. Wolfram Media, Champaign (2003)

\bibitem[Zotos (2016a)]{zot16a} Zotos, E.E.: Fractal basins of attraction in the planar circular restricted three-body problem with oblateness and radiation pressure. Astrophys. Space Sci. \textbf{361}, 181 (2016a)
    
\bibitem[Zotos (2016b)]{zot16b} Zotos, E.E.: Escape and collision dynamics in the planar equilateral restricted four-body problem. International Journal of Non-Linear Mechanics \textbf{86}, 66--82 (2016b)

\bibitem[Zotos (2017a)]{zot17a} Zotos, E.E.: Revealing the basin of convergence in the planar equilateral restricted four-body problem. Astrophys. Space Sci. \textbf{362}, 2 (2017a)

\bibitem[Zotos (2017b)]{zot17b} Zotos, E.E.: Determining the Newton-Raphson basins of attraction in the electromegnetic Copenhagen problem.  International Journal of Non-Linear Mechanics \textbf{90}, 111--123 (2017b)

\bibitem[Zotos (2017c)]{zot17c} Zotos, E.E.: Basins of convergence of equilibrium points in the pseudo-Newtonian planar circular restricted three-body problem.  Astrophys. Space Sci. \textbf{362}, 195 (2017c)
    
\bibitem[Zotos (2017d)]{zot17d} Zotos, E.E.: Comparing the fractal basins of attraction in the Hill problem with oblateness and  radiation.   Astrophys. Space Sci. \textbf{362}, 190 (2017d)
    
\end{thebibliography}
\end{document}